\documentclass{article} 
\pdfoutput=1
\usepackage[T1]{fontenc}
\usepackage{lmodern}
\usepackage{slantsc}
\usepackage[protrusion=true,expansion=true]{microtype}
\usepackage{breakcites}
\usepackage{multirow}
\usepackage{amsmath,amssymb,amsfonts,amsthm}
\usepackage{subcaption}
\usepackage{graphicx}
\usepackage{fullpage}
\usepackage{setspace}
\usepackage[backref=page]{hyperref}
\usepackage[nameinlink]{cleveref}
\usepackage{color}
\usepackage{wrapfig}
\usepackage{tikz}
\usetikzlibrary{decorations.pathreplacing}
\usepackage{float}
\usepackage{algorithm}
\usepackage[noend]{algpseudocode}

\usepackage[framemethod=tikz]{mdframed}
\usepackage{xspace}
\usepackage{pgfplots}
\usepackage{framed}
\usepackage{thmtools}
\makeatletter
\IfFormatAtLeastTF{2024-11-01}{
  \renewcommand*\@addtoreset[2]{%
    \bgroup
      \edef\aliasctr@@truelist{\aliasctr@follow{#2}}%
      \let\@elt\relax
      \expandafter\@cons\aliasctr@@truelist{{#1}}%
    \egroup
    \expandafter\xdef\csname theH#1\endcsname{%
      \expandafter\noexpand\csname theH#2\endcsname.%
      \noexpand\the\noexpand\value{#1}}%
  }
}{}
\makeatother
\usepackage{thm-restate}
\usepackage{tabu}
\usepackage{fancyhdr}
\pgfplotsset{compat=1.5}

\newtheorem{theorem}{Theorem}[section]
\newtheorem{corollary}[theorem]{Corollary}
\newtheorem{lemma}[theorem]{Lemma}

\newtheorem{remark}[theorem]{Remark}

\newtheorem{observation}[theorem]{Observation}

\newenvironment{proofof}[1]{\begin{trivlist} \item {\bf Proof
#1:~~}}
  {\qed\end{trivlist}}

\newcommand{\namedref}[2]{\hyperref[#2]{#1~\ref*{#2}}}

\def\OPT{\mathcal{C}_{\mathrm{OPT}}}

\definecolor{mahogany}{rgb}{0.75, 0.25, 0.0}
\definecolor{darkblue}{rgb}{0.0, 0.0, 0.55}
\definecolor{darkpastelgreen}{rgb}{0.01, 0.75, 0.24}
\definecolor{darkgreen}{rgb}{0.0, 0.2, 0.13}
\definecolor{darkgoldenrod}{rgb}{0.72, 0.53, 0.04}
\definecolor{darkred}{rgb}{0.55, 0.0, 0.0}
\definecolor{forestgreenweb}{rgb}{0.13, 0.55, 0.13}
\definecolor{greencss}{rgb}{0.0, 0.5, 0.0}
\definecolor{bleudefrance}{rgb}{0.19, 0.55, 0.91}

\newcommand{\ignore}[1]{}

\newif\ifnotes\notestrue 
\ifnotes
\newcommand{\samson}[1]{\textcolor{blue}{{\bf (Samson:} {#1}{\bf ) }} \marginpar{\tiny\bf
             \begin{minipage}[t]{0.5in}
               \raggedright S:
            \end{minipage}}}
\newcommand{\samsonz}[1]{\textcolor{red}{{\bf (Samson:} {#1}{\bf ) }} \marginpar{\tiny\bf
             \begin{minipage}[t]{0.5in}
               \raggedright S:
            \end{minipage}}}
\newcommand{\neeraj}[1]{\textcolor{purple}{{\bf (Neeraj:} {#1}{\bf ) }} \marginpar{\tiny\bf
             \begin{minipage}[t]{0.5in}
               \raggedright N:
            \end{minipage}}} 
\newcommand{\katherine}[1]{\textcolor{forestgreenweb}{{\bf (Katherine:} {#1}{\bf ) }} \marginpar{\tiny\bf
             \begin{minipage}[t]{0.5in}
               \raggedright K:
            \end{minipage}}} 
\else
\newcommand{\samson}[1]{}
\newcommand{\samsonz}[1]{}
\newcommand{\neeraj}[1]{}
\newcommand{\katherine}[1]{}
\fi

\makeatletter
\renewcommand*{\@fnsymbol}[1]{\textcolor{mahogany}{\ensuremath{\ifcase#1\or *\or \dagger\or \ddagger\or
 \mathsection\or \triangledown\or \mathparagraph\or \|\or **\or \dagger\dagger
   \or \ddagger\ddagger \else\@ctrerr\fi}}}
\makeatother

\providecommand{\email}[1]{\href{mailto:#1}{\nolinkurl{#1}\xspace}}

\hypersetup{
     colorlinks   = true,
     citecolor    = mahogany,
	 linkcolor	  = darkpastelgreen
}

\fancypagestyle{pg}
{
\lhead{}
\rhead{}
\cfoot{--\ \thepage\ --}

}

\AtBeginDocument{%
  \DeclareFontShape{T1}{lmr}{m}{scit}{<->ssub*lmr/m/scsl}{}%
}
\usepackage{natbib}
\usepackage{booktabs}

\title{Online Geometric Optimization via Projection, Hierarchical Discretization, and Convex Relaxation}

\author{
Neeraj Gogate\thanks{Texas A\&M University.
E-mail: \email{neerajgogate@tamu.edu}.}
\and
Alisa Lu\thanks{Texas A\&M University.
E-mail: \email{alisayiranlu@tamu.edu}.}
\and
David P. Woodruff\thanks{Carnegie Mellon University and Google Research. 
E-mail: \email{dwoodruf@andrew.cmu.edu}. 
David P. Woodruff is supported in part by Office of Naval Research award number N000142112647 and a Simons Investigator Award.}
\and
Samson Zhou\thanks{Texas A\&M University.
E-mail: \email{samsonzhou@gmail.com}. 
Samson Zhou is supported in part by NSF CCF-2335411. 
}
}

\begin{document}
\maketitle
\begin{abstract}
We study online optimization problems in which a learner must choose a compressed representation before observing each incoming data point. We focus on two settings: online $k$-means clustering and online weighted low-rank approximation (WLRA). For online $k$-means, we revisit multiplicative-weights methods over discretized experts and show how marginal Johnson-Lindenstrauss projections reduce the effective dimension without requiring uniform distance preservation over the entire stream. We further decouple projected Voronoi assignment from ambient center recovery by maintaining Follow-the-Leader centroids directly in $\mathbb{R}^d$, avoiding pseudo-inverse distortion. For online WLRA, we formulate the rank-$k$ comparator class as the Grassmannian $\mathrm{Gr}(k,d)$ and develop a hierarchical region decomposition coupled with tree-structured multiplicative weights. This yields sublinear regret against the best fixed rank-$k$ subspace, while making explicit the exponential cost of Grassmannian discretization. A second Grassmannian algorithm obtains a $(1+\varepsilon)$ comparison using a number of regions independent of the horizon. We then introduce a Fantope relaxation that gives polynomial-time online updates and sublinear regret for a weighted projection surrogate against a reduced fractional comparator. Together, these results give a unified view of online approximation under nonconvex geometric structure, supported by empirical evaluations on online low-rank approximation.
\end{abstract}
\section{Introduction}
Modern applications generate unlabeled data at massive scales, driven in part by the shift of everyday activities to digital platforms. In \emph{unsupervised learning}, the goal is to discover useful patterns directly from these observations: which points are similar, what features vary together, and whether a simpler representation can capture the structure of the data~\citep{ghahramani2003unsupervised}. Such representations can make large datasets easier to explore and provide a basis for subsequent analysis. The challenge grows however when data arrives continuously. Rather than examining a complete dataset at once, a learner must use the observations available so far to build a representation that can accommodate new ones as they arrive~\citep{aggarwal2003framework}. This makes efficient, continually updated representations particularly valuable.

Two fundamental approaches for unsupervised learning are $k$-means clustering and low rank approximation. 
By representing observations with a small set of centers, $k$-means provides a simple way to group similar observations together~\citep{macqueen1967methods, lloyd1982least}.
Low-rank approximation instead captures shared structure in a small number of factors, significantly reducing the storage and computation needed to work with high-dimensional data ~\citep{eckart1936approximation}. 
These advantages have made both methods pervasive across machine learning: $k$-means-style clustering has supported applications ranging from image segmentation, where similar pixels are grouped into coherent regions~\citep{achanta2012slic}, to signal quantization, where samples are represented by a small set of prototype values~\citep{lloyd1982least}; 
while low-rank structure has powered applications ranging from recommendation systems, e.g. Netflix Prize~\citep{bell2007bellkor,bell2008bellkor,koren2009bellkor,bell2010all}, to modern foundation models, where LoRA factors weight updates into low-rank matrices to reduce the number of trainable parameters~\citep{HuSWALWWC22,XuXG0CZC0024,wu2024dlora,liloftq24}.

In this work, we study $k$-means and a weighted variant of low-rank approximation (WLRA) in an \emph{online} setting. At each round, the learner selects a representation before observing the next data point and incurring a loss based on that choice. Performance of the learner is then measured by the \emph{regret}: the difference between the learner's cumulative loss and that of the best fixed representation chosen with knowledge of the full stream.  

Formally, in the online $k$-means setting, the learner chooses a set of centers
$C_t=\{c_{t,1},\ldots,c_{t,k}\}\subseteq\mathbb{R}^d$
before observing $x_t\in\mathbb{R}^d$, and then incurs loss
$$\ell_t(C_t) := \min_{c\in C_t}\|x_t-c\|_2^2.$$ The goal is to achieve small regret relative to the best fixed set of $k$ centers chosen after the entire stream is known. This differs from the classical streaming model, where an algorithm processes the data sequentially but is evaluated only on the final clustering it produces. In the online-learning setting, the learner incurs loss at every round.

In online WLRA, a learner must choose a rank-$k$ subspace $U_t\in\mathrm{Gr}(k,d)$ before observing a vector
$x_t\in\mathbb{R}^d$ and a weight matrix $W_t$. Writing
$X_{U_t}\in\mathbb{R}^{d\times k}$ for some matrix having columns spanning $U_t$, its loss is the best weighted reconstruction error within the selected subspace:

$$\ell_t(U_t) := \min_{a\in\mathbb{R}^k} \bigl\|W_t^{1/2}(x_t-X_{U_t}a)\bigr\|_2^2.$$

The comparator is the best fixed rank-$k$ subspace in hindsight. The weights allow different coordinates to have different importance or confidence, as in applications of weighted low-rank models to collaborative filtering and computer vision~\citep{SrebroJ03,GillisG11}, but they also make the problem computationally difficult. In particular, offline WLRA is NP-hard to solve to inverse-polynomial additive accuracy, even in the rank-one case~\citep{GillisG11}. This creates a tension between competing with the exact rank-$k$ optimum and obtaining an efficient online algorithm.

A natural approach for seeking low regret in online problems is to view each candidate solution as an expert and apply multiplicative weights with techniques such as the Multiplicative Weights Update Algorithm (MWUA)~\citep{arora2012multiplicative}, incurring regret $O(\sqrt{T\log n})$, where $n$ is the number of experts~\citep{GravinPS17}. The difficulty with this approach is that the space of possible choices is continuous and nonconvex in both problems, so there are infinitely many candidate experts. One can discretize these classes, but a naive discretization may contain exponentially many experts, with the exponent depending on the ambient dimension, the intrinsic dimension of the hypothesis space, or the desired approximation scale. This leads to the central question of the paper: how can geometric structure be used to obtain meaningful online regret guarantees without paying the full cost of a naive discretization?

Our results address this question differently in the two settings. For online $k$-means, we build on the discretization and multiplicative-weights framework of Cohen-Addad et al.~\citep{Cohen-AddadGKR21}, in which candidate center sets are treated as experts. We combine this framework with dimension reduction while maintaining the actual centers directly in the ambient space. A marginal Johnson-Lindenstrauss argument shows that, at each round, it is sufficient to preserve the distances from the incoming point to the $k$ optimal centers. The projected dimension therefore depends on $k$ and the approximation parameter. Projected grid points determine Voronoi assignments, while ambient Follow-the-Leader updates maintain the corresponding centers in $\mathbb{R}^d$.

For online WLRA, we first compete with the best fixed rank-$k$ subspace in hindsight. The natural comparator space is the Grassmannian $\mathrm{Gr}(k,d)$, whose points represent subspaces independently of their choice of basis. We cover this space with bounded coordinate charts, refine the charts into dyadic regions, and apply Mass Tree MWUA to rank-$k$ representatives. A second-order analysis at the optimal comparator permits resolution $T^{-1/4}$ rather than $T^{-1/2}$, reducing the number of experts from $T^{k(d-k)/2}$ to $T^{k(d-k)/4}$ while retaining sublinear expected regret. We also obtain a $(1+\varepsilon)$-comparator guarantee by running a moving learner within each sufficiently small region and using MWUA to select among them. Every iterate remains a rank-$k$ subspace, and the number of regions is independent of $T$. The expected cumulative loss is at most $(1+\varepsilon)$ times the optimal rank-$k$ loss, plus $\varepsilon W\sqrt{T}$ and a sublinear online-learning term.

Both Grassmannian algorithms remain exponential in the intrinsic dimension $k(d-k)$. An online-to-offline reduction shows that a polynomial-time online algorithm with sufficiently small regret against the exact comparator would yield a randomized polynomial-time algorithm for an NP-hard offline accuracy target. The reduction also applies to a uniform $(1+\varepsilon)$ guarantee at inverse-polynomial $\varepsilon$, but does not establish hardness for fixed constant $\varepsilon$. To obtain polynomial-time updates, we instead optimize over a reduced Fantope. This method uses a weighted projection loss in place of the standard weighted reconstruction loss and competes with fractional projectors supported on an anchor subspace. Its sublinear regret guarantee is therefore for a different loss and comparator class.

\textbf{Paper organization.}
\Cref{contributions} summarizes our theoretical and empirical contributions.
\Cref{sec:kmeans-marginal-jl} presents the online $k$-means algorithm, including the marginal JL argument and ambient Follow-the-Leader tracking.
\Cref{sec:online-wlra} presents the Grassmannian exact-comparator and
$(1+\varepsilon)$-comparator algorithms, followed by the polynomial-time
Fantope surrogate; the appendix gives the corresponding computational
barrier for standard WLRA.
\Cref{experiments} evaluates the low-rank methods on MNIST and MovieLens,
with additional synthetic weighted experiments in the appendix.
\section{Our Contributions}
\label{contributions}

We make the following contributions for online geometric approximation, combining multiplicative weights with projection, hierarchical discretization, and convex relaxation.

\textbf{Online $k$-means via marginal JL projections.}
Building on the discretized-expert framework of Cohen-Addad et al.~\citep{Cohen-AddadGKR21}, we use a marginal Johnson-Lindenstrauss argument to construct the expert class in a reduced dimension.  At fixed
$\varepsilon$, this replaces $d$ in the exponent of a direct $kd$-dimensional center grid with a projected dimension depending only on $k$ and $\varepsilon$; the expert class has no explicit dependence on $d$ or $T$, although its updates still process
$d$-dimensional points. \Cref{thm:projected-kmeans} gives the resulting expected-loss guarantee.

\textbf{Online WLRA via Grassmannian algorithms.}
We give two algorithms for standard weighted low-rank approximation that compete with rank-$k$ subspaces in $\mathrm{Gr}(k,d)$. Bounded graph charts let us discretize $k(d-k)$ coordinates rather than all $dk$ entries of a basis matrix. Our first algorithm refines these charts into dyadic regions and applies Mass Tree MWUA to fixed
representatives. A second-order analysis at the cumulative optimum permits resolution $T^{-1/4}$ instead of the $T^{-1/2}$ given by a first-order bound. This reduces the horizon-dependent expert count from $T^{k(d-k)/2}$ to $T^{k(d-k)/4}$ while achieving sublinear expected regret against the exact comparator (\Cref{thm:grassmannian-hrd}).

Our second algorithm runs a moving learner within each sufficiently small chart region and uses MWUA to select among regions. For positive diagonal weights, a locally convex quadratic surrogate yields expected loss at most $(1+\varepsilon)$ times that of the best fixed rank-$k$
subspace, plus sublinear additive terms (\Cref{thm:grassmannian-approx}). Its number of regions is independent of $T$ at fixed $\varepsilon$. Both algorithms play rank-$k$ subspaces, but their cost remains exponential in $k(d-k)$ .

\textbf{Online-to-offline computational barrier.}
Our reduction shows that a polynomial-time online algorithm with exact-comparator regret
$cT^{1-\alpha}(\log(eT))^b$, where $\alpha, b>0$ are fixed and $c$ is horizon-independent and polynomial in the instance parameters, would yield a randomized polynomial-time algorithm for inverse-polynomial additive accuracy in offline WLRA (\Cref{thm:offline-to-online,cor:wlra-computational-barrier}). The positive-weight hardness result of Gillis and Glineur~\citep{GillisG11} rules this out unless $\mathrm{NP}\subseteq\mathrm{BPP}$. The barrier also applies to uniform $(1+\varepsilon)$ guarantees at inverse-polynomial $\varepsilon$.

\textbf{Polynomial-time surrogate for online WLRA.}
Motivated by this barrier, we replace the standard weighted reconstruction loss with a weighted projection loss and the exact rank-$k$ comparator with a fractional projection model over the Fantope. Restricting the fractional projector to an anchor subspace of dimension $m$ gives an exact reduced expression for the surrogate loss and its gradient. This leads to a polynomial-time online gradient method with per-round runtime $O(m^2d+m^3)$ and sublinear regret against the best fixed fractional comparator in the reduced Fantope (\Cref{thm:fantope-ogd}). Its guarantee concerns this surrogate loss and comparator class, rather than the standard-WLRA optimum.

\textbf{Empirical evaluations.}
We evaluate implementations of both Grassmannian
algorithms and Fantope OGD on real and synthetic streams. On MNIST
and MovieLens, both Grassmannian implementations have slightly higher
mean reconstruction loss than FTRL, with differences below $0.71\%$,
and lower mean loss than the tested untuned online-PCA baselines.
On WeightedImportanceCycleNoisy and WeightedShift, they reduce mean
weighted fitting loss by approximately $26\%$ and $47\%$,
respectively, relative to approximate weighted FTRL. In a separate
heteroscedastic subspace-tracking experiment, rounded Fantope OGD
reduces mean weighted fitting loss by about $10\%$ relative to
approximate weighted FTRL and $58\%$ relative to its weight-ignoring
ablation. Additional recovery diagnostics examine a fixed generating
subspace. These findings concern the reported implementations and
selected experimental settings.

\section{Online \texorpdfstring{$k$}{k}-Means} 
\label{sec:kmeans-marginal-jl}

We first present a finite-dimensional version of our approach in the online $k$-means setting. Following the online $k$-means formulation of \cite{Cohen-AddadGKR21}, at each round $t$ the learner chooses a set of centers $C_t=\{c_{t,1},\ldots,c_{t,k}\}\subset \mathbb{R}^d$ before observing $x_t$, and then incurs $\ell_t(C_t) := \min_{c\in C_t}\|x_t-c\|_2^2.$

Let $C^\star=\{c_1^\star,\ldots,c_k^\star\}$ denote the optimal fixed set of centers in hindsight, and write $\mathrm{\OPT} := \sum_{t=1}^T \min_{j\in[k]}\|x_t-c_j^\star\|_2^2.$
We assume throughout this section that $\|x_t\|_2\le 1$ and
$\|c_j^\star\|_2\le 1$. We also assume the stream is chosen independently of the random projection used below.
In particular, both $x_{1:T}$ and its hindsight comparator $C^\star$
are fixed before the projection is sampled. The same analysis applies
to a random stream whose distribution is independent of the projection,
by conditioning on the realized stream.

Discretizing $k$ centers directly in $\mathbb{R}^d$ gives an expert class exponential in $kd$. A Johnson-Lindenstrauss (JL) projection
offers a smaller space in which to specify the experts, but requiring
one projection to preserve all relevant distances simultaneously over
$T$ rounds introduces a $\log T$ term in its dimension.  Here we need
a weaker guarantee. The projected optimal centers need to lie in the grid box, an event
whose probability can be controlled using the $k$ fixed centers.
For each fixed round, we additionally compare the $k$ distances from
$x_t$ to those centers. Neither bound requires a union bound over the
$T$ rounds.

\subsection{Projected Voronoi Experts with Ambient FTL Tracking}

Fix $0<\varepsilon\le 1/2$. Let $\Pi\in\mathbb{R}^{m\times d}$ have independent entries distributed as
$\mathcal{N}(0,1/m)$, where
$m = \lceil C\varepsilon^{-2}\log\frac{k}{\varepsilon}\rceil$
for a sufficiently large universal constant $C$. Let $S_m$ be an $\varepsilon$-net of the box
$[-2,2]^m$ with mesh width $\varepsilon/\sqrt{m}$, so that every projected point in the box is
within Euclidean distance at most $\varepsilon$ of some grid point. We define each expert as a tuple
$Y=(y_1,\ldots,y_k)\in S_m^k.$
The grid points $y_j$ are not centers in $\mathbb{R}^d$, rather, they only define
Voronoi cells in the projected space. Each expert maintains ambient statistics
$\bigl(N_{t,j}^{Y},\mu_{t,j}^{Y}\bigr)_{j=1}^k,$
where $N_{t,j}^{Y}$ is the number of points previously routed to cluster $j$ by expert $Y$, and
$\mu_{t,j}^{Y}\in\mathbb{R}^d$ is their empirical mean.

At round $t$, MWUA samples an expert $Y_t$ from its current distribution and outputs the current
means, $C_t(Y_t) := \{\mu_{t,1}^{Y_t},\ldots,\mu_{t,k}^{Y_t}\}.$
After observing $x_t$, every expert $Y$ assigns $x_t$ according to projected Voronoi distance, $a_t^Y := \arg\min_{j\in[k]}\|\Pi x_t-y_j\|_2^2.$
The expert is then charged for predicting with the \emph{ambient} mean of the assigned cluster, $\widetilde{\ell}_t(Y)
    :=
    \|x_t-\mu_{t,a_t^Y}^{Y}\|_2^2.$
This charge upper-bounds the learner's actual loss when $Y$ is
sampled, because $\ell_t(C_t(Y)) =\min_{j\in[k]}\|x_t-\mu_{t,j}^Y\|_2^2 \leq\widetilde{\ell}_t(Y).$
Finally, the expert updates the count and mean of cluster $a_t^Y$:
\[
    N_{t+1,a_t^Y}^{Y}=N_{t,a_t^Y}^{Y}+1,
    \qquad
    \mu_{t+1,a_t^Y}^{Y}
    =
    \mu_{t,a_t^Y}^{Y}
    +
    \frac{x_t-\mu_{t,a_t^Y}^{Y}}{N_{t+1,a_t^Y}^{Y}},
\]
and all other cluster statistics remain unchanged. Thus the projected grid controls only the
assignments, while the actual centers are recovered natively in $\mathbb{R}^d$ through
Follow-the-Leader means.

\begin{algorithm}[t]
\caption{Projected Voronoi FTL for online $k$-means}
\label{alg:projected-voronoi-ftl}
\begin{algorithmic}[1]
\State Sample $\Pi\in\mathbb{R}^{m\times d}$ and construct grid $S_m\subset[-2,2]^m$.
\State Let $\mathcal{E}=S_m^k$. Initialize $w_1(Y)=1$ for all $Y\in\mathcal{E}$.
\State Initialize $N_{1,j}^{Y}=0$ and $\mu_{1,j}^{Y}=0$ for all $Y\in\mathcal{E}$ and $j\in[k]$.
\For{$t=1,\ldots,T$}
    \State Sample $Y_t$ with probability proportional to $w_t(Y)$.
    \State Output $C_t=\{\mu_{t,1}^{Y_t},\ldots,\mu_{t,k}^{Y_t}\}$.
    \State Observe $x_t$ and compute $z_t=\Pi x_t$.
    \For{each $Y=(y_1,\ldots,y_k)\in\mathcal{E}$}
        \State $a_t^Y\gets \arg\min_{j\in[k]}\|z_t-y_j\|_2^2$.
        \State $\widetilde{\ell}_t(Y)\gets \|x_t-\mu_{t,a_t^Y}^{Y}\|_2^2$.
        \State $w_{t+1}(Y)\gets w_t(Y)\exp(-\eta \widetilde{\ell}_t(Y))$.
        \State Update $N_{t,a_t^Y}^{Y}$ and $\mu_{t,a_t^Y}^{Y}$ using the ambient mean update.
    \EndFor
\EndFor
\end{algorithmic}
\end{algorithm}

\subsection{Analysis}

The guarantee follows from three simple components, whose full statements and
proofs appear in Appendix~\ref{app:kmeans}. First, discretizing the
projected optimal centers gives an expert whose routing decisions
have small expected excess assignment cost. For each fixed $t$, the
marginal JL bound controls the $k$ distances
$\|x_t-c_j^\star\|_2$ without a union bound over the rounds. Second,
for any expert's routing sequence, Follow-the-Leader means compete
with fixed ambient centers for the induced clusters up to
$O(k\log T)$.Finally, MWUA ensures the learner competes with the resulting expert up to $O(\sqrt{T\log|\mathcal{E}|})$.

\begin{restatable}{theorem}{projectedKMeans}
\label{thm:projected-kmeans}
Run Algorithm~\ref{alg:projected-voronoi-ftl} with
$\eta=\Theta(\sqrt{\log|\mathcal{E}|/T})$. Then
\[
    \mathbb{E}_{\Pi,\mathrm{alg}}
    \left[
        \sum_{t=1}^T \ell_t(C_t)
    \right]
    \le
    (1+O(\varepsilon))\OPT
    +O(\varepsilon T)
    +O\left(\sqrt{T\log|\mathcal{E}|}\right)
    +O(k\log T).
\]
Moreover,
$\log|\mathcal{E}| = O\left(km\log\frac{\sqrt{m}}{\varepsilon}\right) $ and $ m=O\left(\varepsilon^{-2}\log\frac{k}{\varepsilon}\right),$
so the expert complexity is independent of both the ambient dimension $d$ and the time horizon
$T$.
\end{restatable}
The horizon independence should be interpreted at fixed approximation
scale $\varepsilon$. For any fixed $\varepsilon$, the projected expert
class has size depending only on $(k,\varepsilon)$, not on $T$ or $d$.
If one subsequently tunes $\varepsilon$ as a function of $T$ to optimize
the additive discretization term, this dependence re-enters only through
the chosen accuracy parameter, not through a JL union bound over the stream.
\begin{remark}
For $T \geq 16$, setting $\varepsilon = T^{-1/4}$ balances the $O(\varepsilon T)$ and $O(\sqrt{T \log |\mathcal{E}|})$ terms, since $\log |\mathcal{E}| = O(k \varepsilon^{-2} \log(k/\varepsilon))$ grows as $\varepsilon^{-2}$. This gives an overall regret bound of $O(T^{3/4} \cdot \mathrm{polylog}(T))$. For smaller $T$, one can take $\varepsilon = 1/2$.
\end{remark}

\textbf{Dependence on the stream.} Independence from $\Pi$ is essential to the marginal JL argument:
once $x_t$ or $C^\star$ can depend on the sampled projection, their
difference vectors are no longer fixed when the JL bound is applied.
There is also a direct obstruction to extending this analysis to
projection-dependent data. When $m<d$, choose a unit vector
$v\in\ker(\Pi)$ and let the stream alternate between $v$ and $-v$.
Every projected point is then zero. For each expert, the routing
decision is the same on every round, so its two distinct ambient
clusters cannot track the two values, even though a fixed pair of
centers $\{v,-v\}$ has zero loss. An adversary that learns $\Pi$, or
chooses points using information about it revealed by the learner,
can therefore defeat this projected-routing argument. The theorem
does not claim a guarantee against such adaptive adversaries; handling
them would require a different projection guarantee or algorithm.

\textbf{Computational cost.} The flat implementation updates every expert on every round and has per-round time $O\bigl(md+|\mathcal{E}|(km+d)\bigr) = O\bigl(d\cdot C(k,\varepsilon)\bigr),$ where $C(k,\varepsilon)$ may be exponential in $k$ and $\varepsilon^{-1}$ but is independent of
$d$ and $T$ except for the linear cost of applying the projection. Thus the role of this section is conceptual rather than practical as marginal JL removes the horizon-dependent discretization barrier by separating the low-dimensional routing rule from the ambient
representation of the centers.

\section{Online Weighted Low-Rank Approximation}
\label{sec:online-wlra}
For standard WLRA, a rank-$k$ hypothesis is a subspace rather than a particular basis, making the Grassmannian $\mathrm{Gr}(k,d)$ the natural hypothesis space. We first build a hierarchy over this space to compete with the best fixed rank-$k$ subspace in hindsight. We then allow a $(1+\varepsilon)$ comparison by placing a convex learner inside each sufficiently small region, obtaining a number of regions independent of the horizon. Both algorithms retain a cost exponential in the intrinsic dimension $k(d-k)$, which motivates the Fantope surrogate we introduce next for efficient online updates against a relaxed fractional comparator.
\subsection{Grassmannian HRD for Online WLRA}
\label{subsec:grassmann-hrd}

 Because weighted reconstruction loss depends on the subspace rather than its choice of basis, we identify each subspace with its orthogonal projector. The resulting hypothesis class is
\[
  \mathcal P_k(d)=\{P\in\mathbb R^{d\times d}:
  P=P^\top,\ P^2=P,\ \operatorname{tr}P=k\}.
\]
For $P\in\mathcal P_k(d)$, let $X_P$ be any full-column-rank matrix whose
columns span its range. On round $t$, the learner chooses $P_t$ before seeing
$(x_t,W_t)$ and incurs the standard weighted reconstruction loss
$\ell_t(P_t)=\min_{a\in\mathbb R^k} \|W_t^{1/2}(x_t-X_{P_t}a)\|_2^2.$
We assume throughout that $1\le k<d$, $\|x_t\|_2\le1$, and
$\mu I_d\preceq W_t\preceq WI_d$ for $0<\mu\le W$. Set $\kappa=W/\mu$
and assume the input sequence is oblivious. Our benchmark is
\[
  P^\star\in\arg\min_{P\in\mathcal P_k(d)}\sum_{t=1}^T\ell_t(P),
  \qquad C_{\mathrm{WLRA}}=\sum_{t=1}^T\ell_t(P^\star).
\]
The positive lower bound on the weights is used to control how this loss
changes between nearby subspaces. The exact-comparator algorithm below allows
any $W_t$ satisfying the spectral bounds; the locally convexified algorithm
that follows assumes diagonal $W_t$.

\textbf{Exact rank-$k$ comparator.}
Our first algorithm constructs a finite collection of representative
subspaces before the stream begins. It treats these representatives as
experts: on each round, it samples one to play and then updates their
weights using the observed WLRA losses. Its regret has two sources:
approximating $P^\star$ by one representative and learning which
representative to play.

We construct the representatives using
$\binom dk$ graph charts of $\mathrm{Gr}(k,d)$. A chart selects $k$ reference coordinates and normalizes
those coordinates to the identity. The remaining coordinates form a
matrix $Y\in[-1,1]^{(d-k)\times k}$ describing the subspace.
Such a chart exists for every subspace by choosing a largest-determinant
$k\times k$ row minor of a basis. The appendix gives the precise construction
and resolves overlaps between these charts.

In each chart, the coordinate box is divided into dyadic cubes. The center of a
cube specifies a rank-$k$ representative subspace, and halving each coordinate
gives $2^{k(d-k)}$ children. We refine all charts to a common depth
$L$ before the first round and use the resulting leaf centers as
experts. MTMW assigns prior mass to the leaves, samples a leaf on each
round, and updates their masses after observing $(x_t,W_t)$. The
learner may play different leaves over time, but the leaf used to
compare with $P^\star$ has one fixed representative.

We choose $L$ to control the approximation cost. For subspaces $U,V$ with principal angles
$\theta_1,\ldots,\theta_k$, the chordal distance is
  $d_G(U,V)=\left(\sum_{i=1}^k\sin^2\theta_i\right)^{1/2}
  =\frac{1}{\sqrt2}\|P_U-P_V\|_F.$
A depth-$j$ cube has chordal radius at most
$\sqrt{k(d-k)}\,2^{-j}$. We set $L$ so that the leaf representative
$P_{p^\star}$ associated with $P^\star$ satisfies
$\|P_{p^\star}-P^\star\|_F\le\kappa^{-1}T^{-1/4}$.
This scale follows from a property of the \emph{cumulative} loss
$F(P)=\sum_t\ell_t(P)$: its first derivative vanishes at the minimizer
$P^\star$. The appendix proves the resulting second-order estimate $F(P)-F(P^\star) \le\frac{3\pi^2}{2}W\kappa^2T\|P-P^\star\|_F^2.$
Hence replacing $P^\star$ by the fixed leaf representative costs at most
$\frac{3\pi^2}{2}W\sqrt T$. The cancellation occurs only for the entire
sum of losses. It would not justify using a different representative on each
round, which explains the up-front refinement.

The remaining cost comes from selecting among the leaves. MTMW gives the
virtual root mass $1$ and splits mass equally first among charts and then
among children. There are
$N_{\mathrm{ex}}=\binom dk2^{k(d-k)L}$ leaves, each with mass $1/N_{\mathrm{ex}}$.
Its online-learning guarantee therefore adds $W\sqrt{T\log N_{\mathrm{ex}}/2}$ to the
loss of the fixed comparator leaf. Combining this term with the preceding
approximation gives the exact-comparator result.

\begin{restatable}{theorem}{ghrdregret}
\label{thm:grassmannian-hrd}
Under the preceding assumptions, \Cref{alg:ghrd-regret-min} satisfies

$$ \mathbb E\sum_{t=1}^T\ell_t(P_t)-C_{\mathrm{WLRA}}
 \le O\!\left(W\sqrt{k(d-k)T\log(\kappa k(d-k)T)}\right).$$

In the exact-real arithmetic model, with scalar exponentiation and
finite-distribution sampling counted as primitive operations, its total
running time is
$ O\!\left(T
 (4\kappa\sqrt{2k(d-k)}\,T^{1/4})^{k(d-k)}
 \operatorname{poly}(d)\right).$
\end{restatable}

This guarantee uses the standard WLRA loss above and compares with an actual rank-$k$
subspace. The cost of doing so is computational: the number of fixed leaf
experts grows as $T^{k(d-k)/4}$ for fixed $d,k,\kappa$. The appendix supplies
the depth choice, the chart and loss estimates, and the MTMW analysis.

\textbf{$(1+\varepsilon)$ rank-$k$ comparator.}
The preceding construction needs finer cells as the horizon grows because each leaf plays only its center. We can remove this dependence on $T$ from the
number of cells by letting an online learner move inside each cell.
This also gives a relative comparison; if the optimum has zero loss, a nearby
fixed center could have positive loss, so an additive net approximation alone
cannot establish a $(1+\varepsilon)$ guarantee.

We use the same bounded graph charts. In chart $J$, order the coordinates as
$(J,J^c)$, put $X_J(Y)=\binom{I_k}{Y}$, and let $P_J(Y)$ project onto its
columns. For diagonal $W_t$, we write $r_{t,J}(Y)=x_{t,J^c}-Yx_{t,J},$ and $S_{t,J}(Y)=W_{t,J^c}^{-1}+YW_{t,J}^{-1}Y^\top.$
Eliminating the weighted least squares coefficients yields $\ell_t(P_J(Y))=r_{t,J}(Y)^\top S_{t,J}(Y)^{-1}r_{t,J}(Y).$
This expression is exact, but the matrix $S_{t,J}(Y)$ varies with $Y$.
Within a sufficiently small cube $c$ centered at $\bar Y_c$, we hold that matrix
fixed and optimize the surrogate $q_{t,J,c}(Y)=r_{t,J}(Y)^\top S_{t,J}(\bar Y_c)^{-1}r_{t,J}(Y).$
The residual is affine in $Y$, making $q_{t,J,c}$ a convex quadratic.
Moreover, the true loss and surrogate satisfy, for every $Y\in c$ and
every round,
\begin{equation}
 \frac{q_{t,J,c}(Y)}{1+\varepsilon/4}
 \le\ell_t(P_J(Y))
 \le\frac{q_{t,J,c}(Y)}{1-\varepsilon/4}.
 \label{eq:grass-local-sandwich}
\end{equation}
A cube side length $O(\varepsilon/(\kappa k(d-k)))$ suffices, independently
of $T$. Because the comparison is relative, it also holds at zero loss.

Each cube runs projected gradient descent on its own surrogate, with every
iterate representing a genuine rank-$k$ projector. After each round, MWUA
updates the cubes using their \emph{true} WLRA losses; the full-information
setting lets us update all of their inner learners as well. For the cube
containing $P^\star$, gradient descent and both sides of
\eqref{eq:grass-local-sandwich} give a $(1+\varepsilon)$ comparison plus
$\varepsilon W\sqrt T$ additive loss. MWUA adds the cost of selecting that
cube. Thus the number of cells is independent of the horizon, even though
the algorithm still visits every cell on each round.

\begin{restatable}[Horizon-independent discretization]{theorem}{grassApproxRegret}
\label{thm:grassmannian-approx}
For $0<\varepsilon\le1$ and diagonal $W_t$ under the preceding spectral
bounds, \Cref{alg:grass-local-MWUA} satisfies
\[
 \mathbb E\sum_{t=1}^T\ell_t(P_t)
 \le(1+\varepsilon)C_{\mathrm{WLRA}}
 +\varepsilon W\sqrt T
 +W\sqrt{\frac{T}{2}k(d-k)\log\frac{96\kappa k(d-k)}{\varepsilon}}.
\]
Its total running time in the same exact-real model is
 $O\!\left(T
 (48\kappa k(d-k)/\varepsilon)^{k(d-k)}
 (dk^2+k^3)\right).$
\end{restatable}

The appendix specifies the cell depth and learning rates and proves the
relative comparison and both online-learning bounds. In its notation, the
number of cells is $N_\varepsilon=\binom dk2^{k(d-k)D}$, independent of $T$;
memory is $O(N_\varepsilon k(d-k))$. Both constructions remain exponential in
the intrinsic dimension $k(d-k)$, but they use that cost differently: the
exact-comparator algorithm keeps one fixed expert per leaf, whereas the
$(1+\varepsilon)$ algorithm runs one moving learner per cell.

\subsection{Fantope Relaxation and Polynomial-Time Online WLRA}
\label{subsec:fantope}
Given the disappointing runtime of the Grassmannian-HRD algorithms, one may wonder whether there is a way to avoid explicitly storing a weight for each of the exponentially many experts and speed up the MWUA algorithm. 
The result in \Cref{app:lower_bound} gives evidence that it is unlikely that a significant speed-up is possible with the stated sublinear regret guarantees, unless $\mathrm{NP}\subseteq\mathrm{BPP}$. We therefore study a surrogate for online WLRA. First, we replace
the standard weighted reconstruction loss, which optimizes the reconstruction
coefficients within each subspace, by the weighted error of an orthogonal
projection. Second, we allow fractional projectors, but restrict them to a
fixed $m$-dimensional anchor subspace. These changes give a convex objective
with polynomial-time online updates. 

A rank-$k$ orthogonal projector $P_U$ satisfies
$
    P_U^\top=P_U, P_U^2=P_U,\mathrm{Tr}(P_U)=k.
$
Its convex hull is the Fantope
$$
    \mathcal{F}_k(d) = \{P\in\mathbb{R}^{d\times d}: P=P^\top,\; 0\preceq P\preceq I_d,\; \mathrm{Tr}(P)=k\}.
$$
The eigenvalues of $P\in\mathcal{F}_k(d)$ lie in $[0,1]$ and sum to $k$.
Thus $P$ may be viewed as a fractional rank-$k$ projection, assigning
partial mass to multiple directions.

\textbf{Reduced anchor subspace.}
Optimizing directly in $\mathcal{F}_k(d)$ would still require manipulating
$d\times d$ matrices. We therefore restrict the fractional projector to an
anchor subspace. Let
$V\in\mathbb{R}^{d\times m},V^\top V=I_m,$ where $k\leq m\ll d$. The reduced Fantope is therefore $\mathcal{F}_k(m)$. 
Each reduced fractional projector $P\in\mathcal{F}_k(m)$ is lifted to the
ambient space by $\nu(P)=VPV^\top$. This lifted matrix is a valid fractional
projector in the ambient dimension: it is symmetric, positive semidefinite,
bounded above by $I_d$, and has trace $k$:
$
    \mathrm{Tr}(VPV^\top) = \mathrm{Tr}(PV^\top V) = \mathrm{Tr}(P) = k.
$
Thus the algorithm remains an online low-rank method in the ambient space,
but all optimization occurs in dimension $m$.

The surrogate loss of a lifted fractional projector is
$\widetilde{\ell}_t(P) := \|W_t^{1/2}(I_d-VPV^\top)x_t\|_2^2.$
Define the reduced sufficient statistics
$$    \alpha_t = x_t^\top W_t x_t,\quad
    \bar{x}_t = V^\top x_t,\quad
    \bar{z}_t = V^\top W_t x_t,\quad
    \widetilde{W}_t = V^\top W_tV,\quad
    \bar{X}_t = \bar{x}_t\bar{x}_t^\top.$$
Expanding this surrogate gives the exact reduced expression
$\widetilde{\ell}_t(P) = \alpha_t - \mathrm{Tr}\left( P(\bar{z}_t\bar{x}_t^\top+\bar{x}_t\bar{z}_t^\top) \right) + \mathrm{Tr}\left(P\widetilde{W}_tP\bar{X}_t\right).$
This is an exact expression for the ambient surrogate loss using
$m$-dimensional quantities. The anchor $V$ restricts the comparator class;
the reduction introduces no further approximation when evaluating
$\widetilde{\ell}_t(P)$. For nonuniform weights, even an integral orthogonal
projector generally has a different loss from the standard WLRA loss, because standard WLRA chooses the best
weighted reconstruction within the subspace.

The first term $\alpha_t$ is constant in $P$, and the second term is
affine. The only nontrivial term is $H_t(P) = \mathrm{Tr}(P\widetilde{W}_tP\bar{X}_t)$.
Since $\bar{X}_t=\bar{x}_t\bar{x}_t^\top$, we can rewrite this as $H_t(P) = \bar{x}_t^\top P\widetilde{W}_tP\bar{x}_t = \|\widetilde{W}_t^{1/2}P\bar{x}_t\|_2^2.$
Because $\widetilde{W}_t\succeq 0$, this is the squared norm of a linear
map in $P$, and is therefore convex. Hence $\widetilde{\ell}_t(P)$ is
convex over $\mathcal{F}_k(m)$. The gradient is
$
    \nabla \widetilde{\ell}_t(P)
    = -(\bar{z}_t\bar{x}_t^\top+\bar{x}_t\bar{z}_t^\top)
    + \widetilde{W}_tP\bar{X}_t + \bar{X}_tP\widetilde{W}_t.
$
Thus each online update can be performed using standard Online Gradient
Descent followed by projection back to the Fantope.

\begin{algorithm}[H]
\caption{Polynomial-Time OGD over the Reduced Fantope}
\label{alg:fantope}
\begin{algorithmic}[1]
\Require Sequence $\{x_t, W_t\}_{t=1}^T$, anchor $V \in \mathbb{R}^{d \times m}$, step size $\eta$.
\State Initialize $\tilde{P}_1 = \frac{k}{m} I_m \in \mathcal{F}_k(m)$. 
\For{$t = 1$ \textbf{to} $T$}
    \State $\tilde{x}_t \gets V^\top x_t$
    \State $\tilde{z}_t \gets V^\top W_t x_t$
    \State $\tilde{W}_t \gets V^\top W_t V$
    \State $\nabla \tilde{\ell}_t(\tilde{P}_t) \gets -(\tilde{x}_t \tilde{z}_t^\top + \tilde{z}_t \tilde{x}_t^\top) + \tilde{W}_t \tilde{P}_t \tilde{x}_t \tilde{x}_t^\top + \tilde{x}_t \tilde{x}_t^\top \tilde{P}_t \tilde{W}_t$
    \State $M_t \gets \tilde{P}_t - \eta \nabla \tilde{\ell}_t(\tilde{P}_t)$
    \State Compute eigendecomposition $M_t = Q \Lambda Q^\top$
    \State Project eigenvalues $\Lambda$ onto $\{\lambda \in [0,1]^m \mid \sum_i \lambda_i = k\}$ to get $\hat{\Lambda}$
    \State $\tilde{P}_{t+1} \gets Q \hat{\Lambda} Q^\top$
\EndFor
\end{algorithmic}
\end{algorithm}

\textbf{Projection onto the Fantope.}
The projection step has a simple spectral form. Since $\mathcal{F}_k(m)$ is
spectral, projecting $M_t$ onto $\mathcal{F}_k(m)$ reduces to projecting
the eigenvalue vector $\lambda$ onto the capped simplex $\Delta_k^{[0,1]} = \left\{\mu\in[0,1]^m: \sum_{i=1}^m \mu_i=k\right\}.$ If $\widehat{\lambda}$ is this Euclidean projection, then
$\Pi_{\mathcal{F}_k(m)}(M_t) = Q\,\mathrm{diag}(\widehat{\lambda})\,Q^\top$.
The dominant optimization cost is the eigendecomposition in $m$ dimensions.

The Fantope has a bounded diameter: for any $P,Q\in\mathcal{F}_k(m)$,
$\|P-Q\|_F^2 \leq 2k$. Under the assumptions $\|x_t\|_2\leq 1$ and
$\|W_t\|_{\mathrm{op}}\leq W$, the gradients satisfy
$\|\nabla \widetilde{\ell}_t(P)\|_F \leq O(W)$.
Applying the standard OGD regret theorem gives the following guarantee.

\begin{restatable}[Polynomial-time fractional WLRA regret]{theorem}{fantopethm}
\label{thm:fantope-ogd}
Let $V$ be fixed before round $1$, let $\|x_t\|_2\le1$, and let
$W_t$ be diagonal with $0\preceq W_t\preceq WI_d$, where $W>0$.
Running Algorithm~\ref{alg:fantope} with
$\eta = \Theta\left(\frac{\sqrt{k}}{W\sqrt{T}}\right)$
guarantees that for every fixed fractional comparator
$P^\star\in\mathcal{F}_k(m)$,
$
    \sum_{t=1}^T \widetilde{\ell}_t(P_t) - \sum_{t=1}^T \widetilde{\ell}_t(P^\star)
    \leq O(W\sqrt{kT}).
$
Furthermore, the per-round runtime is $O(m^2d+m^3)$.
\end{restatable}
This bound compares surrogate losses within the reduced Fantope; it makes
no multiplicative or additive comparison to $C_{\mathrm{WLRA}}$. The term $O(m^2d)$ comes from forming $\widetilde{W}_t=V^\top W_tV$; since
$W_t$ is diagonal, this can be done without forming any dense $d\times d$
matrix. The term $O(m^3)$ comes from the eigendecomposition needed for
Fantope projection.

\section{Experiments}

\label{experiments}

We compare the Grassmannian methods, Fantope OGD, and online-PCA baselines
on MNIST and MovieLens, conduct a weighted Air Quality experiment, and examine feature importance, cyclic ordering,
and subspace drift in three synthetic weighted streams. Protocols, numerical tables, and additional
experiments appear in \Cref{app:other-experiments}. Source code is available in our
\href{https://github.com/alisayiranlu/Online-Low-Rank-Approximation}{GitHub repository}.

\textbf{MNIST and MovieLens.}
We use $d=50$, $k\in\{10,15,20\}$, $T=5000$ for
MNIST~\citep{lecun1998mnist}, and $T=2000$ for MovieLens
20M~\citep{harper2015movielens}, with ten paired stream orders.
\Cref{alg:ghrd-regret-min,alg:grass-local-MWUA} have slightly higher
mean reconstruction loss than FTRL in all six configurations, with
differences below $0.71\%$, and lower mean loss than the Loss MEG and GD
online-PCA baselines of Nie et al.~\citep{nie2016online}
(\Cref{tab:realdata-experiments,tab:realdata-excess}).
The Grassmannian methods use tuned $\eta=1000$ and finite-step MCMC for computational efficiency;
the PCA baselines use untuned horizon-based rates and isotropic
initialization. These comparisons concern the stated settings. Fantope reports its native fractional
projection loss here, with a different decision class and comparator.

\textbf{Air Quality.}
We evaluate Fantope OGD and approximate weighted FTRL on
$1512$ chronological UCI Air Quality~\citep{vito2008airquality} observations, with $d=12$, $k=2$,
observed-entry weight $1$, and missing-entry weight $0.01$.
Each method's subspace is scored before updating using weighted
least-squares fitting loss. Fantope OGD attains cumulative loss $238.52$,
versus $463.87$ and $462.48$ for weighted FTRL with three and thirty
ALS iterations, respectively: a $48.43\%$ reduction relative to the
stronger baseline. Increasing the ALS budget tenfold reduces FTRL's
loss by only $0.30\%$. This comparison supports that the Fantope OGD algorithm can run better than existing FTLR algorithms and can thus act as a practical algorithm for WLRA. \Cref{fig:air-quality} shows cumulative losses;
preprocessing, full results, and interpretation appear in
\Cref{subsec:air-quality}.

\textbf{WeightedImportanceCycleNoisy.}
Our main synthetic dataset uses $T=2000$, $d=6$, $k=2$, and
strictly positive feature-importance weights with $\kappa=64$.
In this expeirment, a frequent, low-weight distractor competes with a common signal direction
and two directions whose contributions balance within each cycle.
Their order produces temporary changes in cumulative directional
importance. Gaussian perturbations with standard deviation $0.005$
are added before unit normalization. Five fresh validation seeds follow
development screening. All methods are scored by weighted rank-two fitting
loss, including ablations that ignore weights during learning.

\begin{figure}[H]
\centering
  \includegraphics[width=\linewidth]{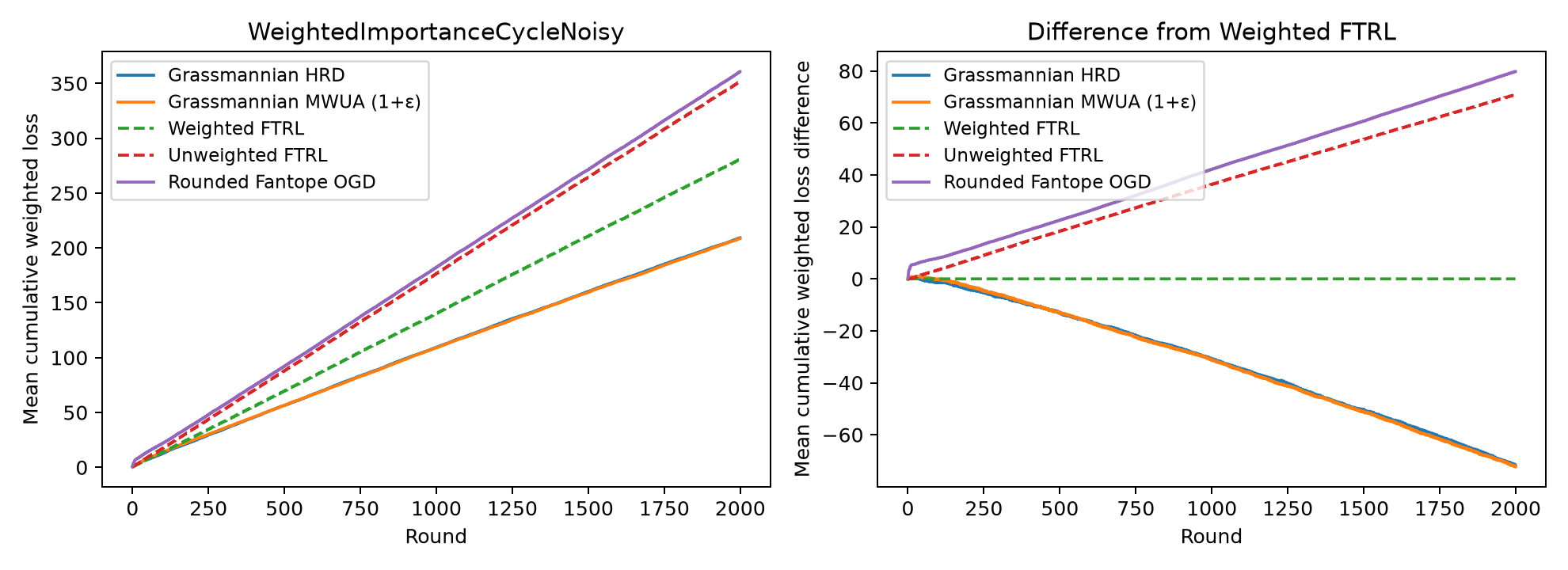}
\caption{WeightedImportanceCycleNoisy over five validation seeds.
Left: mean cumulative weighted fitting loss. Right: mean paired difference
from approximate weighted FTRL (three ALS iterations); negative values
favor the compared method. Weight-ignoring ablations are scored using
the actual weights. Fantope uses its rounded rank-two subspace.
Final losses and sample SDs appear in
\Cref{tab:weighted-importance-cycle}.}
\label{fig:weighted-importance-cycle}. 
\end{figure}

\Cref{fig:weighted-importance-cycle} compares the methods.
\Cref{alg:ghrd-regret-min,alg:grass-local-MWUA}, using
$\eta_4=0.357$ and $\eta_5=0.387$, attain losses
$209.25\pm4.66$ and $208.69\pm6.23$, versus $281.03\pm3.74$ for
approximate weighted FTRL (mean $\pm$ sample SD).
Both have lower loss on all five seeds. Increasing the FTRL solver budget
from three to thirty ALS iterations gives $282.43\pm3.82$.
Ignoring weights during Grassmannian updates increases the losses to
$353.84\pm8.72$ and $355.90\pm4.94$.
These results support the value of feature weights in this construction
and favorable performance of the Grassmannian algorithms relative to the implemented approximate FTRL
solver. \Cref{subsec:weightedshift-details} reports a
noiseless cyclic example, while \Cref{subsec:weighted-tracking} examines
rounded Fantope OGD under heterogeneous noise and subspace drift. 

\section*{AI Disclosure}
The authors developed the research direction and methodology for the algorithms in Section~\ref{subsec:grassmann-hrd}. ChatGPT models (Sol 5.6 and Sol 6.0) helped develop the refinement from a $T^{-1/2}$ to a $T^{-1/4}$ discretization scale and were used to update the proofs to incorporate this refinement and subsequent algorithmic revisions. The results in Section~\ref{sec:kmeans-marginal-jl} and Section~\ref{subsec:fantope} were developed with assistance from Gemini DeepThink.

Initial section drafts were written by the authors, before the ChatGPT models were used to revise and shorten Sections~\ref{sec:kmeans-marginal-jl},~\ref{sec:online-wlra}, and~\ref{experiments}, and the appendices, which included updates to reflect algorithmic improvements and iterations. ChatGPT and Claude also assisted in refining and checking the proofs in Section~\ref{sec:online-wlra} and Appendix~\ref{app:lower_bound}, locating relevant references, and developing and debugging the experimental code. These models also assisted in checking grammar throughout the manuscript. 

The authors refined the final arguments and independently reviewed and verified the proofs, algorithms, references, code, experiments, and claims. The authors independently reviewed all revisions and take full responsibility for the manuscript.

\newpage
\def\shortbib{0}
\bibliographystyle{alpha}
\bibliography{refs}
\newpage
\appendix
\section{Related Work}
\label{related_work}

\textbf{Clustering and online $k$-means.}
The $k$-means objective is one of the most widely studied formulations of clustering, with origins in the work of \cite{macqueen1967methods} and the least-squares quantization procedure of \cite{lloyd1982least}. 
Given a dataset, the goal is to choose $k$ centers minimizing the sum of squared distances from each point to its nearest center. 
Although the objective is simple and widely used, computing an optimal solution is computationally difficult, motivating a large literature on approximation algorithms, initialization schemes, and data summaries. 
The $k$-means++ seeding method of \cite{arthur2007kmeans} gives a simple randomized initialization with logarithmic approximation guarantees, while local-search methods provide constant-factor approximation guarantees for the offline problem~\citep{kanungo2004local}. 
Coresets have also played a central role in scalable clustering, allowing a small weighted summary of the data to preserve clustering costs for all candidate center sets~\citep{harpeled2004coresets}.

The online version of $k$-means is more subtle because the learner must output centers before observing the next point and incurs loss immediately. 
This differs from classical streaming clustering, where the algorithm may process points sequentially but is only required to output a final solution after the stream has ended. 
The regret-based online $k$-means framework of \cite{Cohen-AddadGKR21} is closest to our setting: the learner maintains $k$ candidate centers at each time step and competes with the best fixed $k$-means solution in hindsight. 
Their work shows that multiplicative weights over a discretized grid can achieve sublinear regret information-theoretically, while efficient exact regret minimization is obstructed by online-to-offline hardness. 
It also develops an incremental coreset, adaptive hierarchical region decomposition, and Mass Tree MWUA method for a $(1+\varepsilon)$-approximate regret benchmark.
Our $k$-means contribution builds on this regret perspective but uses projected Voronoi assignments and ambient running means.
At fixed $\varepsilon$, the projection dimension and expert-class size have no explicit dependence on $d$ or $T$, although updates still involve $d$-dimensional data and tuning $\varepsilon$ with $T$ introduces an indirect horizon dependence.

\textbf{Low-rank approximation.}
Low-rank approximation has a rich history in both linear algebra and machine learning, serving as a foundational technique for data compression, dimensionality reduction, and uncovering latent structure. 
The seminal result of \cite{eckart1936approximation} established that the optimal rank-$k$ approximation of a matrix in the Frobenius norm is obtained via truncated singular value decomposition (SVD). 
Building on this, a wide range of classical algorithms have been developed for efficiently computing low-rank approximations of static datasets. 
Randomized SVD methods~\citep{halko2011} exploit random projections to accelerate computation while preserving accuracy, and principal component analysis (PCA)~\citep{jolliffe2002pcal} remains a widely used method for identifying directions of maximal variance in high-dimensional data. 
These techniques are central to many modern applications, including collaborative filtering, natural language processing, and computer vision, where they reduce storage requirements, accelerate downstream computations, and improve interpretability.

The online or streaming setting, in which data arrives sequentially, introduces additional challenges. 
Unlike static datasets, where the full matrix is known prior, online low-rank approximation must maintain an accurate representation of the matrix as new rows or columns arrive, without recomputing the full SVD at each step. 
Incremental SVD updates~\citep{brand2006fast} address this by algebraically modifying existing decompositions to account for additions or deletions of rows and columns, enabling efficient updates while retaining the low-rank structure. 
Similarly, online subspace identification from incomplete observations~\citep{balzano2010online} generalizes these ideas to sparse streaming matrices, which are common in recommendation systems and user-item interactions. 
These streaming methods address representation updates, whereas our online-learning protocol requires a subspace to be chosen before each observation and evaluates cumulative loss against a fixed hindsight comparator.

Online PCA extends these ideas further, enabling incremental dimensionality reduction in applications with continuously arriving high-dimensional data. 
For instance, \cite{ross2008incremental} apply online PCA to visual tracking, processing frames sequentially, projecting each onto the current subspace, and updating mean, covariance, and principal components. 
Subsequent work~\citep{feng2013online} addresses the limitations of classical online PCA methods by incorporating robustness against outliers and corrupted observations, which frequently arise in real-world streaming datasets. 
These approaches demonstrate the importance of maintaining adaptability and resilience in online low-rank models, particularly when the data distribution is non-stationary or contains noise.

Collectively, these contributions highlight both the theoretical foundations and practical algorithms for online low-rank approximation. 
They motivate the need for methods that balance computational efficiency with adaptive accuracy, especially in high-dimensional, streaming, and noisy environments. 
Our work studies this problem from a regret-minimization perspective, distinguishing exact rank-$k$ comparison, $(1+\varepsilon)$ rank-$k$ comparison, and optimization of a tractable fractional surrogate.

\textbf{Online PCA and online subspace learning.}
The unweighted version of online low-rank approximation is closely related to online PCA, where the learner selects a low-dimensional subspace and suffers either compression loss or reconstruction gain on the next vector. 
This problem has been studied extensively, and several works obtain sublinear regret without discretizing the Grassmannian into a finite expert class. 
\cite{warmuth2008randomized} introduced randomized online PCA algorithms with regret bounds logarithmic in the ambient dimension, using matrix multiplicative-weights style updates over density matrices. 
\cite{nie2016online} later gave optimal regret guarantees for online PCA, analyzing gradient descent and exponentiated-gradient methods in the appropriate projection-matrix parameterization. 
For the rank-one case and related spectral objectives, \cite{garber2015online} studied online learning of eigenvectors and gave efficient regret-minimization algorithms. 
These results show that, in the ordinary unweighted PCA setting, one can often avoid an explicit finite net of subspaces by working directly with convex spectral relaxations.

Our work differs from this as the low-rank results focus on the weighted reconstruction loss, where the diagonal weights vary across coordinates and the corresponding offline problem is NP-hard. 
Thus, the main role of Grassmannian HRD is not to rederive known online PCA guarantees, but to provide an exact rank-$k$ comparator framework for the weighted, nonconvex setting and to clarify why polynomial-time algorithms require a relaxed comparator.

\textbf{Weighted low-rank approximation and convex relaxations.}
Weighted low-rank approximation generalizes classical low-rank approximation by assigning different importance weights to different entries of the matrix. 
Given a target matrix $A$ and nonnegative weights $W_{ij}$, the objective is to find a low-rank matrix $L$ minimizing
\[
    \sum_{i,j} W_{ij}(A_{ij}-L_{ij})^2.
\]
This formulation captures heterogeneous confidence, missing entries, and coordinate-dependent noise, and appears naturally in collaborative filtering, matrix completion, computer vision, and structure-from-motion problems. 
\cite{SrebroJ03} introduced weighted low-rank approximation as a general framework and proposed an EM-style algorithm for solving such problems in practice. 
In contrast with ordinary Frobenius-norm low-rank approximation, the weighted problem is computationally hard: \cite{GillisG11} showed that computing an optimal weighted low-rank approximation is NP-hard, even for rank one and strictly positive weights, including hardness at a prescribed inverse-polynomial accuracy scale.
Our online-to-offline reduction uses this accuracy scale to constrain sufficiently strong online guarantees against the standard WLRA comparator; it does not establish a barrier for every fixed constant $\varepsilon$.

A related line of work uses convex relaxations for low-rank learning and matrix completion, including trace-norm and spectral relaxations. 
These approaches replace an explicit rank constraint with a convex surrogate, enabling efficient optimization under appropriate assumptions. 
For example, trace-norm regularization has been used extensively in matrix completion and collaborative filtering~\citep{srebro2005rank,salakhutdinov2010weighted,shamir2011collaborative,shamir2014matrix}. 
Our Fantope relaxation follows the same broad philosophy, but the object being relaxed is the set of rank-$k$ projection matrices rather than the set of low-rank matrices themselves. 
The resulting feasible set is the convex hull of rank-$k$ projectors, namely the set of positive semidefinite matrices with eigenvalues in $[0,1]$ and trace $k$. 
This connects our polynomial-time online WLRA algorithm to the online PCA literature, while also making clear why the guarantee is against a fractional comparator rather than the exact Grassmannian optimum.

\textbf{Grassmannian geometry and subspace methods.}
The Grassmann manifold $\mathrm{Gr}(k,d)$, the space of $k$-dimensional linear subspaces of $\mathbb{R}^d$, provides the natural geometric setting for low-rank approximation and subspace learning. 
Working directly on the Grassmannian eliminates basis dependence and enables invariant formulations of subspace tracking and approximation problems. 
The geometry of the Grassmannian is characterized by \emph{principal angles} between subspaces~\citep{bjorck1973numerical}, which quantify subspace similarity and induce natural distance metrics such as the chordal distance and geodesic distance. 
These geometric structures have been extensively studied in the context of optimization algorithms on matrix manifolds~\citep{edelman1998geometry, absil2008optimization}, where gradient-based methods exploit the manifold structure for efficient computation. 
In machine learning and computer vision, Grassmannian methods have been applied to problems including subspace clustering, video analysis, and face recognition, where data naturally lies in or near low-dimensional subspaces. 
Our work uses bounded graph charts of the Grassmannian in two ways: to construct a fixed hierarchy of rank-$k$ representatives for exact-comparator learning, and to define cells within which a rank-$k$ learner moves for the $(1+\varepsilon)$ comparison.

\textbf{Online optimization on manifolds.}
Another related direction is online optimization over Riemannian manifolds. 
For low-rank matrix manifolds, \cite{shalit2010online} proposed online learning algorithms based on Riemannian gradient steps and retractions, allowing low-rank structure to be maintained throughout the optimization process. 
More recently, \cite{wang2023online} developed online gradient and bandit algorithms over Riemannian manifolds, giving regret bounds for geodesically convex and strongly geodesically convex losses in appropriate manifold settings. 
These works are relevant because the Grassmannian is itself a Riemannian manifold, and they provide general tools for online optimization beyond Euclidean domains.

Our setting differs in the role played by the manifold.
We do not assume that standard weighted reconstruction loss is geodesically convex over the Grassmannian or use Riemannian gradient descent to track a single subspace.
For the exact-comparator result, we refine bounded graph charts to a fixed depth before the stream and run Mass Tree MWUA over their leaf representatives.
For the $(1+\varepsilon)$ result, projected gradient descent runs on a convex surrogate within each fixed chart cell, while MWUA selects among cells using their true weighted reconstruction losses.
The latter gradient updates take place in chart coordinates, and every iterate still represents a rank-$k$ subspace.

\textbf{Online learning with experts.}
Online learning with experts and related sequential decision-making problems have been extensively studied in the machine learning and theoretical computer science communities~\citep{cesa2006prediction}. 
In this framework, an algorithm repeatedly selects actions or ``experts'' from a predefined set, receives feedback in the form of losses, and aims to minimize its cumulative loss relative to the best fixed expert in hindsight. 
Early work assumed the existence of a ``perfect expert'' whose predictions were always correct. 
In this idealized scenario, a folklore ``halving'' algorithm achieves at most $\log_2 n$ mistakes by retaining the set of experts that have not yet erred and predicting based on majority vote.

Recognizing that a perfect expert rarely exists in practice, \cite{LittlestoneW94} introduced the randomized weighted majority (RWM) algorithm. 
Here, predictions are sampled according to a probability distribution proportional to the experts' weights, which are updated multiplicatively after each round. 
This yields $O(\sqrt{T \log n})$ regret and is asymptotically optimal~\citep{cover1966behavior}. MWUA, also called multiplicative weights in this setting, applies the same exponential-weighting idea to a finite class of experts.
More generally, the Multiplicative Weights Update (MWU) framework extends these ideas to arbitrary loss functions and has found applications in boosting (e.g., AdaBoost~\citep{freund1997decision}), approximately solving zero-sum games~\citep{freund1999adaptive}, and efficiently approximating linear and semi-definite programs~\citep{Clarkson95,PlotkinST95,GargK07}. 
Variants such as Follow the Perturbed Leader (FTPL) and Follow the Regularized Leader (FTRL) introduce random perturbations or regularization to improve stability and computational efficiency in structured problems~\citep{KalaiV05,McMahan11a}. 
For an in-depth survey of MWUA and its widespread applications in online optimization, game theory, and combinatorial learning, see \cite{arora2012multiplicative}. 

A particularly relevant extension is the work on online learning with low-rank experts by \cite{HazanKLM16}, which exploits hidden low-rank structure in the expert loss matrix. 
In this setting, although the nominal number of experts may be very large, the effective dimensionality of the loss matrix is small, allowing regret bounds to scale with the rank rather than the total number of experts. 
This perspective is related to our use of structured hypothesis classes, but the technical setting is different: our losses arise from geometric approximation problems, and the relevant hypothesis spaces are center tuples, Grassmannian subspaces, or convex spectral relaxations.

Variants of MWUA have also been developed to handle hierarchical or tree-structured expert sets. 
Notably, the Mass Tree MWUA (MTMW) framework of \cite{Cohen-AddadGKR21} introduces a hierarchical organization of experts, allowing multiplicative updates to be applied over adaptively revealed tree-structured classes. 
This hierarchical approach provides both computational advantages and refined control over exploration across different scales of the expert set.

The expert-based algorithms in this paper use these tools in different ways.
Online $k$-means uses MWUA over a flat class of projected Voronoi experts, each of which tracks ambient running means.
Exact-comparator WLRA uses MTMW over fixed Grassmannian leaf representatives.
The $(1+\varepsilon)$ WLRA algorithm uses MWUA over chart cells containing moving projected-gradient learners.
The reduced Fantope method takes a separate route, applying online gradient descent to a convex weighted projection surrogate.
\section{Preliminaries}
\label{preliminaries}

We collect the notation and online-learning tools used in the proofs. The
two WLRA objectives in the paper are defined separately below: the
Grassmannian algorithms use standard weighted reconstruction loss, while
the Fantope algorithm uses a weighted projection surrogate.

\textbf{Notation and online protocol.}
Write $[n]=\{1,\ldots,n\}$. For a matrix $A$, let $\|A\|_F$ and
$\|A\|_{\mathrm{op}}$ denote its Frobenius and operator norms. We write
$A\succeq 0$ when $A$ is positive semidefinite and use $I_n$ for the
$n\times n$ identity matrix.

On each round $t\in[T]$, the learner chooses a hypothesis $h_t$ before
observing the round's data and incurring loss $\ell_t(h_t)$. For a fixed
comparator class $\mathcal H$, its regret is
\[
  \sum_{t=1}^T\ell_t(h_t)
  -
  \min_{h\in\mathcal H}\sum_{t=1}^T\ell_t(h).
\]
Our randomized guarantees take expectation over the learner's choices
and, in the $k$-means result, the random projection. The comparator
class and even the loss being compared differ between the standard
WLRA and Fantope results, so we specify each benchmark explicitly.

\textbf{Online $k$-means.}
The learner chooses $k$ centers
$C_t=\{c_{t,1},\ldots,c_{t,k}\}\subseteq\mathbb R^d$ before observing
$x_t\in\mathbb R^d$ and incurs
\[
  \ell_t(C_t)=\min_{c\in C_t}\|x_t-c\|_2^2.
\]
For a best fixed center set
$C^\star=\{c_1^\star,\ldots,c_k^\star\}$ in hindsight, define
\[
  \OPT
  :=
  \sum_{t=1}^T\min_{j\in[k]}\|x_t-c_j^\star\|_2^2.
\]
We assume $\|x_t\|_2\leq 1$ and $\|c_j^\star\|_2\leq 1$. The stream is
chosen independently of the Gaussian projection used by the algorithm;
in particular, $x_{1:T}$ and $C^\star$ can be fixed before that projection
is sampled.

\textbf{Marginal projection and ambient center tracking.}
The $k$-means algorithm samples a matrix
$\Pi\in\mathbb R^{m\times d}$ with independent
$\mathcal N(0,1/m)$ entries. At any fixed round $t$, its analysis asks
$\Pi$ to preserve only the $k$ distances from $x_t$ to the centers in
$C^\star$. A Gaussian Johnson-Lindenstrauss bound therefore gives
\[
  m=O\!\left(\varepsilon^{-2}
      \log\frac{k}{\varepsilon}\right)
\]
without a union bound over all $T$ rounds. This is a marginal guarantee:
it does not assert that every distance is preserved simultaneously
throughout the stream.

A projected grid tuple defines an expert's Voronoi assignments. Each
expert separately maintains, for every assigned cluster, its count and
running mean in $\mathbb R^d$. Those means are Follow-the-Leader (FTL)
predictions for the points assigned to that cluster. Thus projected
coordinates determine routing, while the predicted centers and their
losses remain in the ambient space.

\textbf{Standard online WLRA.}
Let $1\leq k<d$ and let $\mathrm{Gr}(k,d)$ be the set of
$k$-dimensional subspaces of $\mathbb R^d$. We identify a subspace
$U$ with its orthogonal projector $P_U$ and write
\[
  \mathcal P_k(d)
  =
  \{P\in\mathbb R^{d\times d}:
    P=P^\top,\ P^2=P,\ \operatorname{tr}(P)=k\}.
\]
For $P\in\mathcal P_k(d)$, let $X_P$ be any full-column-rank matrix
spanning its range. The learner chooses $P_t$ before observing
$(x_t,W_t)$ and incurs the \emph{standard weighted reconstruction loss}
\[
  \ell_t(P_t)
  =
  \min_{a\in\mathbb R^k}
  \bigl\|W_t^{1/2}(x_t-X_{P_t}a)\bigr\|_2^2.
\]
This definition is independent of the choice of basis $X_P$. The
exact rank-$k$ benchmark is
\[
  C_{\mathrm{WLRA}}
  :=
  \min_{P\in\mathcal P_k(d)}
  \sum_{t=1}^T\ell_t(P).
\]

For the Grassmannian results, we assume $\|x_t\|_2\leq 1$ and
\[
  \mu I_d\preceq W_t\preceq WI_d,
  \qquad 0<\mu\leq W,
  \qquad \kappa:=W/\mu.
\]
The exact-comparator algorithm permits any weight matrix satisfying
these spectral bounds. The locally convexified
$(1+\varepsilon)$-comparator algorithm additionally requires $W_t$ to
be diagonal. In both cases, $0\leq\ell_t(P)\leq W$.

\textbf{Graph charts and Grassmannian distance.}
For a $k$-element set $J\subseteq[d]$, order the coordinates as
$(J,J^c)$. A subspace whose basis has an invertible minor in rows $J$
can be represented by
\[
  X_J(Y)=
  \begin{pmatrix}I_k\\Y\end{pmatrix},
  \qquad
  Y\in\mathbb R^{(d-k)\times k},
\]
with the coordinate ordering reversed when forming the ambient
projector. Its projector is
\[
  P_J(Y)
  =
  X_J(Y)(I_k+Y^\top Y)^{-1}X_J(Y)^\top.
\]
Choosing a largest-magnitude $k\times k$ row minor ensures that the
entries of $Y$ lie in $[-1,1]$. Hence the $\binom dk$ bounded chart
boxes $[-1,1]^{(d-k)\times k}$ cover $\mathrm{Gr}(k,d)$, although they
may overlap. Each box has $k(d-k)$ coordinates. The Grassmannian
algorithms partition these boxes into cubes: the exact-comparator
algorithm plays fixed cube representatives, and the
$(1+\varepsilon)$ algorithm moves within a cube.

For subspaces $U,V$, we use the chordal distance
\[
  d_G(U,V)
  :=
  \frac{1}{\sqrt{2}}\|P_U-P_V\|_F
  =
  \left(\sum_{i=1}^k\sin^2\theta_i\right)^{1/2},
\]
where $\theta_1,\ldots,\theta_k$ are their principal angles. This
distance connects chart resolution to approximation error between
rank-$k$ projectors.

\textbf{Expert and online convex-learning bounds.}
For $N$ experts with losses in $[0,\rho]$, MWUA, or equivalently
multiplicative weights, gives expected loss at most
\[
  \min_{e\in[N]}\sum_{t=1}^T\ell_t(e)
  +O\!\left(\rho\sqrt{T\log N}\right).
\]
We use this bound for projected Voronoi experts in $k$-means and to
select among chart cubes in the locally convexified WLRA algorithm.
An expert may maintain its own state, such as running cluster means
or a within-cube learner; its prediction is determined before the
current data are revealed.

The exact-comparator WLRA algorithm organizes fixed representatives
as leaves of a tree and uses Mass Tree MWUA. Starting with unit mass
at the root and splitting each node's mass among its children assigns
a prior mass $M(p)$ to each leaf path $p$. Its comparison term scales as
\[
  O\!\left(\rho\sqrt{T\log\frac{1}{M(p)}}\right).
\]
The hierarchy in that algorithm is refined to its chosen depth before
the first round. The locally convexified algorithm instead uses
projected gradient descent within each fixed cube. We also use
projected gradient descent over the Fantope below; for convex losses
with bounded gradients on a bounded domain, its standard regret
bound is sublinear in $T$.

\textbf{Fantope surrogate and reduced comparator.}
The convex hull of $\mathcal P_k(d)$ is the Fantope
\[
  \mathcal F_k(d)
  =
  \{Q\in\mathbb R^{d\times d}:
    Q=Q^\top,\ 0\preceq Q\preceq I_d,\
    \operatorname{tr}(Q)=k\}.
\]
Its elements are fractional projectors: their eigenvalues lie in
$[0,1]$ and sum to $k$. To obtain smaller updates, fix an anchor
matrix $V\in\mathbb R^{d\times m}$ with $V^\top V=I_m$ and
$k\leq m\leq d$. The algorithm chooses
$Q\in\mathcal F_k(m)$ and lifts it to $VQV^\top\in\mathcal F_k(d)$.

The Fantope algorithm uses the \emph{weighted projection surrogate}
\[
  \widetilde{\ell}_t(Q)
  :=
  \bigl\|W_t^{1/2}(I_d-VQV^\top)x_t\bigr\|_2^2
\]
and compares with the best fixed $Q\in\mathcal F_k(m)$ under this
same loss. Its objective is convex in $Q$ and admits an exact
expression using $m$-dimensional sufficient statistics, developed in
\Cref{app:fantope}.

Even when $Q$ is an integral projector, this surrogate generally
differs from standard WLRA for nonuniform $W_t$: standard WLRA
minimizes over the reconstruction coefficients within the selected
subspace, whereas the surrogate applies the orthogonal projector
directly. They agree for scalar weights $W_t=w_tI_d$ and an integral
projector. The Fantope guarantee therefore concerns both a different
loss and a reduced fractional comparator class; it is not a regret
bound against $C_{\mathrm{WLRA}}$.
\section{Online \texorpdfstring{$k$}{k}-Means via Marginal JL}
\label{app:kmeans}

We use the notation of \Cref{sec:kmeans-marginal-jl}. The aim of this section is to justify the analysis used in the proof of \Cref{thm:projected-kmeans}:
the marginal Johnson-Lindenstrauss comparison, the ambient FTL tracking
bound, and the final MWUA comparison.

The analysis can be viewed as the $k$-means analogue of the loss-decoupling
argument used for the fantope relaxation. In the fantope setting, the ambient
weighted loss is rewritten exactly as a reduced convex objective. Here, the
projected anchors are used only for routing, while the representative centers
are accumulated directly in $\mathbb{R}^d$. Thus the proof separates the
low-dimensional assignment rule from the ambient representation error.

We first prove the marginal JL comparison used in
\Cref{lem:marginal-jl-kmeans}. The important point is that the projection is
not required to preserve all distances over the full stream simultaneously.
For each fixed time $t$, it only needs to preserve the $k$ distances from
$x_t$ to the fixed offline centers.
\begin{restatable}{lemma}{marginalJlKMeans}
\label{lem:marginal-jl-kmeans}
For each realization of $\Pi$, define
$Y^\star=(y_1^\star,\ldots,y_k^\star)\in S_m^k$ as follows.
If $\|\Pi c_j^\star\|_2\le 2$ for every $j\in[k]$, choose each
$y_j^\star$ within distance $\varepsilon$ of $\Pi c_j^\star$.
Otherwise, set $Y^\star$ to any fixed element of $S_m^k$. Define
\[
    i_t := \arg\min_{j\in[k]}\|x_t-c_j^\star\|_2^2,
    \qquad
    a_t := \arg\min_{j\in[k]}\|\Pi x_t-y_j^\star\|_2^2.
\]
Then, taking expectation over the random projection $\Pi$,
\[
    \mathbb{E}_{\Pi}\bigl[\|x_t-c_{a_t}^\star\|_2^2\bigr]
    \le
    (1+O(\varepsilon))\|x_t-c_{i_t}^\star\|_2^2+O(\varepsilon).
\]
\end{restatable}
\begin{proof}
Fix a time $t$. Let
\[
    i_t
    =
    \arg\min_{j\in[k]}\|x_t-c_j^\star\|_2^2,
    \qquad
    a_t
    =
    \arg\min_{j\in[k]}\|\Pi x_t-y_j^\star\|_2^2.
\]
Let $G$ be the event that $\|\Pi c_j^\star\|_2\le 2$ for every
$j\in[k]$. A Gaussian JL bound applied to the $k$ fixed centers gives
$\Pr(G^c)\le\varepsilon$: indeed, preservation of their squared norms
within $(1\pm\varepsilon)$ implies
$\|\Pi c_j^\star\|_2\le\sqrt{1+\varepsilon}<2$.

Let $D_t$ be the event that $\Pi$ preserves the squared norms of
$x_t-c_1^\star,\ldots,x_t-c_k^\star$ within $(1\pm\varepsilon)$.
A Gaussian JL bound and a union bound over these $k$ fixed vectors give
$\Pr(D_t^c)\le\varepsilon$ for the stated choice of $m$.
Set $E_t=G\cap D_t$. Then $\Pr(E_t^c)\le2\varepsilon$.
No union bound over the rounds is taken.
Condition on $E_t$. Since $a_t$ minimizes the projected distance to the grid
anchors,
\[
    \|\Pi x_t-y_{a_t}^\star\|_2
    \le
    \|\Pi x_t-y_{i_t}^\star\|_2.
\]
Using the grid approximation
$\|y_j^\star-\Pi c_j^\star\|_2\le\varepsilon$ and the triangle inequality,
\[
\begin{aligned}
    \|\Pi x_t-\Pi c_{a_t}^\star\|_2
    &\le
    \|\Pi x_t-y_{a_t}^\star\|_2+\varepsilon  \\
    &\le
    \|\Pi x_t-y_{i_t}^\star\|_2+\varepsilon  \\
    &\le
    \|\Pi x_t-\Pi c_{i_t}^\star\|_2+2\varepsilon .
\end{aligned}
\]
Applying the JL inequalities on $E_t$ gives
\[
    \sqrt{1-\varepsilon}\,\|x_t-c_{a_t}^\star\|_2
    \le
    \sqrt{1+\varepsilon}\,\|x_t-c_{i_t}^\star\|_2
    +2\varepsilon .
\]
Squaring both sides and using
$\|x_t\|_2,\|c_j^\star\|_2\le 1$, so that all relevant distances are at most
$2$, yields
\[
    \|x_t-c_{a_t}^\star\|_2^2
    \le
    (1+O(\varepsilon))\|x_t-c_{i_t}^\star\|_2^2
    +O(\varepsilon).
\]
On the complement $E_t^c$, the squared distance is at most $4$, and
$\Pr(E_t^c)\leq 2\varepsilon$, contributing an additional $O(\varepsilon)$ term.
Taking expectation over $\Pi$ proves
\[
    \mathbb{E}_{\Pi}
    \bigl[\|x_t-c_{a_t}^\star\|_2^2\bigr]
    \le
    (1+O(\varepsilon))\|x_t-c_{i_t}^\star\|_2^2
    +O(\varepsilon).
\]
\end{proof}

We next justify the ambient tracking term. Once a projected expert fixes a
routing rule, each cluster receives a subsequence of ambient points. The
running mean on each such subsequence competes with the best fixed center for
that subsequence up to a logarithmic penalty.
\begin{restatable}{lemma}{ambientFtlKMeans}
\label{lem:ambient-ftl-kmeans}
Let $v_1,\ldots,v_n\in B_d$, and let $\mu_i=\frac{1}{i-1}\sum_{r<i}v_r$ for $i\ge 2$, with
$\mu_1=0$. Then
\[
    \sum_{i=1}^n\|v_i-\mu_i\|_2^2
    \le
    \min_{c\in\mathbb{R}^d}\sum_{i=1}^n\|v_i-c\|_2^2
    +O(\log n).
\]
Consequently, for any expert-induced partition into $k$ clusters, the total FTL tracking penalty is
$O(k\log T)$.
\end{restatable}
\begin{proof}
Let
\[
    E_i
    :=
    \min_{c\in\mathbb{R}^d}
    \sum_{r=1}^i \|v_r-c\|_2^2
    =
    \sum_{r=1}^i \|v_r-\bar v_i\|_2^2,
    \qquad
    \bar v_i:=\frac1i\sum_{r=1}^i v_r .
\]
For $i\ge 2$, the prediction $\mu_i$ is exactly the empirical mean of the
previous $i-1$ points. The standard variance update identity gives
\[
    E_i-E_{i-1}
    =
    \frac{i-1}{i}\|v_i-\mu_i\|_2^2.
\]
Thus
\[
    \|v_i-\mu_i\|_2^2
    =
    \left(1+\frac{1}{i-1}\right)(E_i-E_{i-1}).
\]
Since $v_i,\mu_i\in B_d$, we have
$\|v_i-\mu_i\|_2^2\le 4$, and hence
$E_i-E_{i-1}\le 4$. Summing over $i\ge 2$,
\[
\begin{aligned}
    \sum_{i=2}^n \|v_i-\mu_i\|_2^2
    &=
    \sum_{i=2}^n (E_i-E_{i-1})
    +
    \sum_{i=2}^n \frac{E_i-E_{i-1}}{i-1}  \\
    &\le
    E_n
    +
    4\sum_{i=2}^n \frac1{i-1}  \\
    &\le
    E_n + O(\log n).
\end{aligned}
\]
The first-round loss is bounded by $\|v_1-\mu_1\|_2^2\le 1$, so it is absorbed
into the logarithmic term. Therefore,
\[
    \sum_{i=1}^n \|v_i-\mu_i\|_2^2
    \le
    \min_{c\in\mathbb{R}^d}
    \sum_{i=1}^n \|v_i-c\|_2^2
    +O(\log n).
\]
Applying this bound separately to the subsequence assigned to each of the
$k$ clusters gives the total penalty $O(k\log T)$.
\end{proof}

\projectedKMeans*
\begin{proof}[Proof of \Cref{thm:projected-kmeans}]
Fix the projection $\Pi$. For the surrogate losses
\[
    \widetilde{\ell}_t(Y)
    =
    \|x_t-\mu_{t,a_t^Y}^{Y}\|_2^2,
\]
the standard MWUA bound gives
\[
    \mathbb{E}_{\mathrm{alg}}
    \left[
        \sum_{t=1}^T \widetilde{\ell}_t(Y_t)
    \right]
    \le
    \sum_{t=1}^T \widetilde{\ell}_t(Y^\star)
    +
    O\left(\sqrt{T\log|\mathcal{E}|}\right),
\]
where $Y^\star$ is the projection-dependent grid expert defined in
\Cref{lem:marginal-jl-kmeans}.
The learner's actual loss is no larger than the surrogate loss of the sampled
expert, since
\[
    \ell_t(C_t(Y_t))
    =
    \min_{j\in[k]}\|x_t-\mu_{t,j}^{Y_t}\|_2^2
    \le
    \|x_t-\mu_{t,a_t^{Y_t}}^{Y_t}\|_2^2
    =
    \widetilde{\ell}_t(Y_t).
\]
Thus it remains to bound the cumulative surrogate loss of $Y^\star$.

Let
\[
    a_t
    :=
    \arg\min_{j\in[k]}\|\Pi x_t-y_j^\star\|_2^2
\]
be the projected assignment induced by $Y^\star$. This assignment partitions
the stream into $k$ subsequences. Applying \Cref{lem:ambient-ftl-kmeans} to
each induced subsequence gives
\[
    \sum_{t=1}^T \widetilde{\ell}_t(Y^\star)
    \le
    \sum_{j=1}^k
    \min_{c\in\mathbb{R}^d}
    \sum_{t:a_t=j}\|x_t-c\|_2^2
    +
    O(k\log T).
\]
For each cluster $j$, the offline center $c_j^\star$ is a feasible fixed
center for the corresponding induced subsequence. Therefore,
\[
    \sum_{t=1}^T \widetilde{\ell}_t(Y^\star)
    \le
    \sum_{t=1}^T \|x_t-c_{a_t}^\star\|_2^2
    +
    O(k\log T).
\]
Taking expectation over $\Pi$ and applying
\Cref{lem:marginal-jl-kmeans} at each time step,
\[
\begin{aligned}
    \mathbb{E}_{\Pi}
    \left[
        \sum_{t=1}^T \|x_t-c_{a_t}^\star\|_2^2
    \right]
    &=
    \sum_{t=1}^T
    \mathbb{E}_{\Pi}
    \left[
        \|x_t-c_{a_t}^\star\|_2^2
    \right] \\
    &\le
    (1+O(\varepsilon))
    \sum_{t=1}^T
    \min_{j\in[k]}\|x_t-c_j^\star\|_2^2
    +
    O(\varepsilon T) \\
    &=
    (1+O(\varepsilon))\mathrm{OPT}
    +
    O(\varepsilon T).
\end{aligned}
\]
Combining the MWUA comparison, the surrogate-to-actual loss inequality, and
the ambient tracking bound proves
\[
    \mathbb{E}_{\Pi,\mathrm{alg}}
    \left[
        \sum_{t=1}^T \ell_t(C_t)
    \right]
    \le
    (1+O(\varepsilon))\mathrm{OPT}
    +O(\varepsilon T)
    +O\left(\sqrt{T\log|\mathcal{E}|}\right)
    +O(k\log T).
\]

It remains only to bound the size of the expert class. The grid satisfies
\[
    |S_m|
    \le
    \left(\frac{C\sqrt m}{\varepsilon}\right)^m
\]
for a universal constant $C$. Since an expert is a $k$-tuple of projected
grid points,
\[
    |\mathcal{E}|\le |S_m|^k,
    \qquad
    \log|\mathcal{E}|
    =
    O\left(km\log\frac{\sqrt m}{\varepsilon}\right).
\]
Substituting
\[
    m=O\left(\varepsilon^{-2}\log\frac{k}{\varepsilon}\right)
\]
shows that the expert complexity has no dependence on $T$ or on the ambient
dimension $d$.
\end{proof}

The preceding guarantee should be interpreted as removing the
horizon-dependent and ambient-dimension-dependent discretization barrier from
the regret analysis. The projected expert class still has size exponential in
$k$ and $\varepsilon^{-1}$, but its logarithm is independent of both $T$ and
$d$. Thus the result plays the same conceptual role for online $k$-means that
the reduced fantope formulation plays for WLRA: it separates the statistical
comparison from the original ambient geometry. The difference is that the
fantope relaxation replaces the discrete expert class by a convex feasible
region, while the projected $k$-means result remains an experts-based
discretization.

\section{Grassmannian HRD for Exact-Comparator Online WLRA}
\label{sec:app-ghrd}

We prove \Cref{thm:grassmannian-hrd} for the standard WLRA loss. The argument
has three parts. First, bounded graph charts give a finite collection of
rank-$k$ representatives, including one close to the cumulative optimum
$P^\star$. Second, the cumulative loss at that representative differs from
$C_{\mathrm{WLRA}}$ by $O(W\sqrt T)$. Third, MTMW competes with this fixed
representative using its path mass. We give the chart construction and the
algorithm before proving the two loss bounds and combining them.

\subsection{Chart decomposition of the Grassmannian}

To partition the Grassmannian into small regions, we need coordinates for
each subspace. We obtain them by choosing $k$ reference rows of a basis
matrix and writing the remaining rows as a matrix $Y$. The first lemma
shows that every subspace has such a representation with entries of $Y$
in $[-1,1]$. The second shows that nearby coordinate matrices give nearby
subspaces, measured by the distance between their orthogonal projectors.

\begin{lemma}[Bounded graph charts]
\label{lem:charts}
For each $k$-element subset $J\subseteq[d]$, permute the rows in $J$
to the top and define, for $Y\in[-1,1]^{(d-k)\times k}$,
\[
    X_J(Y)=\begin{pmatrix}I_k\\Y\end{pmatrix},
    \qquad
    P_J(Y)=X_J(Y)\bigl(I_k+Y^\top Y\bigr)^{-1}X_J(Y)^\top,
\]
with the row permutation reversed in the resulting ambient projector.
Then:
\begin{enumerate}
    \item the $\binom{d}{k}$ charts cover $\mathcal{P}_k(d)$;
    \item if $P=P_J(Y)$, then $P_{J,J}$ is invertible and
    \[
        Y=P_{J^c,J}(P_{J,J})^{-1};
    \]
    \item every $P_J(Y)$ has rank exactly $k$, and is rational whenever
    $Y$ is rational.
\end{enumerate}
\end{lemma}

\begin{proof}
Take a full-column-rank basis matrix of a $k$-dimensional subspace.
Choose a $k\times k$ row minor with maximum absolute determinant and let
$J$ be its row indices. This minor is nonzero. Multiplying the basis on
the right by its inverse makes the selected rows equal to $I_k$ without
changing the column space. By Cramer's rule, each remaining entry is,
up to sign, the ratio of another $k\times k$ minor to the selected one.
Its magnitude is therefore at most $1$, proving the chart cover.

The blocks of $P_J(Y)$ are
\[
    P_{J,J}=(I_k+Y^\top Y)^{-1},
    \qquad
    P_{J^c,J}=Y(I_k+Y^\top Y)^{-1}.
\]
The first block is invertible, and multiplying the second by its inverse
recovers $Y$.

Finally, the columns of $X_J(Y)$ contain an identity minor, so they
are linearly independent and $P_J(Y)$ has rank $k$. If $Y$ is rational,
then $I_k+Y^\top Y$ and its inverse are rational, so $P_J(Y)$ is rational.
\end{proof}

\begin{lemma}[Stability within a graph chart]
\label{lem:chart-stability}
For every $k$-element subset $J\subseteq[d]$ and all
$Y,Z\in[-1,1]^{(d-k)\times k}$,
\[
    \|P_J(Y)-P_J(Z)\|_F
    \le \sqrt{2}\,\|Y-Z\|_F.
\]
\end{lemma}

\begin{proof}
Write $X=X_J(Y)$, $X'=X_J(Z)$, and
$U=X(X^\top X)^{-1/2}$. Then $U^\top U=I_k$ and
$UU^\top=P_J(Y)$. Since
\[
    X^\top X=I_k+Y^\top Y\succeq I_k,
\]
we have $\|(X^\top X)^{-1/2}\|_{\mathrm{op}}\le1$. Also,
$(I_d-P_J(Z))X'=0$, so
\[
\begin{aligned}
    \|(I_d-P_J(Z))U\|_F
    &=\|(I_d-P_J(Z))(X-X')(X^\top X)^{-1/2}\|_F\\
    &\le\|X-X'\|_F
     =\|Y-Z\|_F.
\end{aligned}
\]
For any two rank-$k$ projectors $P=UU^\top$ and $Q$,
\[
    \|P-Q\|_F^2
    =2k-2\operatorname{tr}(U^\top QU)
    =2\|(I_d-Q)U\|_F^2.
\]
Taking $Q=P_J(Z)$ proves the claim.
\end{proof}
\textbf{Regions and representatives.}
The chart boxes can overlap, so we first assign each projector to one
chart. Fix an ordering of the $k$-element subsets of $[d]$. A chart $J$
is eligible for $P\in\mathcal P_k(d)$ if $P_{J,J}$ is invertible and
$$Y_J(P):=P_{J^c,J}(P_{J,J})^{-1} \in[-1,1]^{(d-k) \times k}.$$
Every projector has an eligible chart by \Cref{lem:charts}(i),(ii).
We let $J(P)$ be the first eligible chart in the fixed ordering.

We now divide each chart box into dyadic cubes, viewing a coordinate
matrix as a vector of its $k(d-k)$ entries. The tree has a virtual
root with one child for each chart. At depth $h\ge0$ below a chart,
its cubes have side length $2^{1-h}$. We use a fixed half-open
endpoint convention, including the endpoint $1$, so that each
coordinate matrix belongs to exactly one cube at each depth.

A chart $J$ and cube $c$ define the region
$$R_{J,c} :=\{P\in\mathcal P_k(d):J(P)=J,\ Y_J(P)\in c\}.$$
Its representative is $P_{J,c}:=P_J(\bar Y_c)$, where $\bar Y_c$
is the center of $c$, reshaped as a $(d-k)\times k$ matrix. Each
projector belongs to exactly one region at every depth. Refinement
divides a cube into $2^{k(d-k)}$ children. Some regions are empty, but we retain these to keep the tree's branching factor uniform.

The representative $P_{J,c}$ need not itself belong to $R_{J,c}$:
the projector at the cube center might be assigned to an earlier
eligible chart. We only use $P_{J,c}$ to approximate projectors that
belong to $R_{J,c}$.

\begin{lemma}[Partition and radius]
\label{lem:grass-diam-upper}
For every depth $h$, the regions $R_{J,c}$ partition the space $\mathcal{P}_k(d)$ with every child region being contained in its parent. Additionally, for every $P\in R_{J,c}$ with $c$ at depth $h$,

$$\|P-P_{J,c}\|_F\le\sqrt{2k(d-k)}\;2^{-h},
\qquad
d_G(P,P_{J,c})\le\sqrt{k(d-k)}\;2^{-h}.$$
\end{lemma}

\begin{proof}
The fixed chart ordering assigns each projector $P$ a unique chart $J(P)$. Within that chart, the half-open cubes at depth $h$ partition the coordinate box, so $Y_{J(P)}(P)$ belongs to exactly one cube. Thus $P$ belongs to exactly one region at depth $h$. Each child cube lies inside its parent cube, so the corresponding child region lies inside its parent region. 

Now let $P\in R_{J,c}$, where $c$ has depth $h$. The side length of $c$
is $2^{1-h}$, so each entry of $Y_J(P)$ differs from the corresponding
entry of the center $\bar Y_c$ by at most $2^{-h}$. There are $k(d-k)$
entries, giving
$$\|Y_J(P)-\bar Y_c\|_F \leq \sqrt{k(d-k)}\,2^{-h}.$$
Since $P=P_J(Y_J(P))$ and $P_{J,c}=P_J(\bar Y_c)$,
\Cref{lem:chart-stability} gives
$$\|P-P_{J,c}\|_F \leq \sqrt{2k(d-k)}\,2^{-h}.$$
Finally, $d_G(P,Q)=\|P-Q\|_F/\sqrt2$ for rank-$k$ projectors, which
gives the chordal-distance bound.
\end{proof}

\textbf{Depth and learning rate.}
To make the quadratic comparator error of
\Cref{lem:quadratic} at most $\frac{3\pi^2}{2}W\sqrt T$, we use the following depth and number of leaves
$$L=\left\lceil\log_2\!\left(\kappa\sqrt{2k(d-k)}\,T^{1/4}\right)\right\rceil,
 \qquad N_{\mathrm{ex}}=\binom dk2^{k(d-k)L}.$$
All charts are built to this depth before the first round. For MTMW set
$K=\log N_{\mathrm{ex}}$ and $\eta=\sqrt{8K/T}$. These are the parameters of
\Cref{alg:ghrd-regret-min}; the next corollary verifies the claimed
resolution and leaf count.

\begin{corollary}[Depth and number of regions]
\label{lem:grass-max-regions}
Every leaf at depth $L$ satisfies
\[
    \|P-P_{J,c}\|_F
    \le \delta_L
    :=\sqrt{2k(d-k)}\,2^{-L}
    \le \kappa^{-1}T^{-1/4}
\]
for all $P$ in its region. The number of leaves is
\[
    N_{\mathrm{ex}} = \binom{d}{k}2^{k(d-k)L}
    \le
    \binom{d}{k}
    \Bigl(2\kappa\sqrt{2k(d-k)}\;T^{1/4}\Bigr)^{k(d-k)},
\]
and the whole tree has at most $2N+1$ nodes.
\end{corollary}

\begin{proof}
By the choice of $L$,
$2^L\ge\kappa\sqrt{2k(d-k)}\,T^{1/4}$, which gives
$\delta_L\le\kappa^{-1}T^{-1/4}$ by \Cref{lem:grass-diam-upper}.
Since $L$ is the smallest such integer and
$\kappa\sqrt{2k(d-k)}\,T^{1/4}>1$, also
$2^L<2\kappa\sqrt{2k(d-k)}\,T^{1/4}$, which gives the bound on
$N_{\mathrm{ex}}$. Each internal chart node has $2^{k(d-k)}\ge2$ children, so each
chart subtree has at most twice as many nodes as leaves; adding the virtual
root gives the final bound.
\end{proof}

\begin{algorithm}[H]
  \caption{Grassmannian Chart Decomposition}
  \label{alg:ghrd_update}
  \begin{algorithmic}
    \Require Dimensions $d,k$, depth $L$
    \Ensure Leaves $\mathcal{R}$ with representatives and masses
    \State $\mathcal{R}\gets\emptyset$
    \For{each $k$-subset $J\subseteq[d]$, in the fixed order}
      \For{each dyadic cube $c\subseteq[-1,1]^{k(d-k)}$ of side $2^{1-L}$}
        \State Add $R_{J,c}$ with basis $X_J(\bar Y_c)$, projector $P_J(\bar Y_c)$, and mass $\binom{d}{k}^{-1}2^{-k(d-k)L}$ to $\mathcal{R}$
      \EndFor
    \EndFor
    \State \Return $\mathcal{R}$
  \end{algorithmic}
\end{algorithm}

\begin{algorithm}[H]
  \caption{Grassmannian HRD Regret Minimization}
  \label{alg:ghrd-regret-min}
  \begin{algorithmic}
    \Require Horizon $T$, rank $k$, dimension $d$, bounds $0<\mu\le W$
    \Ensure Sequence of rank-$k$ projectors $P_1,\ldots,P_T$
    \State Set $L$, $N_{\mathrm{ex}}$, $K$, and $\eta$ as defined above
    \State Build the leaves $\mathcal{R}$ with \Cref{alg:ghrd_update}; set $w(R)\gets M(R)$ for each $R\in\mathcal{R}$
    \For{$t \gets 1$ to $T$}
      \State Sample $R_t$ with probability $w(R_t)/\sum_{R}w(R)$ and play $P_t=P_{R_t}$
      \State Receive $(x_t,W_t)$ and incur loss $\ell_t(P_t)$
      \For{each $R\in\mathcal{R}$}
        \State $w(R)\gets w(R)\exp\bigl(-\eta\,\ell_t(P_R)/W\bigr)$
      \EndFor
    \EndFor
  \end{algorithmic}
\end{algorithm}

\begin{remark}[Why the depth is fixed in advance]
\label{rem:fixed-depth}
The quadratic bound of \Cref{lem:quadratic} compares the cumulative loss at
$P^\star$ with the cumulative loss at one \emph{fixed} representative. It
does not transfer to a time-varying sequence of representatives chosen by a
loss-dependent refinement rule: the first-order errors on different
intervals need not cancel. We retain the nested tree construction because it makes the geometry
of the discretization explicit and may provide a starting point for
future adaptive refinements. We therefore refine every chart to depth $L$
before round $1$. The general MTMW statement in \Cref{sec:app-ghrd-mtmw}
also permits adaptive refinement, but the exact-comparator proof here uses
this fixed tree.
\end{remark}

\subsection{Loss stability and quadratic approximation}

For an orthonormal basis $U$ of a subspace, write
\[
    \ell_t(U)
    :=
    \min_{a\in\mathbb{R}^k}
    \|W_t^{1/2}(x_t-Ua)\|_2^2 .
\]
This notation is basis-invariant.

The exact-comparator analysis uses a second-order estimate for the whole
horizon, rather than a pointwise estimate for each round. We first prove that
estimate and then compare it with the first-order alternative.

\begin{lemma}[Quadratic approximation at the cumulative optimum]
\label{lem:quadratic}
Let $F(P)=\sum_{t=1}^T\ell_t(P)$ and let
$P^\star\in\arg\min_{P\in\mathcal{P}_k(d)}F(P)$.  Then every
$P\in\mathcal{P}_k(d)$ satisfies
\[
    F(P)-F(P^\star)
    \le
    \frac{3\pi^2}{2}\,
    W\kappa^2T\,\|P-P^\star\|_F^2.
\]
\end{lemma}

\begin{proof}
We connect $P^\star$ to $P$ by a Grassmannian geodesic and bound the second
derivative of the loss along it. Let $U(0)$ and $U(1)$ be principal-vector
orthonormal bases for their ranges. Along the principal-angle geodesic, the
$i$th basis vector rotates through angle $s\theta_i$. With
$\theta_{\max}:=\max_i\theta_i$, this gives
\[
    \|U'(s)\|_{\mathrm{op}}\le\theta_{\max},
    \qquad
    \|U''(s)\|_{\mathrm{op}}\le\theta_{\max}^2.
\]
Fix one round and abbreviate $W_t$ by $\widetilde W$. The normal matrix
$G(s)$ and right-hand side $b(s)$ describe weighted least squares in the
moving basis; $z(s)$ is the minimizing coefficient vector. Define
\[
    G(s):=U(s)^\top\widetilde WU(s),
    \qquad
    b(s):=U(s)^\top\widetilde Wx_t,
    \qquad
    z(s):=G(s)^{-1}b(s).
\]
Then
\[
    \ell_t(U(s))
    =
    x_t^\top\widetilde Wx_t-b(s)^\top z(s).
\]
The bound $G(s)\succeq\mu I_k$ prevents this least squares problem from
becoming singular. Because the fitted vector is a weighted orthogonal
projection, $\|z(s)\|_2\le\sqrt\kappa$. Differentiating the normal
matrix and right-hand side along the geodesic gives
\[
\begin{array}{lll}
    \|G'(s)\|_{\mathrm{op}}\le2W\theta_{\max}, &
    \|G''(s)\|_{\mathrm{op}}\le4W\theta_{\max}^2,\\[2mm]
    \|b'(s)\|_2\le W\theta_{\max}, &
    \|b''(s)\|_2\le W\theta_{\max}^2,
\end{array}
\]
Differentiating $G(s)z(s)=b(s)$ then gives
\[
    z'(s)=G(s)^{-1}\bigl(b'(s)-G'(s)z(s)\bigr),
    \qquad
    \|z'(s)\|_2
    \le
    \kappa\theta_{\max}(1+2\sqrt\kappa).
\]
The loss is a constant minus $q(s):=b(s)^\top z(s)$. Differentiating this
quadratic term, using $Gz=b$ to collect the terms involving $z''$, yields
\[
    q''(s)
    =
    2b''(s)^\top z(s)
    +2b'(s)^\top z'(s)
    -2z'(s)^\top G'(s)z(s)
    -z(s)^\top G''(s)z(s).
\]
Each displayed term is now controlled by the bounds for $G,b,z$ and their
derivatives. Since $\kappa\ge1$, they imply
\[
    |q''(s)|
    \le
    W\theta_{\max}^2
    \bigl(2\sqrt\kappa+6\kappa+8\kappa^{3/2}+8\kappa^2\bigr)
    \le
    24W\kappa^2\theta_{\max}^2.
\]
Thus
$|\frac{d^2}{ds^2}\ell_t(U(s))|
\le24W\kappa^2\theta_{\max}^2$.

This is a per-round curvature bound; stationarity enters only after the
rounds have been summed. Let $g(s):=F(P_{U(s)})$. The geodesic is feasible
for all real $s$, and $P^\star$ minimizes $F$, so $g'(0)=0$. Taylor's
theorem with integral remainder therefore gives
\[
    F(P)-F(P^\star)
    =
    \int_0^1(1-s)g''(s)\,ds
    \le
    12W\kappa^2T\theta_{\max}^2.
\]
Finally, convert the maximum principal angle into Frobenius distance
between the two projectors:
\[
    \|P-P^\star\|_F^2
    =
    2\sum_i\sin^2\theta_i,
    \qquad
    \theta_{\max}^2
    \le
    \frac{\pi^2}{8}\|P-P^\star\|_F^2.
\]
Substituting $\theta_{\max}^2\le(\pi^2/8)\|P-P^\star\|_F^2$ into the preceding Taylor
bound proves the claim.
\end{proof}

\begin{corollary}[Comparator path discretization error]
\label{cor:grass-path-approx}
Let $P_{p^\star}$ be the representative of the depth-$L$ leaf containing
$P^\star$.  Then
\[
    \sum_{t=1}^T\ell_t(P_{p^\star})
    \le
    C_{\mathrm{WLRA}}+\frac{3\pi^2}{2}W\sqrt T.
\]
\end{corollary}

\begin{proof}
By \Cref{lem:grass-max-regions},
$\|P_{p^\star}-P^\star\|_F\le
\kappa^{-1}T^{-1/4}$. Apply \Cref{lem:quadratic}: the factor
$W\kappa^2T$ multiplies a squared distance of at most
$\kappa^{-2}T^{-1/2}$, leaving $W\sqrt T$ and the stated constant.
\end{proof}

\begin{remark}[Why positive weights matter]
The assumption $\mu I_d\preceq W_t$ gives the uniform bound
\[
    \bigl\|(U^\top W_tU)^{-1}\bigr\|_{\mathrm{op}}
    \le \frac{1}{\mu}.
\]
With zero weights the standard WLRA loss
can fail to be continuous on the Grassmannian.  For example, take
$W_{\mathrm{ex}}=\operatorname{diag}(1,0)$, $x=e_1$, and $k=1$.  Every line
$U_\theta=\operatorname{span}((\cos\theta,\sin\theta))$ with
$\cos\theta\ne0$ has loss zero, whereas $U_{\pi/2}$ has loss one.  Thus a
uniform Grassmannian net cannot approximate the loss without a quantitative
positive lower bound.
\end{remark}

\subsection{Mass Tree MWUA}
\label{sec:app-ghrd-mtmw}

The geometry controls the cost of the comparator leaf. To control the
learner's additional loss, we assign that leaf a prior mass and apply a
tree version of multiplicative weights. For each node $v$, define
$M(\mathrm{root})=1$ and
$M(v)=M(\mathrm{par}(v))/\deg(\mathrm{par}(v))$. The mass of a path is
the mass of its terminal node. The mass identity below ensures that the
leaf priors sum to one. We state MTMW for trees that may be refined between
rounds: splitting an active leaf also splits its current weight equally.
Our fixed-depth algorithm is the special case with no subsequent splits.

\begin{lemma}
\label{lem:mass-region}
Let $v$ be a node and $\tilde{\mathcal T}$ a finite subtree rooted at $v$ that contains all children of each of its internal nodes. If $\tilde V$ is its leaf set, then $M(v)=\sum_{v'\in\tilde V}M(v')$.
\end{lemma}

\begin{proof}
By induction on the height of $\tilde{\mathcal T}$. If $\tilde{\mathcal T}=\{v\}$, then $\tilde V=\{v\}$. Otherwise let $\mathcal U$ be the children of $v$. By the induction hypothesis applied to the subtree below each $u\in\mathcal U$,
\[
    \sum_{v'\in\tilde V}M(v') = \sum_{u\in\mathcal U}M(u) = \sum_{u\in\mathcal U}\frac{M(v)}{|\mathcal U|} = M(v). \qedhere
\]
\end{proof}

\begin{theorem}
\label{thm:mtmwregret}
Consider a tree whose active leaves are the experts, initialized with weights equal to their masses, possibly refined adaptively between rounds with weights split equally among children. Suppose the losses $m_t(R)$ of the active representatives lie in $[0,1]$, and update $w(R)\gets w(R)e^{-\eta m_t(R)}$ with $\eta>0$. For a root-to-leaf path $p$, let $m_t(p)$ be the loss of its active node on round $t$. Then the mixture losses $\hat m_t=\sum_Rw(R)m_t(R)/\sum_Rw(R)$ satisfy
\[
    \sum_{t=1}^T\hat m_t-\sum_{t=1}^Tm_t(p)\le\frac{\log(1/M(p))}{\eta}+\frac{\eta T}{8}.
\]
For $K\ge\log(1/M(p))$ and $\eta=\sqrt{8K/T}$, the right-hand side is at most $\sqrt{TK/2}$. For an oblivious input sequence, the same bound holds for the expected loss of a leaf sampled with probability proportional to its weight. Each round takes time linear in the number of active leaves plus the cost of evaluating their losses.
\end{theorem}

\begin{proof}
The proof compares an upper bound on the total active weight with a lower
bound from any one comparator path. Let $\Omega_t$ be that total before round
$t$; by \Cref{lem:mass-region}, $\Omega_1=1$. A split preserves it because
the children's weights sum to their parent's. Applying Hoeffding's lemma
to a leaf drawn from the normalized weights gives
\[
    \Omega_{t+1}=\Omega_t\,\mathbb{E}_{R}\bigl[e^{-\eta m_t(R)}\bigr]\le \Omega_t\exp\Bigl(-\eta\hat m_t+\frac{\eta^2}{8}\Bigr),
    \qquad\text{so}\qquad
    \log \Omega_{T+1}\le-\eta\sum_t\hat m_t+\frac{\eta^2T}{8}.
\]
For the lower bound, follow the active ancestors of the terminal node of
$p$. Each split multiplies that branch's weight by the reciprocal of the
parent's degree, giving its path mass $M(p)$ overall. Each multiplicative
update on the branch contributes $e^{-\eta m_t(p)}$. Because this branch's
final weight is part of the total, we obtain
\[
    M(p)\exp\!\left(-\eta\sum_t m_t(p)\right)
    \le \Omega_{T+1}.
\]
Taking logarithms of the lower bound and combining it with the upper bound
on $\log \Omega_{T+1}$ gives the claimed regret inequality, even when refinements
depend on past inputs. The choice of $\eta$ gives
$K/\eta+\eta T/8=\sqrt{TK/2}$. For an oblivious sequence, the weights and
refinements do not depend on the learner's samples, so the expected sampled
loss on each round equals $\hat m_t$. Normalizing, sampling, and updating are
linear scans over the active leaves.
\end{proof}

\begin{corollary}
\label{cor-mtmw-regret-low-rank}
For standard online WLRA with $\|x_t\|_2\le1$ and $\mu I_d\preceq W_t\preceq WI_d$, running MTMW on the normalized losses $\ell_t/W$ with $\eta=\sqrt{8K/T}$ gives, for any path $p$ with $\log(1/M(p))\le K$,
\[
    \mathbb{E}\sum_{t=1}^T\ell_t(P_t)\le\sum_{t=1}^T\ell_t(P_{p,t})+W\sqrt{TK/2},
\]
where $P_{p,t}$ is the representative of the active node of $p$ on round $t$.
\end{corollary}

\begin{proof}
Since $0\le\ell_t\le W$, the normalized losses lie in $[0,1]$. Apply \Cref{thm:mtmwregret} and multiply by $W$.
\end{proof}

\subsection{Putting the components together}

\begin{restatable}[Comparator path mass]{lemma}{grassPathMass}
\label{lem:grass-path-mass}
For charts refined to depth $L$, every root-to-leaf path $p$ has
\[
  \log\frac1{M(p)}=\log\binom dk+k(d-k)L\log2.
\]
\end{restatable}

\begin{proof}
Every root-to-leaf path first selects one of the $\binom{d}{k}$ charts and
then makes $L$ choices among $2^{k(d-k)}$ children. Its prior mass is
\[
    M(p)=\binom{d}{k}^{-1}2^{-k(d-k)L}.
\]
\end{proof}

The preceding mass calculation gives $K=\log N_{\mathrm{ex}}$ for the comparator leaf.
Its loss is bounded by \Cref{cor:grass-path-approx}; combining the two
estimates proves the main theorem.

\ghrdregret*

\begin{proof}
All leaves are at depth $L$, so
$\log(1/M(p^\star))=\log N_{\mathrm{ex}}=K$.  The tree is fixed before the first round,
and hence $p^\star$ uses one representative throughout the horizon.
By \Cref{cor-mtmw-regret-low-rank,cor:grass-path-approx},
\[
\begin{split}
    \mathbb{E}\sum_{t=1}^T\ell_t(P_t)
    &\le
    \sum_{t=1}^T\ell_t(P_{p^\star})+W\sqrt{TK/2}\\
    &\le
    C_{\mathrm{WLRA}}
    +\frac{3\pi^2}{2}W\sqrt T
    +W\sqrt{TK/2}.
\end{split}
\]
Moreover, \Cref{lem:grass-max-regions} gives
\[
    K
    \le
    \log\binom{d}{k}
    +
    k(d-k)
    \log\!\left(
      2\kappa\sqrt{2k(d-k)}\,T^{1/4}
    \right).
\]
For $1\le k<d$, the first term is
$O(k(d-k)\log(k(d-k)))$.  Consequently,
\[
    K
    =
    O\!\left(
      k(d-k)\log\!\bigl(\kappa k(d-k)T\bigr)
    \right),
\]
which proves the regret bound.

For the running time, the algorithm evaluates all $N_{\mathrm{ex}}$ representatives on
every round. Store both the chart basis
$X_R=X_J(\bar Y_c)$ and its projector for each representative.  For any positive definite $W_t$ under the stated spectral bounds, evaluating one expert uses
\[
    G_R=X_R^\top W_tX_R,
    \qquad
    c_R=X_R^\top W_tx_t,
    \qquad
    \ell_t(P_R)=x_t^\top W_tx_t-c_R^\top G_R^{-1}c_R,
\]
and costs $O(d^2k+dk^2+k^3)$ arithmetic operations for a general dense
positive definite $W_t$ (or $O(dk^2+k^3)$ when $W_t$ is diagonal).
Sampling and updating take linear time in the $N_{\mathrm{ex}}$ leaves. Thus the total
cost is
$O(TN_{\mathrm{ex}}\operatorname{poly}(d))$. Since $1\le k<d$, we have
$\binom dk\le 2^{d-1}\le 2^{k(d-k)}$. Hence from the bound in \Cref{lem:grass-max-regions},
\[
N_{\mathrm{ex}}
\le
\bigl(4\kappa\sqrt{2k(d-k)}\,T^{1/4}\bigr)^{k(d-k)},
\]
which gives the stated running time.
\end{proof}

\section{Locally Convexified Grassmannian Algorithm}
\label{sec:app-grass-local}

We now prove \Cref{thm:grassmannian-approx}. The exact-comparator algorithm plays a fixed projector for each chart cube, whereas here each cube instead runs an online learner that can move within it. We first express the standard
WLRA loss in chart coordinates. On a small cube, freezing one matrix in
that expression gives a convex loss that approximates the true loss
multiplicatively. We then analyze the learner in one cube and use MWUA
to choose among cubes. Throughout, $W_t$ is diagonal with
$\mu I_d\preceq W_t\preceq WI_d$, $\kappa=W/\mu$, and
$0<\varepsilon\le1$. We use the same chart notation of
\Cref{sec:app-ghrd}.

\subsection{The loss in chart coordinates}

By \Cref{lem:charts}, every rank-$k$ projector has a representation
$P_J(Y)$ with $Y\in[-1,1]^{(d-k)\times k}$. We will also use the
coordinate bound
\[
    \|Y\|_{\mathrm{op}}\le\|Y\|_F\le\sqrt{k(d-k)},
    \qquad Y\in[-1,1]^{(d-k)\times k}.
\]
The last inequality follows by counting the $k(d-k)$ entries of $Y$.

The next formula removes the minimization over fitting coefficients
from the standard WLRA loss. It is the reason that freezing a matrix
will be useful.

\begin{lemma}[Chart formula]
\label{lem:grass-local-chart-formula}
For a chart $J$, a coordinate matrix $Y$, and a round $t$, define
\[
    r_{t,J}(Y):=x_{t,J^c}-Yx_{t,J},
    \qquad
    S_{t,J}(Y):=W_{t,J^c}^{-1}+YW_{t,J}^{-1}Y^\top.
\]
Then
\[
    \ell_t(P_J(Y))
    =r_{t,J}(Y)^\top S_{t,J}(Y)^{-1}r_{t,J}(Y).
\]
The loss and $S_{t,J}(Y)^{-1}r_{t,J}(Y)$ can each be computed in
$O(dk^2+k^3)$ arithmetic operations.
\end{lemma}

\begin{proof}
Fix $t$ and $J$, and abbreviate
$A=W_{t,J}$, $B=W_{t,J^c}$, and $r=r_{t,J}(Y)$.
In the basis $X_J(Y)$, a fitted vector has coordinates $(a,Ya)$.
Set $e=a-x_{t,J}$. Its residual has blocks $-e$ and $r-Ye$,
respectively, so
\[
    \ell_t(P_J(Y))
    =\min_{e\in\mathbb R^k}
       \bigl\{e^\top Ae+(r-Ye)^\top B(r-Ye)\bigr\}.
\]
The minimizing coefficient is
$e^\star=(A+Y^\top BY)^{-1}Y^\top Br$. Substitution gives
\[
    \ell_t(P_J(Y))
    =r^\top Br-r^\top BY(A+Y^\top BY)^{-1}Y^\top Br.
\]
Woodbury's identity identifies the matrix in this quadratic form as
$(B^{-1}+YA^{-1}Y^\top)^{-1}=S_{t,J}(Y)^{-1}$. It also gives
\[
    S_{t,J}(Y)^{-1}r
    =Br-BY(A+Y^\top BY)^{-1}Y^\top Br.
\]
Since $A$ and $B$ are diagonal, forming $Y^\top BY$ costs
$O(dk^2)$ operations, and solving the $k\times k$ system costs
$O(k^3)$.
\end{proof}

\subsection{A convex surrogate on each cube}

The residual $r_{t,J}(Y)$ is affine in $Y$, but the matrix
$S_{t,J}(Y)^{-1}$ changes with $Y$. For a fixed matrix $Z$, define
\[
    q_{t,J}^{Z}(Y)
    :=r_{t,J}(Y)^\top S_{t,J}(Z)^{-1}r_{t,J}(Y).
\]
This is a convex quadratic in $Y$: its matrix is fixed, positive
definite, and applied to an affine residual. Our goal is to choose
cubes small enough that replacing $S_{t,J}(Y)$ by its value at the
cube center changes the loss by only a factor close to one.

\begin{lemma}[Frozen-metric sandwich]
\label{lem:grass-local-sandwich}
Fix a chart $J$ and coordinate matrices $Y,Z$. Set
\[
    \rho=\|Y-Z\|_{\mathrm{op}},
    \qquad
    \theta=\kappa\bigl(2\|Z\|_{\mathrm{op}}\rho+\rho^2\bigr).
\]
If $\theta<1$, then
\[
    \frac{q_{t,J}^{Z}(Y)}{1+\theta}
    \le \ell_t(P_J(Y))
    \le \frac{q_{t,J}^{Z}(Y)}{1-\theta}
    \qquad\text{for every round }t.
\]
\end{lemma}

\begin{proof}
Write $E=Y-Z$. Expanding $S_{t,J}(Y)$ around $Z$ gives
\[
    S_{t,J}(Y)-S_{t,J}(Z)
    =EW_{t,J}^{-1}Z^\top
     +ZW_{t,J}^{-1}E^\top
     +EW_{t,J}^{-1}E^\top.
\]
Since $\|W_{t,J}^{-1}\|_{\mathrm{op}}\le1/\mu$,
$$\|S_{t,J}(Y)-S_{t,J}(Z)\|_{\mathrm{op}} \leq \frac{2\|Z\|_{\mathrm{op}}\rho+\rho^2}{\mu}.$$
Also $S_{t,J}(Z)\succeq W_{t,J^c}^{-1}\succeq W^{-1}I_{d-k}$.
The last two bounds show, for every vector $v$, that
$$\bigl|v^\top(S_{t,J}(Y)-S_{t,J}(Z))v\bigr| \leq \theta\,v^\top S_{t,J}(Z)v.$$
Consequently,
$$(1-\theta)S_{t,J}(Z) \preceq S_{t,J}(Y) \preceq(1+\theta)S_{t,J}(Z).$$
These matrices are positive definite when $\theta<1$. Inverting the
inequalities, evaluating them at $r_{t,J}(Y)$, and using the chart formula gives the claim.
\end{proof}

We now fix the cube resolution used by the algorithm. Let
\[
    D=\left\lceil\log_2
       \frac{12\kappa k(d-k)}{\varepsilon}\right\rceil,
    \qquad s=2^{1-D}.
\]
Let $\mathcal C$ contain every pair $(J,c)$ consisting of a chart
$J$ and a closed dyadic cube $c$ of side $s$ in its coordinate box.
Closed cubes may share boundary points; this does not affect the
algorithm because each cube maintains a separate learner. Write
$\bar Y_c$ for the center of $c$ and
$q_{t,J,c}:=q_{t,J}^{\bar Y_c}$.

\begin{corollary}[Cube resolution]
\label{cor:grass-local-cells}
For every $(J,c)\in\mathcal C$, every $Y\in c$, and every round $t$,
\[
    \frac{q_{t,J,c}(Y)}{1+\varepsilon/4}
    \le \ell_t(P_J(Y))
    \le \frac{q_{t,J,c}(Y)}{1-\varepsilon/4}.
\]
The number of chart--cube pairs is
\[
    N_\varepsilon:=|\mathcal C|
    =\binom dk 2^{k(d-k)D}
    \le\binom dk
       \left(\frac{24\kappa k(d-k)}{\varepsilon}
       \right)^{k(d-k)}.
\]
\end{corollary}

\begin{proof}
Each entry of $Y-\bar Y_c$ has magnitude at most $s/2=2^{-D}$.
Thus
\[
    \rho=\|Y-\bar Y_c\|_{\mathrm{op}}
    \le\sqrt{k(d-k)}\,2^{-D},
    \qquad
    \|\bar Y_c\|_{\mathrm{op}}\le\sqrt{k(d-k)}.
\]
For $Z=\bar Y_c$, the parameter in \Cref{lem:grass-local-sandwich}
therefore satisfies
\[
    \theta
    \le\kappa k(d-k)\bigl(2^{1-D}+4^{-D}\bigr)
    \le3\kappa k(d-k)2^{-D}
    \le\frac{\varepsilon}{4}.
\]
Apply that lemma to obtain the loss comparison. Finally,
$2^D<24\kappa k(d-k)/\varepsilon$, and each chart contains
$2^{k(d-k)D}$ cubes, giving the stated count.
\end{proof}

\subsection{The algorithm and its two regret bounds}

Each cube runs projected gradient descent on its frozen-metric
surrogate. MWUA treats the current projector from each cube as an
expert and updates its weights using the \emph{true} WLRA losses.
The full-information protocol allows both updates for every cube
after each round.

\begin{algorithm}[H]
  \caption{Locally Convexified Grassmannian MWUA}
  \label{alg:grass-local-MWUA}
  \begin{algorithmic}
    \Require Horizon $T$, $1\le k<d$, $0<\mu\le W$, and
             $0<\varepsilon\le1$
    \Ensure Rank-$k$ plays $P_1,\ldots,P_T$
    \State Set $\kappa=W/\mu$ and construct $D,s,\mathcal C,N_\varepsilon$
           as above
    \State Set $\eta=\sqrt{8\log N_\varepsilon/T}$,
           $G_{\mathrm{grad}}=2W\sqrt{1+k(d-k)}$, and
           $\Delta_{\mathrm{cube}}=s\sqrt{k(d-k)}$
    \For{each $(J,c)\in\mathcal C$}
      \State Set $Y_1^{J,c}=\bar Y_c$ and $w(J,c)=1$
    \EndFor
    \For{$t=1,\ldots,T$}
      \State Sample $(J,c)$ proportionally to $w(J,c)$ and play
             $P_t=P_J(Y_t^{J,c})$
      \State Observe $(x_t,W_t)$
      \For{each $(J,c)\in\mathcal C$}
        \State $\ell_t^{J,c}\gets\ell_t(P_J(Y_t^{J,c}))$
        \State $w(J,c)\gets w(J,c)
               \exp(-\eta\ell_t^{J,c}/W)$
        \State $g_t^{J,c}\gets
               -2S_{t,J}(\bar Y_c)^{-1}
                  r_{t,J}(Y_t^{J,c})x_{t,J}^\top$
        \State $Y_{t+1}^{J,c}\gets
               \operatorname{proj}_c\!\left(
               Y_t^{J,c}
               -\frac{\Delta_{\mathrm{cube}}}
                     {G_{\mathrm{grad}}\sqrt t}\,g_t^{J,c}
               \right)$
      \EndFor
    \EndFor
  \end{algorithmic}
\end{algorithm}

Here $\operatorname{proj}_c$ clips each matrix entry to its interval
in the closed cube $c$. In particular, every iterate stays inside
its cube and every play remains a rank-$k$ projector.

We first bound one cube's regret for its surrogate. The small cube
diameter makes this bound proportional to $\varepsilon$.

\begin{lemma}[Projected gradient descent in one cube]
\label{lem:grass-local-ogd}
Fix $(J,c)\in\mathcal C$ and abbreviate
$Y_t=Y_t^{J,c}$ and $q_t=q_{t,J,c}$. For every $Y^\circ\in c$,
\[
    \sum_{t=1}^Tq_t(Y_t)-\sum_{t=1}^Tq_t(Y^\circ)
    \le\frac32G_{\mathrm{grad}}\Delta_{\mathrm{cube}}\sqrt T
    \le\frac{3\varepsilon}{4}W\sqrt T.
\]
\end{lemma}

\begin{proof}
The cube is convex and has Frobenius diameter
$\Delta_{\mathrm{cube}}$. Each $q_t$ is convex, with gradient
\[
    \nabla q_t(Y)
    =-2S_{t,J}(\bar Y_c)^{-1}r_{t,J}(Y)x_{t,J}^\top.
\]
The spectral bounds give
$\|S_{t,J}(\bar Y_c)^{-1}\|_{\mathrm{op}}\le W$.
For $Y\in c$, the coordinate and data bounds give
\[
    \|r_{t,J}(Y)\|_2
    \le\sqrt{1+\|Y\|_{\mathrm{op}}^2}\,\|x_t\|_2
    \le\sqrt{1+k(d-k)}.
\]
Therefore $\|\nabla q_t(Y)\|_F\le G_{\mathrm{grad}}$.
Projected gradient descent with step size
$\gamma_t=\Delta_{\mathrm{cube}}/
(G_{\mathrm{grad}}\sqrt t)$ has regret at most
\[
    \frac{\Delta_{\mathrm{cube}}^2}{2\gamma_T}
    +\frac{G_{\mathrm{grad}}^2}{2}\sum_{t=1}^T\gamma_t
    \le\frac32G_{\mathrm{grad}}\Delta_{\mathrm{cube}}\sqrt T.
\]
By the choice of $D$,
$s\le\varepsilon/(6\kappa k(d-k))$. As $\kappa\ge1$ and
$k(d-k)\ge1$,
\[
    G_{\mathrm{grad}}\Delta_{\mathrm{cube}}
    =2W s\sqrt{k(d-k)(1+k(d-k))}
    \le\frac{\sqrt2}{3}\varepsilon W
    \le\frac{\varepsilon W}{2}.
\]
This gives the second inequality.
\end{proof}

Next, MWUA compares its sampled plays with the current learner
in any one cube. This comparison uses true losses; it does not
replace them by surrogates.

\begin{lemma}[MWUA over cubes]
\label{lem:grass-local-MWUA}
Let
$\ell_t^{J,c}=\ell_t(P_J(Y_t^{J,c}))\in[0,W]$.
For an oblivious input sequence,
\[
    \mathbb E\sum_{t=1}^T\ell_t(P_t)
    \le
    \min_{(J,c)\in\mathcal C}\sum_{t=1}^T\ell_t^{J,c}
    +W\sqrt{\frac{T\log N_\varepsilon}{2}}.
\]
\end{lemma}

\begin{proof}
Every cube's iterate depends only on earlier data, so the current
loss of each cube is determined before MWUA samples a cube.
Apply the standard MWUA bound to the losses
$\ell_t^{J,c}/W\in[0,1]$ with uniform initial weights.
The learning rate $\eta=\sqrt{8\log N_\varepsilon/T}$ gives
normalized regret at most
$\log N_\varepsilon/\eta+\eta T/8
=\sqrt{T\log N_\varepsilon/2}$.
Conditional on the data, the sampled play has expected loss equal
to MWUA's weighted average loss. Multiplying the regret bound by
$W$ proves the claim.
\end{proof}

We now move to a full proof of \Cref{thm:grassmannian-approx}.

\grassApproxRegret*

\begin{proof}
Choose an optimal projector $P^\star$ with cumulative loss
$C_{\mathrm{WLRA}}$. By \Cref{lem:charts}, it has bounded chart
coordinates: $P^\star=P_{J^\star}(Y^\star)$ for some
$Y^\star\in[-1,1]^{(d-k)\times k}$. Choose a cube $c^\star$
containing $Y^\star$. The algorithm has a learner for this pair
$(J^\star,c^\star)$.

Write $Y_t$ and $q_t$ for that learner's iterates and surrogates,
and put $\theta=\varepsilon/4$. The iterates and $Y^\star$ all lie
in $c^\star$. The cube comparison gives
\[
    \sum_{t=1}^T\ell_t^{J^\star,c^\star}
    \le\frac{1}{1-\theta}\sum_{t=1}^Tq_t(Y_t).
\]
The per-cube regret bound then gives
\[
    \sum_{t=1}^Tq_t(Y_t)
    \le\sum_{t=1}^Tq_t(Y^\star)
       +3\theta W\sqrt T.
\]
Finally, the other side of the cube comparison yields
$q_t(Y^\star)\le(1+\theta)\ell_t(P^\star)$.
Combining these three inequalities,
\[
    \sum_{t=1}^T\ell_t^{J^\star,c^\star}
    \le\frac{1+\theta}{1-\theta}C_{\mathrm{WLRA}}
       +\frac{3\theta}{1-\theta}W\sqrt T.
\]
Because $\theta=\varepsilon/4\le1/4$, the two coefficients
satisfy $(1+\theta)/(1-\theta)\le1+\varepsilon$ and
$3\theta/(1-\theta)\le\varepsilon$. Thus the chosen cube's true
loss is at most
$(1+\varepsilon)C_{\mathrm{WLRA}}+\varepsilon W\sqrt T$.
\Cref{lem:grass-local-MWUA} adds
$W\sqrt{T\log N_\varepsilon/2}$ for choosing that cube.

For the stated logarithmic bound, use
$\binom dk\le2^d$ and $d\le2k(d-k)$ for $1\le k<d$.
\Cref{cor:grass-local-cells} then gives
\[
    \log N_\varepsilon
    \le d\log2
      +k(d-k)\log\frac{24\kappa k(d-k)}{\varepsilon}
    \le k(d-k)\log\frac{96\kappa k(d-k)}{\varepsilon}.
\]
Substitution into the MWUA term proves the regret guarantee.

Each round updates all $N_\varepsilon$ cubes. For one cube,
\Cref{lem:grass-local-chart-formula} computes the true loss and
$S_{t,J}(\bar Y_c)^{-1}r_{t,J}(Y_t^{J,c})$ in
$O(dk^2+k^3)$ operations. The latter vector uses the metric
at the cube center and the residual at the current iterate.
The gradient, clipping, weight update, and amortized sampling
fit within the same per-cube bound. Combining this cost with
the cube count in \Cref{cor:grass-local-cells} proves the running
time bound. The algorithm stores one $(d-k)\times k$ iterate
and one MWUA weight per cube.
\end{proof}
\section{Fantope Relaxation}
\label{app:fantope}

This appendix proves the algebraic and optimization claims used in
\Cref{subsec:fantope}.We assume the following: $\|x_t\|_2\le1$ and $0\preceq W_t\preceq WI_d$.
The anchor $V$ is fixed before the online rounds. Diagonal $W_t$ is
required for the runtime stated in \Cref{thm:fantope-ogd}; the loss
identities and regret proof hold for positive semidefinite $W_t$. In particular, the only new object in this appendix
is the reduced Fantope variable $P\in\mathcal{F}_k(m)$, which is lifted to
the ambient space through a fixed orthonormal anchor
$V\in\mathbb{R}^{d\times m}$ by
\[
    \nu(P)=VPV^\top .
\]
The regret guarantee below is against the best fixed fractional comparator
in the reduced Fantope, not against the exact Grassmannian comparator from
\Cref{subsec:grassmann-hrd}.

\subsection{Lifting and Exact Loss Reduction}

We first record two basic facts used by the algorithm: the lift preserves
Fantope feasibility, and the ambient weighted loss can be written exactly
using reduced quantities.

\begin{lemma}
\label{lem:app-fantope-lift}
If $P\in\mathcal{F}_k(m)$, then $\nu(P)=VPV^\top\in\mathcal{F}_k(d)$.
\end{lemma}

\begin{proof}
Symmetry follows from
\[
    (VPV^\top)^\top=VP^\top V^\top=VPV^\top .
\]
For any $y\in\mathbb{R}^d$,
\[
    y^\top VPV^\top y=(V^\top y)^\top P(V^\top y)\ge 0,
\]
so $VPV^\top\succeq 0$. Since $P\preceq I_m$,
\[
    y^\top VPV^\top y
    \le \|V^\top y\|_2^2
    \le \|y\|_2^2,
\]
and thus $VPV^\top\preceq I_d$. Finally,
\[
    \mathrm{Tr}(VPV^\top)
    =
    \mathrm{Tr}(PV^\top V)
    =
    \mathrm{Tr}(P)
    =
    k.
\]
Thus $VPV^\top\in\mathcal{F}_k(d)$.
\end{proof}

For the loss reduction, define
\[
    \widetilde{\ell}_t(P)
    :=
    \|W_t^{1/2}(I_d-\nu(P))x_t\|_2^2
    =
    \|W_t^{1/2}(I_d-VPV^\top)x_t\|_2^2,
\]
and introduce
\[
    \alpha_t=x_t^\top W_t x_t,\qquad
    \bar{x}_t=V^\top x_t,\qquad
    \bar{z}_t=V^\top W_t x_t,\qquad
    \widetilde{W}_t=V^\top W_tV,\qquad
    \bar{X}_t=\bar{x}_t\bar{x}_t^\top .
\]

\begin{lemma}
\label{lem:app-fantope-decoupling}
For every $P\in\mathcal{F}_k(m)$,
\[
    \widetilde{\ell}_t(P)
    =
    \alpha_t
    -
    \mathrm{Tr}\left(
        P(\bar{z}_t\bar{x}_t^\top+\bar{x}_t\bar{z}_t^\top)
    \right)
    +
    \mathrm{Tr}\left(P\widetilde{W}_tP\bar{X}_t\right).
\]
\end{lemma}

\begin{proof}
Let $\bar{P}=VPV^\top$. Expanding the ambient loss gives
\[
    \ell_t(\bar{P})
    =
    x_t^\top W_t x_t
    -
    2x_t^\top W_t\bar{P}x_t
    +
    x_t^\top \bar{P}W_t\bar{P}x_t .
\]
The first term is $\alpha_t$. For the linear term,
\[
    x_t^\top W_t\bar{P}x_t
    =
    x_t^\top W_tVPV^\top x_t
    =
    \bar{z}_t^\top P\bar{x}_t .
\]
Since $P=P^\top$,
\[
    2\bar{z}_t^\top P\bar{x}_t
    =
    \mathrm{Tr}\left(
        P(\bar{z}_t\bar{x}_t^\top+\bar{x}_t\bar{z}_t^\top)
    \right).
\]
For the quadratic term,
\[
    x_t^\top \bar{P}W_t\bar{P}x_t
    =
    \bar{x}_t^\top P(V^\top W_tV)P\bar{x}_t
    =
    \mathrm{Tr}(P\widetilde{W}_tP\bar{X}_t).
\]
Combining the three identities proves the claim.
\end{proof}

\begin{lemma}
\label{lem:app-fantope-gradient}
The reduced loss $\widetilde{\ell}_t(P)$ is convex over
$\mathcal{F}_k(m)$, and
\[
    \nabla \widetilde{\ell}_t(P)
    =
    -(\bar{z}_t\bar{x}_t^\top+\bar{x}_t\bar{z}_t^\top)
    +
    \widetilde{W}_tP\bar{X}_t
    +
    \bar{X}_tP\widetilde{W}_t.
\]
\end{lemma}

\begin{proof}
The constant and affine terms in
\Cref{lem:app-fantope-decoupling} are convex. It remains to consider
\[
    H_t(P)=\mathrm{Tr}(P\widetilde{W}_tP\bar{X}_t).
\]
Since $\bar{X}_t=\bar{x}_t\bar{x}_t^\top$,
\[
    H_t(P)
    =
    \bar{x}_t^\top P\widetilde{W}_tP\bar{x}_t
    =
    \|\widetilde{W}_t^{1/2}P\bar{x}_t\|_2^2.
\]
Because $\widetilde{W}_t=V^\top W_tV\succeq 0$, this is the squared norm
of a linear function of $P$, and is convex.

For the gradient, set
\[
    C_t=\bar{z}_t\bar{x}_t^\top+\bar{x}_t\bar{z}_t^\top .
\]
The affine term contributes $-C_t$. For the quadratic term, write
$A=\widetilde{W}_t$ and $B=\bar{X}_t$. Then
\[
    d\,\mathrm{Tr}(PAPB)
    =
    \mathrm{Tr}(dPAPB)+\mathrm{Tr}(PAdPB)
    =
    \mathrm{Tr}(dP(APB+BPA)).
\]
Thus
\[
    \nabla H_t(P)
    =
    \widetilde{W}_tP\bar{X}_t+\bar{X}_tP\widetilde{W}_t,
\]
which gives the stated expression.
\end{proof}

The Euclidean projection onto $\mathcal{F}_k(m)$ is spectral. If
$M=M^\top$ has eigendecomposition
\[
    M=Q\mathrm{diag}(\lambda)Q^\top,
\]
then
\[
    \Pi_{\mathcal{F}_k(m)}(M)
    =
    Q\mathrm{diag}(\widehat{\lambda})Q^\top,
\]
where $\widehat{\lambda}$ is the projection of $\lambda$ onto the capped
simplex
\[
    \Delta_k^{[0,1]}
    =
    \left\{
        \mu\in[0,1]^m:
        \sum_{i=1}^m \mu_i=k
    \right\}.
\]
Equivalently, there exists a threshold $\tau$ such that
\[
    \widehat{\lambda}_i
    =
    \min\{1,\max\{0,\lambda_i-\tau\}\},
    \qquad
    \sum_{i=1}^m\widehat{\lambda}_i=k.
\]
Thus projection reduces to an eigendecomposition and a capped-simplex
projection.

\begin{lemma}
\label{lem:app-fantope-bounds}
The reduced Fantope satisfies
\[
    \sup_{P,Q\in\mathcal{F}_k(m)}\|P-Q\|_F\le \sqrt{2k}.
\]
Moreover, for every $P\in\mathcal{F}_k(m)$,
\[
    \|\nabla \widetilde{\ell}_t(P)\|_F\le 4W.
\]
\end{lemma}

\begin{proof}
For $P,Q\in\mathcal{F}_k(m)$,
\[
    \|P-Q\|_F^2
    =
    \mathrm{Tr}(P^2)+\mathrm{Tr}(Q^2)-2\mathrm{Tr}(PQ).
\]
Since $0\preceq P,Q\preceq I_m$ and both have trace $k$,
\[
    \mathrm{Tr}(P^2)\le k,
    \qquad
    \mathrm{Tr}(Q^2)\le k.
\]
Also $\mathrm{Tr}(PQ)\ge 0$, so $\|P-Q\|_F^2\le 2k$.

For the gradient bound, the affine term satisfies
\[
    \|\bar{z}_t\bar{x}_t^\top+\bar{x}_t\bar{z}_t^\top\|_F
    \le
    2\|\bar{z}_t\|_2\|\bar{x}_t\|_2
    \le
    2W.
\]
For the quadratic terms,
\[
    \|\widetilde{W}_tP\bar{X}_t\|_F
    \le
    \|\widetilde{W}_t\|_{\mathrm{op}}
    \|P\|_{\mathrm{op}}
    \|\bar{X}_t\|_F
    \le
    W,
\]
and the same bound holds for
$\|\bar{X}_tP\widetilde{W}_t\|_F$. Summing the three contributions gives
$\|\nabla \widetilde{\ell}_t(P)\|_F\le 4W$.
\end{proof}
\fantopethm*
\begin{proof}
Online gradient descent over a convex set of diameter $D$ with gradient
bound $G$ gives, for every fixed
$P^\star\in\mathcal{F}_k(m)$,
\[
    \sum_{t=1}^T \widetilde{\ell}_t(P_t)
    -
    \sum_{t=1}^T \widetilde{\ell}_t(P^\star)
    \le
    \frac{D^2}{2\eta}
    +
    \frac{\eta G^2T}{2}.
\]
By \Cref{lem:app-fantope-bounds}, take
\[
    D=\sqrt{2k},
    \qquad
    G=4W.
\]
With
\[
    \eta=\frac{\sqrt{2k}}{4W\sqrt{T}},
\]
we obtain
\[
    \sum_{t=1}^T \widetilde{\ell}_t(P_t)
    -
    \sum_{t=1}^T \widetilde{\ell}_t(P^\star)
    \le
    4W\sqrt{2kT}.
\]

For runtime, the reduced statistics
$\bar{x}_t=V^\top x_t$ and $\bar{z}_t=V^\top W_t x_t$ take $O(md)$ time,
while
\[
    \widetilde{W}_t=V^\top W_tV
\]
takes $O(m^2d)$ time since $W_t$ is diagonal. The gradient is computed in
$O(m^2)$ time in the reduced space. The Fantope projection requires an
$m\times m$ eigendecomposition, costing $O(m^3)$, plus capped-simplex
projection. Thus the per-round runtime is
\[
    O(m^2d+m^3).
\]
\end{proof}

\section{Lower Bound}
\label{app:lower_bound}

We distinguish the exact-comparator and $(1+\varepsilon)$-comparator
guarantees. The same offline hardness result constrains both, but its
accuracy parameter is important: it gives hardness at an
\emph{inverse-polynomial} approximation scale, rather than hardness for
an arbitrary fixed constant $\varepsilon$.

For an $m\times d$ data matrix $A$ with rows $a_i^\top$ and a strictly
positive entrywise weight matrix $H=(h_{ij})$, define
\[
  D_i=\operatorname{diag}(h_{i1},\ldots,h_{id}),\qquad
  \Phi(P)=\sum_{i=1}^m\min_{b\in\mathbb R^k}
          \|D_i^{1/2}(a_i-X_Pb)\|_2^2.
\]
The offline WLRA objective
$\min_{\operatorname{rank}(Z)\le k}\sum_{i,j}h_{ij}(A_{ij}-Z_{ij})^2$
equals $\min_{P\in\mathcal P_k(d)}\Phi(P)$: for a fixed subspace,
each row's fitting coefficients can be optimized independently, and
every rank-at-most-$k$ matrix has a row space contained in some
$k$-dimensional subspace. Write $\operatorname{OPT}=\min_P\Phi(P)$.

\begin{theorem}[Positive-weight WLRA hardness; \cite{GillisG11}]
\label{thm:wla:apx}
For rank $k=1$, binary $A\in\{0,1\}^{m\times d}$, and strictly
positive weights $0<h_{ij}\le1$, it is NP-hard to produce a solution
whose objective is within
\[
    \gamma_{m,d}:=2^{-11}(md)^{-6}
\]
of $\operatorname{OPT}$. The hard construction has weights bounded
below by an inverse polynomial in $md$.
\end{theorem}

The cited theorem states an \emph{additive} objective-accuracy gap.
It also yields hardness at a suitable $(1+\varepsilon)$ scale:
because the zero matrix has cost at most $md$ on these instances,
$\operatorname{OPT}\le md$. Set
$\varepsilon_{m,d}=\gamma_{m,d}/(2md)$. Any rank-one solution of cost
at most $(1+\varepsilon_{m,d})\operatorname{OPT}$ would have cost at
most $\operatorname{OPT}+\gamma_{m,d}/2$, contradicting
\Cref{thm:wla:apx}. In particular, the positive-weight problem has no
fully polynomial-time approximation scheme unless $\mathrm P=\mathrm{NP}$.
This argument does not establish NP-hardness for a fixed constant
$\varepsilon>0$.

We next convert a sublinear online regret guarantee into an offline
approximation algorithm. Polynomial running time in the reduction below
means polynomial \emph{bit} complexity for the online updates and for
evaluating the returned solutions; the exact-real arithmetic convention
used to state the upper bounds does not on its own imply that complexity
claim.

\begin{restatable}[Online-to-offline reduction]{theorem}{offlinetoonline}
\label{thm:offline-to-online}
For an offline instance with $m$ rows $a_i$ and positive diagonal weight
matrices $D_i$, write
$\Phi(P)=\sum_{i=1}^m\min_b\|D_i^{1/2}(a_i-X_Pb)\|_2^2$
and $\operatorname{OPT}=\min_P\Phi(P)$. Let
$B=\max\{1,\max_i\|a_i\|_2\}$. Suppose that for every oblivious
stream and a parameter $\sigma\ge0$, an online algorithm satisfies
\[
  \mathbb E\sum_{t=1}^T\ell_t(P_t)
  \le(1+\sigma)\min_P\sum_{t=1}^T\ell_t(P)+R_\sigma(T).
\]
Then a randomized offline algorithm using $T$ online rounds outputs a
rank-$k$ projector $\widehat P$ with
\[
  \mathbb E\,\Phi(\widehat P)
  \le(1+\sigma)\operatorname{OPT}
       +\frac{mB^2}{T}R_\sigma(T).
\]
Suppose the online guarantee is available for every $0<\sigma\le1$ and
\[
R_\sigma(T)\le
c_\sigma T^{1-\alpha}(\log(eT))^b
\]
for fixed constants $\alpha>0$ and $b\ge0$ independent of $\sigma$ and
$T$. Assume $c_\sigma$ is independent of $T$ and is bounded by a
polynomial in the offline input size, $B^2$, the weight condition number,
and $1/\sigma$. Assume also that the per-round bit complexity, including
the representation and evaluation of the output projector, is polynomial
in these parameters and $T$. Then, for any $\varepsilon,\delta>0$, there
is a randomized algorithm, polynomial in the same input parameters and
$1/\varepsilon,1/\delta$, that returns a solution of cost at most
$(1+\varepsilon)\operatorname{OPT}+\delta$ with probability at least
$2/3$. The same conclusion with $\varepsilon=0$ follows if an analogous
$\sigma=0$ guarantee holds with a horizon-independent coefficient.
\end{restatable}

\begin{proof}
Let $B=\max\{1,\max_i\|a_i\|_2\}$. Draw indices
$I_1,\ldots,I_T$ independently and uniformly from $[m]$. At time $t$
feed the online algorithm the normalized example
$(x_t,W_t)=(a_{I_t}/B,D_{I_t})$. The unnormalized objective of a
projector $P$ is $\Phi(P)$, so its normalized expected one-round loss is
\[
  \mathbb E_{I_t}\bigl[\ell_t(P)\bigr]
  =\frac{\Phi(P)}{mB^2}.
\]
The projector $P_t$ is chosen before $I_t$ is drawn. Conditional on the
past and the algorithm's randomness, this identity therefore also
holds with $P=P_t$. Let $P^\star$ minimize $\Phi$. For each realized
stream the online regret guarantee, with its comparator minimum bounded
above by the loss of the \emph{fixed} $P^\star$, gives
\[
 \mathbb E_{\mathcal A}\sum_{t=1}^T\ell_t(P_t)
 \le (1+\sigma)\sum_{t=1}^T\ell_t(P^\star)+R_\sigma(T).
\]
Taking expectation over the sampled indices and applying conditional
independence yields
\[
 \frac1{mB^2}\sum_{t=1}^T\mathbb E\,\Phi(P_t)
 \le \frac{(1+\sigma)T}{mB^2}\operatorname{OPT}
       +R_\sigma(T).
\]
Evaluate every $P_t$ on the entire offline instance and return
$\widehat P\in\arg\min_t\Phi(P_t)$. Since its cost is no greater than
the average of the $T$ costs, this proves
\[
  \mathbb E\,\Phi(\widehat P)
  \le(1+\sigma)\operatorname{OPT}
       +\frac{mB^2}{T}R_\sigma(T).
\]

For the high-probability statement, run the online algorithm with
$\sigma=\min\{1,\varepsilon/3\}$. Choose $T$ so that $mB^2R_\sigma(T)/T\le\delta/3$.
Because $c_\sigma$ is independent of $T$ and polynomially bounded in
the stated parameters,
\[
\frac{R_\sigma(T)}{T}
\le c_\sigma T^{-\alpha}(\log(eT))^b.
\]
For fixed $\alpha>0$ and $b\ge0$, such a $T$ is polynomial in the
stated parameters and $1/\varepsilon,1/\delta$.
The nonnegative random variable
$\Phi(\widehat P)-\operatorname{OPT}$ has expectation at most
$(\varepsilon\operatorname{OPT}+\delta)/3$. Markov's inequality gives
\[
 \Pr\!\left[\Phi(\widehat P)>
       (1+\varepsilon)\operatorname{OPT}+\delta\right]
 \le\frac13.
\]
For an exact-comparator online guarantee, use $\sigma=0$ and the same
argument with $mB^2R_0(T)/T\le\delta/3$. Full evaluation of the
$T$ outputs and the assumed bit-complexity bound keep the total
running time polynomial.
\end{proof}

\begin{corollary}[Barrier for both online guarantees]
\label{cor:wlra-computational-barrier}
Unless $\mathrm{NP}\subseteq\mathrm{BPP}$, there is no online family
with polynomial per-round bit complexity and a sublinear regret bound
as in \Cref{thm:offline-to-online} whose coefficient is polynomial in
the dimensions, the weight condition number, and $1/\varepsilon$
uniformly for every $0<\varepsilon\le1$. The same barrier applies to
a polynomial-time sublinear-regret algorithm against the exact
comparator under the analogous coefficient bound.
\end{corollary}

\begin{proof}
Apply \Cref{thm:offline-to-online} to the rank-one instances of
\Cref{thm:wla:apx}. On these instances $B^2\le d$,
$\operatorname{OPT}\le md$, and the condition number of the positive
weights is polynomial in $md$. Choose offline parameters
$\varepsilon=\gamma_{m,d}/(4md)$ and
$\delta=\gamma_{m,d}/4$. Both inverse parameters are polynomial in
$md$, while the resulting offline cost is at most
\[
 (1+\varepsilon)\operatorname{OPT}+\delta
 \le\operatorname{OPT}+\gamma_{m,d}/2
 <\operatorname{OPT}+\gamma_{m,d}
\]
with probability at least $2/3$. A randomized polynomial-time
algorithm achieving this on the hard instances would imply
$\mathrm{NP}\subseteq\mathrm{BPP}$. For the exact comparator, set
$\delta=\gamma_{m,d}/2$ instead.
\end{proof}

The exact Grassmannian construction and its locally convexified
$(1+\varepsilon)$ variant both have exponential dependence on
$k(d-k)$, so the corollary does not contradict either upper bound.
Its conclusion is about a uniform approximation family with polynomial
dependence on $1/\varepsilon$; it makes no claim that a fixed constant
factor $1+\varepsilon$ is NP-hard for general positive-weight WLRA.

\section{Experimental Details and Additional Results}
\label{app:other-experiments}

\subsection{Study Protocol and Implementations}
\label{subsec:experimental-details}

We compare \Cref{alg:ghrd-regret-min,alg:grass-local-MWUA}, Fantope OGD,
FTRL, and the Loss MEG and GD online-PCA baselines of
Nie et al.~\citep{nie2016online} on MNIST and MovieLens. We also compare Fantope OGD
and weighted FTRL on a chronological Air Quality stream. The synthetic fitting-loss
studies comprise WeightedImportanceCycleNoisy, WeightedShift, and a
heteroscedastic subspace-tracking stream. Methods receive the same
observations and coordinate weights within each trial, and every
prediction is made before the current observation is used for updating.
Runs that ignore weight receive identity weights during learning but
are evaluated against the actual weighted loss metric.

\textbf{Trials and development screening.}
The MNIST and MovieLens results use ten permutations (seeds $0,\ldots,9$) of each
fixed feature dataset, with orders paired across methods. The SDs therefore
measure order variation on the selected data, rather than variation across
independently sampled datasets. Air Quality uses a single chronological
evaluation stream, with no across-trial SD reported. Each synthetic fitting-loss study uses
five fresh validation seeds after development screening. WeightedShift
uses seeds $30,\ldots,34$ after a seed-$0$ design screen.
WeightedImportanceCycleNoisy was selected from six screened cases; the
four rotating-drift cases did not outperform weighted FTRL. The separate
Fantope tracking study also follows development screening. Reported SDs
are sample SDs across the stated trials. Synthetic trajectory figures
show means and paired differences; the endpoint tables provide sample SDs.
These experiments assess performance on selected constructions rather
than estimating average performance over a broad class of streams.

\textbf{Objectives.}
For a rank-$k$ basis $U_t$, Air Quality and all three synthetic fitting-loss
studies report
\[
  L_T=\sum_{t=1}^T\ell_t(U_t),\qquad
  \ell_t(U_t)=\min_{a\in\mathbb R^k}
  \|W_t^{1/2}(x_t-U_ta)\|_2^2.
\]
For rounded Fantope OGD, $U_t$ is the top-$k$ eigenspace of the current
ambient fractional matrix $VP_tV^\top$, extracted before updating.
This fitting loss differs from the fractional surrogate used for Fantope
updates. On MNIST and MovieLens, $W_t=I$ and Fantope instead reports its
native fractional projection loss
\[
  \widetilde\ell_t(P_t)=\|x_t-VP_tV^\top x_t\|_2^2.
\]
The online-PCA methods report the exact expected loss of randomized
rank-$k$ projectors. Offline SVD is a retrospective fixed-rank reference
for the unweighted feature streams; it does not lower-bound fractional
predictions. No offline weighted optimum is reported for Air Quality or the
synthetic fitting-loss studies, so their differences from weighted FTRL are baseline
comparisons rather than measured regret against the best fixed subspace.
The older recovery diagnostics in \Cref{subsec:weighted-synthetic} use
separate projection and subspace-error metrics.

\textbf{Grassmannian implementations.}
\Cref{alg:ghrd-regret-min} predicts fixed chart representatives.
\Cref{alg:grass-local-MWUA} runs projected gradient descent on a
frozen-metric surrogate inside each cube and uses true fitting losses
for multiplicative weights. Both implementations use the prescribed
chart grids with finite-step informed MCMC and proposals based on past
observations. \Cref{alg:grass-local-MWUA} uses $\varepsilon=1$,
corresponding to a theoretical comparator factor of $2$ for the analyzed
algorithm. The experiments use approximate sampling for computational efficiency and therefore serve as a proxy for true implementations of the algorithms with exact sampling.

The MNIST and MovieLens runs use tuned $\eta=1000$ for both methods, with $100$
transitions per round for \Cref{alg:ghrd-regret-min} and two outer
corrections of $50$ inner transitions for
\Cref{alg:grass-local-MWUA}. Their first $100$ rounds use budgets of
$1000$ and $500$ inner transitions, respectively. Both cyclic synthetic
studies use $\eta_4=0.357$ and $\eta_5=0.387$. On WeightedShift,
Algorithm 4 uses $100$ transitions per round; Algorithm 5 uses $25$
inner transitions and proposal scale $128$. Their initial transition
budgets are $500$ and $100$, respectively.

\textbf{Online-PCA baselines.}
We implement Loss MEG (Eq.~(2.6)) and GD (Eq.~(2.5)) of
Nie et al.~\citep{nie2016online}; Loss MEG is the online-PCA update introduced by
\cite{warmuth2008randomized}. Write $M_t=\mathbb E[I-Q_t]$, where $Q_t$
is a random orthogonal rank-$k$ projector, to distinguish the PCA state
from the coordinate weights $W_t$. Both methods maintain
\[
  0\preceq M_t\preceq I_d,\qquad
  \operatorname{tr}(M_t)=d-k,\qquad
  M_1=\frac{d-k}{d}I_d.
\]
Their pre-update loss is evaluated analytically:
\[
  \ell_t^{\mathrm{PCA}}=x_t^\top M_tx_t
  =\mathbb E\bigl[\|x_t-Q_tx_t\|_2^2\bigr].
\]
This is expected rank-$k$ compression loss, rather than the squared
residual of a fractional prediction or a top-$k$ rounding. Its best fixed
comparator coincides with offline rank-$k$ SVD. Analytic evaluation removes
projector-sampling variance; the reported SDs reflect stream-order variation.

Loss MEG forms $\exp(\log M_t-\eta x_tx_t^\top)$ and projects onto the
stated set using quantum relative entropy. GD forms
$M_t-\eta x_tx_t^\top$ and uses Euclidean projection. Both projections
operate on eigenvalues, imposing the cap at one and trace $d-k$.
The untuned horizon-based rates are
\[
  \eta_{\mathrm{MEG}}=
  \log\!\left(1+\sqrt{\frac{2d\log(d/(d-k))}{T}}\right),
  \qquad
  \eta_{\mathrm{GD}}=\sqrt{\frac{k(d-k)}{dT}}.
\]
The MEG rate uses the a priori loss budget $B=T(d-k)/d$; the GD rate
minimizes the $L_1$-instance bound in Appendix B, Eq.~(B.1), of
Nie et al.~\citep{nie2016online}. No evaluated-stream loss is used to choose these
rates. Both PCA baselines start isotropically, whereas MNIST and MovieLens FTRL
starts from the first $k$ feature coordinates. 

\textbf{Fantope}
Fantope OGD updates a fractional matrix in a fixed anchor space after
initialization. The MNIST and MovieLens anchor dimension is $m=\min(d,3k)$ and
the prefix length is $\min\{50,\max\{10,m\}\}$, giving $30$, $45$, and
$50$ observations for $k=10,15,20$. WeightedShift uses $m=d=6$, a
ten-observation prefix, and the default step size without Fantope tuning.
These prefix rounds use zero reconstruction; the fixed-anchor guarantee
concerns the suffix after initialization. The tracking study uses a
constant Fantope step size. Its rounded predictions are evaluated using
weighted fitting loss, which is not the fractional-surrogate guarantee.

\textbf{FTRL.} Unweighted FTRL updates from past second moments. Weighted FTRL uses
warm-started ALS to approximately fit the accumulated weighted objective.
WeightedImportanceCycleNoisy and the tracking study use three ALS
iterations for their main baselines, with checks at thirty and ten
iterations, respectively. Air Quality uses three ALS iterations and a
thirty-iteration check. The comparisons concern these approximate
solvers; they do not establish performance relative to an exact FTRL
oracle. WeightedShift includes one weighted FTRL baseline.

\textbf{Legacy diagnostic control.}
The MNIST and MovieLens trajectory plots include a fixed matrix $B$ with
normalized, nonorthogonal columns, scored by
$\|x_t-BB^\top x_t\|_2^2$. This is a diagnostic control rather than a
uniformly random orthogonal rank-$k$ projector. 

\subsection{MNIST and MovieLens Feature Streams}
\label{subsec:realdata-experiments}

\textbf{Data and comparison design.}
For MNIST, Truncated SVD is fitted to all $70{,}000$ images with $784$
features. We retain $d=50$ coordinates, normalize each row, and stream the
first $5000$ selected observations in each trial's order. For MovieLens
20M, we retain the $1000$ most frequently rated movies and the $2000$ most
active users among them, subtract $3$ from observed ratings, and zero-fill
missing entries. Truncated SVD is fitted to that selected matrix, followed
by row normalization. The feature maps are constructed once and shared
across trials. These experiments evaluate online updates in fixed,
transductively constructed feature spaces, rather than past-only feature
learning or weighted fitting in the original movie coordinates.

\begin{figure}[!htb]
\centering
  \includegraphics[width=\linewidth]{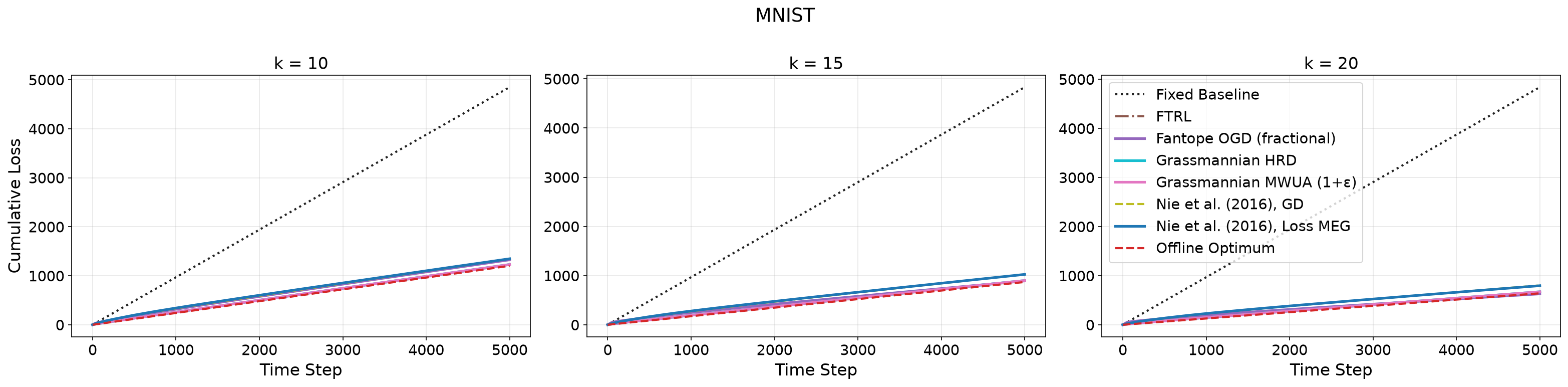}
\caption{MNIST cumulative reported losses over ten trials, with $d=50$
and $T=5000$. Panels show $k=10,15,20$ from left to right; curves are
means and shading shows one sample SD. Fantope uses fractional projection
loss; \Cref{alg:ghrd-regret-min,alg:grass-local-MWUA} and FTRL use rank-$k$ reconstruction loss.
``Grassmannian HRD'' denotes \Cref{alg:ghrd-regret-min} and ``Grassmannian MWUA
$(1+\varepsilon)$'' denotes \Cref{alg:grass-local-MWUA}. The fixed control is described
in the protocol.}
\label{fig:MNIST}
\end{figure}

\begin{figure}[!htb]
\centering
\includegraphics[width=\linewidth]{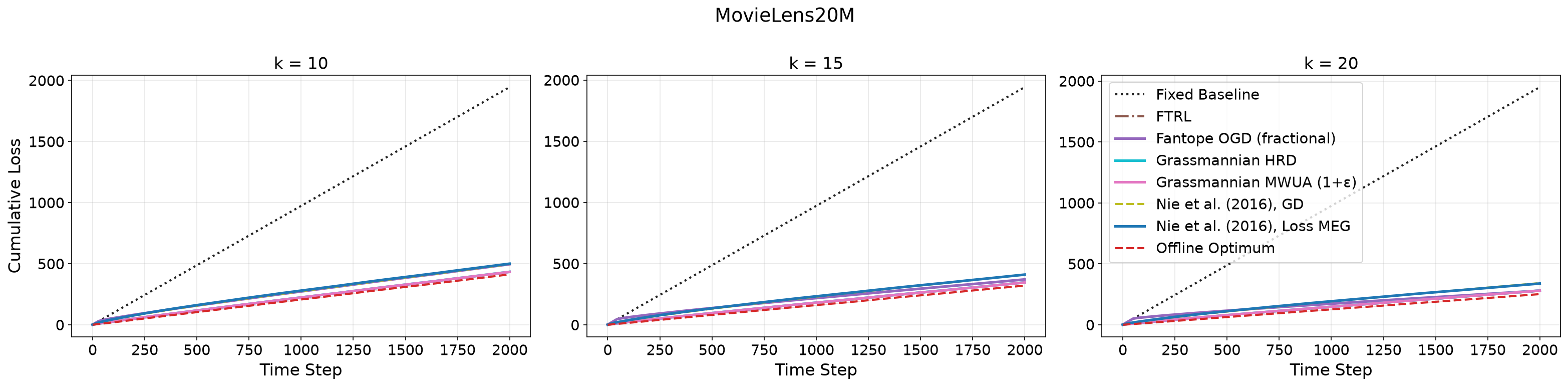}
\caption{MovieLens 20M cumulative reported losses over ten trials,
with $d=50$, $T=2000$, and $k=10,15,20$ from left to right. Curves are
means and shading shows one sample SD. Fantope uses fractional projection
loss; \Cref{alg:ghrd-regret-min,alg:grass-local-MWUA} and FTRL use rank-$k$ reconstruction loss.
Algorithm and control labels follow \Cref{fig:MNIST}.}
\label{fig:MovieLens}
\end{figure}

\textbf{Findings.}
\Cref{tab:realdata-experiments} and \Cref{fig:MNIST,fig:MovieLens} report all tested
ranks. \Cref{alg:ghrd-regret-min,alg:grass-local-MWUA} track FTRL closely, with slightly higher mean
rank-$k$ reconstruction loss in every configuration. \Cref{alg:grass-local-MWUA} has a
smaller mean loss than \Cref{alg:ghrd-regret-min} across these configurations.
Both Grassmannian methods also have lower mean loss than the untuned Loss
MEG and GD baselines at every tested rank on both datasets. These are
comparisons under the reported rates and initialization protocols, not a
claim that the Grassmannian methods improve the optimal online-PCA regret
guarantees or dominate tuned PCA implementations.
Fantope's reported fractional loss on MNIST changes from $1326.36\pm22.17$
at $k=10$ to $892.14\pm6.62$ at $k=15$ and $629.18\pm0.32$ at $k=20$.
The latter two values are below the Grassmannian reconstruction losses;
at $k=20$, its value is also below the fixed-rank offline SVD reference.

On MovieLens, Fantope reports $495.48\pm5.98$, $370.48\pm2.31$, and
$280.11\pm0.30$ at $k=10,15,20$. Its reported values exceed the
Grassmannian reconstruction losses at $k=10,15$ and are close at $k=20$.
The row plots therefore show the dataset and rank dependence of the
fractional relaxation alongside the reconstruction methods.

\textbf{Numerical comparisons.}
\Cref{tab:realdata-experiments} gives final losses under each method's
stated objective. \Cref{tab:realdata-excess} gives mean excess loss over
offline SVD for methods using the rank-$k$ reconstruction objective or
its exact randomized expectation. Because the selected data and SVD
reference are fixed across permutations, each excess has the same sample
SD as the corresponding cumulative loss. Fantope is excluded from this
excess table because its fractional objective differs.

\begin{table}[!htb]
\centering
\small
\setlength{\tabcolsep}{2pt}
\renewcommand{\arraystretch}{1.2}
\begin{tabular}{@{}lcrrrrrrr@{}}
\toprule
Dataset & $k$ & Alg.~4 & Alg.~5 & Fantope$^\dagger$ & FTRL
 & \shortstack{Nie Loss\\MEG} & Nie GD & \shortstack{Offline\\SVD} \\
\midrule
MNIST & 10 & $1230.06\pm0.73$ & $1229.05\pm0.59$ & $1326.36\pm22.17$ & $1228.46\pm0.51$ & $1347.31\pm0.09$ & $1344.88\pm0.59$ & $1203.07$ \\
MNIST & 15 & $899.95\pm0.54$ & $899.93\pm0.56$ & $892.14\pm6.62$ & $898.72\pm0.77$ & $1024.74\pm0.06$ & $1026.68\pm0.44$ & $869.97$ \\
MNIST & 20 & $672.38\pm0.69$ & $672.33\pm0.73$ & $629.18\pm0.32$ & $670.65\pm0.63$ & $796.05\pm0.05$ & $802.42\pm0.27$ & $640.84$ \\
\addlinespace[3pt]
MovieLens & 10 & $433.32\pm0.45$ & $432.92\pm0.45$ & $495.48\pm5.98$ & $431.83\pm0.44$ & $499.91\pm0.06$ & $496.30\pm0.20$ & $412.51$ \\
MovieLens & 15 & $346.21\pm0.64$ & $345.66\pm0.46$ & $370.48\pm2.31$ & $344.18\pm0.54$ & $409.93\pm0.06$ & $410.19\pm0.19$ & $320.23$ \\
MovieLens & 20 & $280.40\pm0.61$ & $280.15\pm0.54$ & $280.11\pm0.30$ & $278.44\pm0.51$ & $338.41\pm0.05$ & $342.08\pm0.19$ & $251.49$ \\
\bottomrule
\end{tabular}
\caption{MNIST and MovieLens: final cumulative reported loss, mean
$\pm$ sample SD over ten paired stream orders. Algorithms 4 and 5 and
FTRL use rank-$k$ reconstruction loss; Nie Loss MEG and GD use its
analytic expectation under randomized rank-$k$ predictions.
$^\dagger$Fantope uses fractional projection loss. Offline SVD is an
order-invariant fixed-rank reference, not a lower bound for fractional
predictions.}
\label{tab:realdata-experiments}
\end{table}
\begin{table}[!htb]
\centering
\small
\setlength{\tabcolsep}{4pt}
\renewcommand{\arraystretch}{1.2}
\begin{tabular}{lrrrrrr}
\toprule
Dataset & $k$ & Alg.~4 & Alg.~5 & FTRL & Nie Loss MEG & Nie GD \\
\midrule
MNIST & 10 & $26.99\pm0.73$ & $25.98\pm0.59$ & $25.39\pm0.51$ & $144.24\pm0.09$ & $141.81\pm0.59$ \\
MNIST & 15 & $29.98\pm0.54$ & $29.96\pm0.56$ & $28.75\pm0.77$ & $154.77\pm0.06$ & $156.71\pm0.44$ \\
MNIST & 20 & $31.54\pm0.69$ & $31.49\pm0.73$ & $29.81\pm0.63$ & $155.21\pm0.05$ & $161.58\pm0.27$ \\
\addlinespace[3pt]
MovieLens & 10 & $20.81\pm0.45$ & $20.41\pm0.45$ & $19.32\pm0.44$ & $87.40\pm0.06$ & $83.79\pm0.20$ \\
MovieLens & 15 & $25.98\pm0.64$ & $25.43\pm0.46$ & $23.95\pm0.54$ & $89.70\pm0.06$ & $89.96\pm0.19$ \\
MovieLens & 20 & $28.91\pm0.61$ & $28.66\pm0.54$ & $26.95\pm0.51$ & $86.92\pm0.05$ & $90.59\pm0.19$ \\
\bottomrule
\end{tabular}
\caption{Mean excess cumulative rank-$k$ reconstruction loss over offline
SVD, with sample SD across ten stream orders. Means are computed from
the reported rounded losses. SVD is order-invariant, so the sample SDs
equal those of the cumulative losses.}
\label{tab:realdata-excess}
\end{table}
\subsection{Air Quality: Chronological Weighted Reconstruction}
\label{subsec:air-quality}

\textbf{Data and preprocessing.}
This exploratory real-data comparison on $T=1512$
chronologically ordered UCI Air Quality ~\citep{vito2008airquality} observations for all methods,
with $d=12$ and target rank $k=2$. Scaling is fitted on the first $256$
raw hours, which are excluded from evaluation. Subsequent observations
are sampled every six hours; missing values are zero-filled after
scaling, and retained rows are unit-normalized. Observed entries receive
weight $1$ and naturally missing entries receive weight $0.01$, so all
weights are strictly positive. The reported losses measure normalized
reconstruction, rather than prediction error in physical sensor units.

\textbf{Comparison design.}
Fantope OGD and both weighted FTRL variants receive the same observations
and weights. Each method's current rank-two subspace is scored before
updating using the weighted least-squares fitting loss defined in
\Cref{subsec:experimental-details}. Fantope uses constant-step gradient
updates; weighted FTRL approximately optimizes the accumulated weighted
objective using three or thirty ALS iterations. Both methods change
their subspaces online. Results are reported for the single chronological
evaluation stream, without an across-trial SD.

\begin{figure}[H]
\centering

\includegraphics[width=0.5\linewidth]{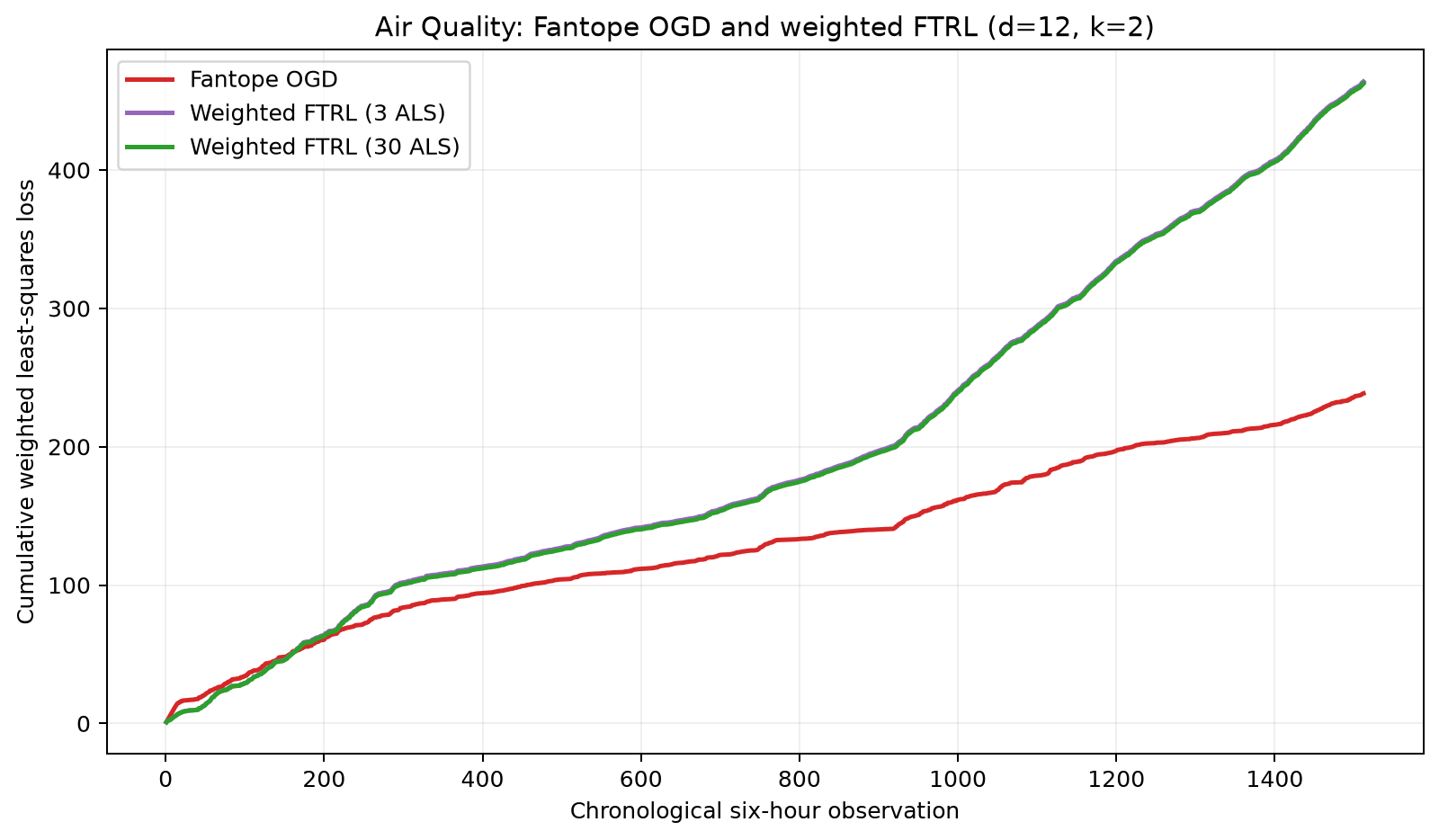}
\caption{Air Quality cumulative weighted fitting losses for Fantope OGD
and weighted FTRL on the chronological stream, with $d=12$, $k=2$, and
$T=1512$. Observed and missing entries receive weights $1$ and $0.01$,
respectively. Each subspace is evaluated before updating on scaled,
unit-normalized observations. Curves show the single evaluation stream;
final losses and ALS iteration budgets appear in \Cref{tab:air-quality}.}
\label{fig:air-quality}
\end{figure}

\textbf{Results and interpretation.}
\Cref{tab:air-quality} gives cumulative and per-observation losses;
\Cref{fig:air-quality} shows the cumulative trajectories.
Fantope OGD attains cumulative loss $238.5191$, compared with $463.8658$
and $462.4832$ for weighted FTRL using three and thirty ALS iterations,
respectively. Its loss is $48.43\%$ lower than the stronger,
thirty-iteration baseline. Increasing the baseline's inner optimization
budget tenfold reduces its loss by only $0.30\%$, so the substantial
gap persists under this check. This check does not certify convergence
to an exact weighted FTRL solution.

The result supports an advantage for Fantope OGD over
these implemented weighted-FTRL baselines in some real-world datasets. Constant-step updates may
respond more effectively to changing structure than approximate
optimization over the entire observation history; the comparison is
consistent with this explanation but does not inheritely isolate it experimentally.

\begin{table}[!htb]
\centering
\small
\setlength{\tabcolsep}{5pt}
\renewcommand{\arraystretch}{1.2}
\begin{tabular}{lrr}
\toprule
Method & \shortstack{Cumulative\\weighted loss}
 & \shortstack{Mean loss per\\observation} \\
\midrule
Fantope OGD & $\mathbf{238.5191}$ & $\mathbf{0.157751}$ \\
Weighted FTRL (3 ALS iterations) & $463.8658$ & $0.306790$ \\
Weighted FTRL (30 ALS iterations) & $462.4832$ & $0.305875$ \\
\bottomrule
\end{tabular}
\caption{Air Quality: pre-update weighted rank-two fitting losses on the
same $1512$ chronological observations, with $d=12$, observed-entry weight
$1$, and missing-entry weight $0.01$. Mean loss is cumulative loss
divided by $1512$. Losses describe normalized reconstruction.}
\label{tab:air-quality}
\end{table}

\subsection{WeightedImportanceCycleNoisy: Construction and Results}
\label{subsec:weighted-importance-cycle}

\textbf{Construction.}
This stream contains $T=2000$ observations in $d=6$, with target rank
$k=2$. A random permutation assigns four coordinate directions to a
common direction $c$, competitors $a,b$, and a distractor $n$. After
six initialization observations giving $a$ a small initial advantage,
the generator repeats the eight-observation sequence
\[
  e_b,\ e_c,\ e_a,\ e_n,\ e_c,\ e_a,\ e_n,\ e_n.
\]
Within each block, draw $q$ uniformly from $\{0.75,1\}$. The common
coordinate has weight $1$, the competitors have weights $0.4q$ and
$0.8q$, and the distractor and remaining coordinates have weight
$1/64$. All weights are strictly positive, with $\kappa=64$.
Before noise is added, the competitors' total weighted energies in
each block balance: $2(0.4q)=0.8q$. Their ordering creates temporary
changes in cumulative directional importance. The distractor occurs
three times per block but has small weight.

Independent Gaussian noise with standard deviation $0.005$ is added to
every coordinate before each observation is normalized to unit length.
The remaining coordinates contain only this added noise. These weights
encode feature importance rather than inverse noise variance. The
construction is independent of algorithm predictions. The cyclic
energy balance describes the noiseless components; perturbation and
normalization need not preserve an exact balance in each realized block.
The six-case development screen and fresh validation protocol are
described in \Cref{subsec:experimental-details}.

\textbf{Results and ablations.}
\Cref{tab:weighted-importance-cycle} gives final weighted fitting losses;
the trajectories appear in the main-body
\Cref{fig:weighted-importance-cycle}.
Both weighted Grassmannian implementations outperform the main
approximate weighted FTRL baseline on every validation seed. Their mean
losses are $209.25$ and $208.69$, versus $281.03$ for weighted FTRL.
Increasing the weighted FTRL budget from three to thirty ALS iterations
gives mean loss $282.43$, so this budget increase does not close the
observed gap. This check does not establish convergence to the exact
weighted FTRL solution.

Ignoring weights during Grassmannian learning increases the mean losses
to $353.84$ and $355.90$. Unweighted FTRL and rounded Fantope OGD give
$351.90$ and $360.82$, respectively. The ablations support the importance
of incorporating feature weights in this constructed stream and the robustness of the Grassmannian algorithms in comparison to FTRL.

\begin{table}[!htb]
\centering
\small
\setlength{\tabcolsep}{5pt}
\renewcommand{\arraystretch}{1.2}
\begin{tabular}{lr}
\toprule
Method & Weighted fitting loss \\
\midrule
\Cref{alg:ghrd-regret-min} & $209.25\pm4.66$ \\
\Cref{alg:grass-local-MWUA} ($\varepsilon=1$) & $208.69\pm6.23$ \\
Weighted FTRL (3 ALS iterations) & $281.03\pm3.74$ \\
Weighted FTRL (30 ALS iterations) & $282.43\pm3.82$ \\
Unweighted FTRL & $351.90\pm4.51$ \\
Rounded Fantope OGD & $360.82\pm5.36$ \\
\Cref{alg:ghrd-regret-min}, ignoring weights & $353.84\pm8.72$ \\
\Cref{alg:grass-local-MWUA}, ignoring weights & $355.90\pm4.94$ \\
\bottomrule
\end{tabular}
\caption{WeightedImportanceCycleNoisy: cumulative weighted fitting loss,
mean $\pm$ sample SD across five fresh validation seeds, each with
$T=2000$ rounds. All methods are scored with the actual weights.
Both weighted FTRL variants use approximate ALS updates; Fantope is
evaluated through its rounded rank-two subspace.}
\label{tab:weighted-importance-cycle}
\end{table}
\subsection{WeightedShift: Construction and Results}
\label{subsec:weightedshift-details}

\textbf{Construction.}
WeightedShift is a noiseless companion to the main-body cyclic
experiment, with $T=4000$, $d=6$, and $k=2$. For each seed, a permutation
of the coordinate axes selects a common direction $c$, competitors $a,b$,
and a nuisance direction $n$. The first four observations are $e_c$.
A symmetric pair $\tfrac12 e_a\pm\tfrac{\sqrt3}{2}e_n$ establishes an
energy offset of $1/2$ between $a$ and $b$. Subsequent observations repeat
$e_b,e_c,e_a,e_c$. Every vector has unit norm.

The common coordinate has weight $1$. In each four-round block, the
competing coordinates share a weight $q\in\{3/4,1\}$ sampled independently
across blocks; other coordinates have weight $1/64$. During the initial
six rounds, the competing weights are $1$. Thus $\kappa=64$ and every
weight is positive. Each complete competing-direction pair adds equal
weighted energy to $a$ and $b$, while the within-pair order changes their
cumulative-energy advantage. Weights encode feature importance, and no
observation noise is added. The stream is independent of learner
predictions, with presets fixed before validation.

\begin{figure}[!htbp]
\centering
\includegraphics[width=\linewidth]{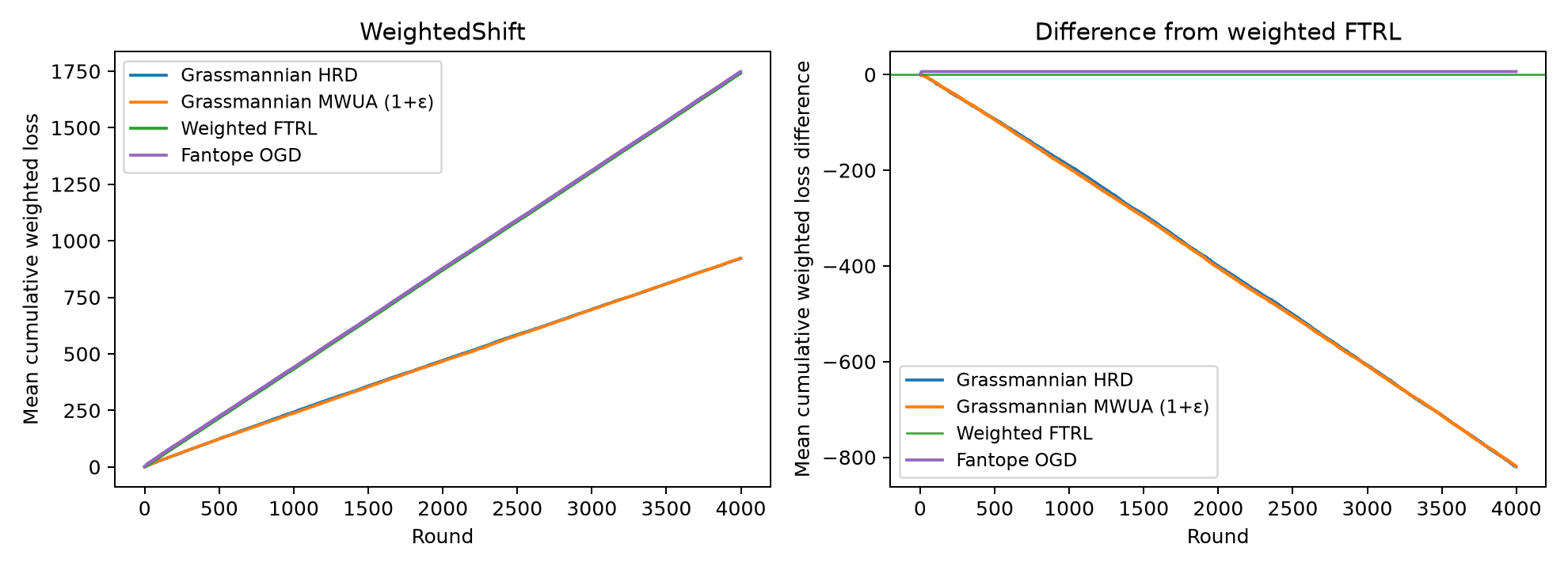}%
\caption{WeightedShift over five validation seeds. Left: mean cumulative
weighted fitting loss. Right: mean paired difference from approximate
weighted FTRL. Fantope uses its rounded rank-two subspace, including
its zero-reconstruction initialization prefix.}
\label{fig:weightedshift}
\end{figure}

\textbf{Results.}
\Cref{tab:weightedshift} and \Cref{fig:weightedshift} compare the methods on the
same weighted fitting loss. The two Grassmannian implementations attain
mean losses $921.96$ and $923.08$, versus $1742.10$ for approximate
weighted FTRL, reductions of $47.08\%$ and $47.01\%$. Rounded Fantope OGD
gives $1748.42$. These findings support the robustness of the Grassmannian algorithms under shifting data when compared to FTRL.

\begin{table}[!htb]
\centering
\small
\setlength{\tabcolsep}{5pt}
\renewcommand{\arraystretch}{1.2}
\begin{tabular}{lr}
\toprule
Method & Weighted fitting loss \\
\midrule
\Cref{alg:ghrd-regret-min} & $921.96\pm9.28$ \\
\Cref{alg:grass-local-MWUA} ($\varepsilon=1$) & $923.08\pm6.61$ \\
Weighted FTRL & $1742.10\pm9.33$ \\
Rounded Fantope OGD & $1748.42\pm8.84$ \\
\bottomrule
\end{tabular}
\caption{WeightedShift: cumulative weighted fitting loss, mean $\pm$
sample SD over five fresh validation seeds. Fantope uses its extracted
rank-two subspace, with zero reconstruction during its initialization
prefix. Only the weighted FTRL baseline is included.}
\label{tab:weightedshift}
\end{table}

\subsection{Weighted Subspace Tracking with Heterogeneous Noise}
\label{subsec:weighted-tracking}

\textbf{Signal and observation model.}
We generate $T=6000$ observations in $d=60$ with target rank $k=20$.
Ten signal directions remain fixed. Ten additional orthonormal directions
rotate in disjoint two-dimensional coordinate planes, each traversing
$180^\circ$ over the stream. Thus the instantaneous signal has rank $20$
and its coordinate support occupies $30$ coordinates over time. For each
round, let $U_t^\star$ contain the current signal directions, draw
$g_t\sim\mathcal N(0,I_{20})$, and form
  $s_t=\frac{U_t^\star g_t}{\|U_t^\star g_t\|_2}.$
Independent Gaussian observation noise has coordinate standard deviation
$0.025$ on the $30$ signal-supporting coordinates. On the other $30$
coordinates, the standard deviation is selected from $\{0.375,0.5\}$
independently for each block of $25$ rounds. Coordinate roles are randomly
permuted for each seed.

For noise standard deviations $\sigma_{t,j}$, set
  $w_{t,j}=\left(\frac{0.025}{\sigma_{t,j}}\right)^2,
  \qquad W_t=\operatorname{diag}(w_{t,1},\ldots,w_{t,60}), $ and $
  \qquad x_t=\frac{s_t+\xi_t}{\|s_t+\xi_t\|_2},$
where $\xi_{t,j}\sim\mathcal N(0,\sigma_{t,j}^2)$ independently.
The weights are $1$ on signal-supporting coordinates and $1/225$ or
$1/400$ on distractors. They encode inverse observation-noise variance
before final unit normalization; the uniform condition-number bound is
$\kappa\le400$. The stream and its weights are generated independently
of algorithm predictions.

\textbf{Comparison design.}
We compare constant-step rounded Fantope OGD with weighted and unweighted
FTRL and a Fantope ablation that ignores weights during updates. All are
scored by the same cumulative weighted rank-$20$ fitting loss, using
the actual weights and predictions made before updating. The main
weighted FTRL baseline uses three ALS iterations; an additional check
uses ten. The study uses five fresh validation seeds after development
screening. It evaluates the rounded Fantope predictions, rather than the
native fractional surrogate or the instantaneous generating subspace
as a regret comparator.

\begin{figure}[H]
\centering
\includegraphics[width=\linewidth]{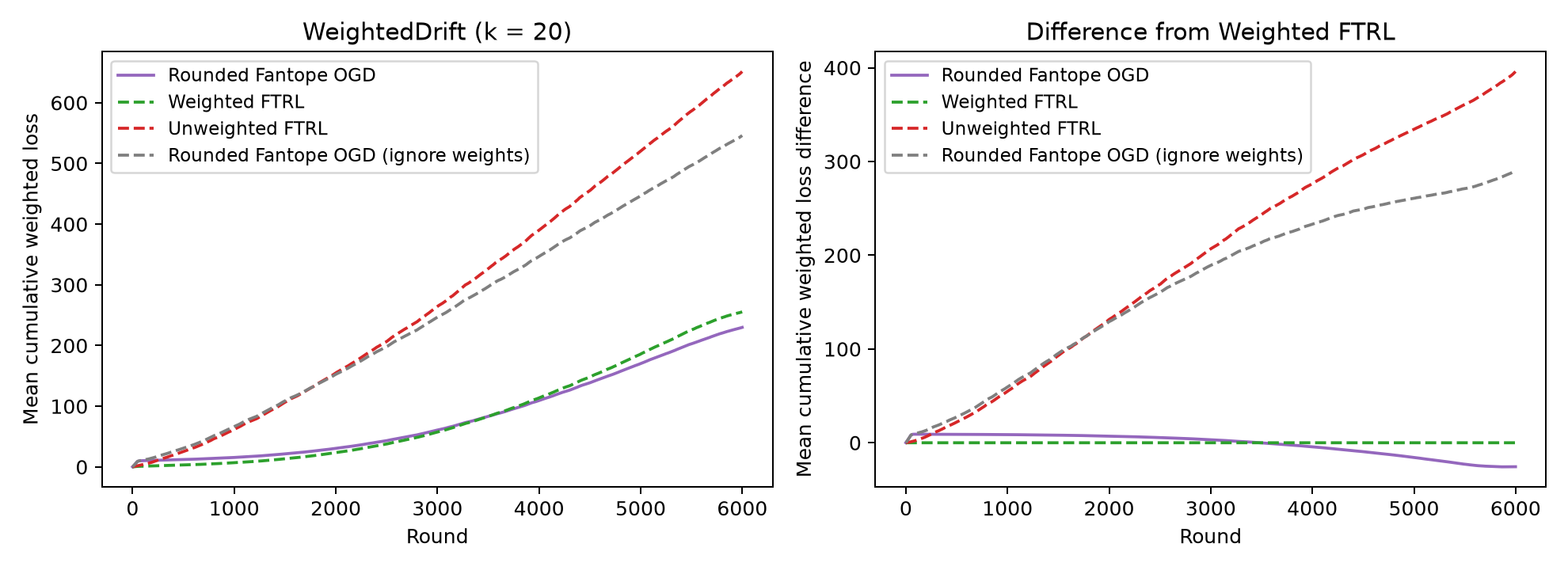}
\caption{Heteroscedastic weighted subspace tracking over five validation
seeds, with $d=60$, $k=20$, and $T=6000$. Left: mean cumulative weighted
fitting loss. Right: mean paired difference from approximate weighted
FTRL (three ALS iterations). Fantope uses its rounded rank-$20$ subspace.
All curves are scored using the actual coordinate weights.}
\label{fig:weighted-tracking}
\end{figure}

\textbf{Results.}
In \Cref{tab:weighted-tracking} and \Cref{fig:weighted-tracking}, rounded Fantope OGD
attains $229.93\pm2.86$, versus $255.50\pm2.86$ for approximate weighted
FTRL and $255.48\pm2.92$ with ten ALS iterations. Its mean loss is about
$10\%$ lower than the main weighted FTRL baseline, and it outperforms
both weighted FTRL variants on all five seeds. Ignoring weights during
Fantope updates gives $546.10\pm2.72$, while unweighted FTRL gives
$651.46\pm5.80$. The weighted Fantope loss is about $58\%$ lower than
its weight-ignoring ablation.

The Fantope ablation supports the value of coordinate reliability in this
construction. Its lower loss than accumulated-history weighted FTRL is
consistent with constant-step adaptation under drift, although these
comparisons do not necessarily isolate drift adaptation as the cause of the gap.
The moving signal does not alter the fixed comparator in the paper's
theorems, and the rounded fitting-loss results do not establish either
dynamic regret or Fantope's fractional-surrogate guarantee. 

\begin{table}[!htb]
\centering
\small
\setlength{\tabcolsep}{5pt}
\renewcommand{\arraystretch}{1.2}
\begin{tabular}{lr}
\toprule
Method & Weighted fitting loss \\
\midrule
Rounded Fantope OGD & $229.93\pm2.86$ \\
Weighted FTRL (3 ALS iterations) & $255.50\pm2.86$ \\
Weighted FTRL (10 ALS iterations) & $255.48\pm2.92$ \\
Unweighted FTRL & $651.46\pm5.80$ \\
Rounded Fantope OGD, ignoring weights & $546.10\pm2.72$ \\
\bottomrule
\end{tabular}
\caption{Weighted subspace tracking with $d=60$, $k=20$, and $T=6000$:
cumulative weighted fitting loss, mean $\pm$ sample SD over five fresh
validation seeds. Fantope is evaluated through its rounded rank-$20$
subspace. Weight-ignoring methods are scored using the actual weights.
Both weighted FTRL baselines use approximate ALS updates.}
\label{tab:weighted-tracking}
\end{table}

\subsection{Synthetic Streams with Unequal Coordinate Reliability}
\label{subsec:weighted-synthetic}

The MNIST and MovieLens runs use unweighted observations. We separately test how
coordinate-dependent weights affect the Fantope implementation on synthetic
streams with a known, fixed generating subspace. These recovery experiments
complement the three synthetic fitting-loss studies above. The implementation
updates a fractional Fantope matrix using the weighted projection surrogate;
the diagnostics measure the rank-$k$ eigenspace extracted before each update.
They do not measure standard WLRA fitting loss or the theorem's cumulative
loss for fractional decisions.

\textbf{Experimental setup.}
We generate $T=2000$ observations in $d=20$ with a rank-$k$ generating
subspace, $k=3$, and use ten independently generated streams (seeds
$0,\ldots,9$). We compare weighted Fantope OGD, the same update with
$W_t=I_d$, and streaming-SVD online PCA. Both Fantope variants use $m=d$ and
build an anchor from their first $20$ observations under the implemented rule.
For the weighted and unweighted variants, let $U_t$ consist of the top $k$
eigenvectors of $VP_tV^\top$ before the current update. We use this extracted
basis, and the streaming-SVD basis, to compute the following diagnostics:
\begin{align*}
 d_{\mathrm G}(U_t,U^\star)
 &=\sqrt{k-\|U_t^\top U^\star\|_F^2}
 =\frac{\|U_tU_t^\top-U^\star(U^\star)^\top\|_F}{\sqrt{2}}\\
 L_t^{\mathrm{clean}}&=\|s_t-U_tU_t^\top s_t\|_2^2,\\
 L_t^{\mathrm{weight}}&=\|W_t^{1/2}(x_t-U_tU_t^\top x_t)\|_2^2.
\end{align*}
Here $s_t$ is the clean signal in the normalized coordinate system and
$U^\star$ is the generating basis. The last metric is a weighted projection
residual of the extracted basis, not the fractional surrogate used to update
$P_t$, and not the standard WLRA fitting loss for nonuniform weights.
During the Fantope anchor prefix, $U_t$ is undefined: its chordal error is
omitted, while its two residual diagnostics use the zero reconstruction
(giving $\|s_t\|_2^2$ and $\|W_t^{1/2}x_t\|_2^2$). Streaming SVD makes a
prediction from its current basis on every round. All three methods observe
the same stream, but their initialization procedures differ.
\begin{figure}[H]
    \centering
    \includegraphics[width=\textwidth]{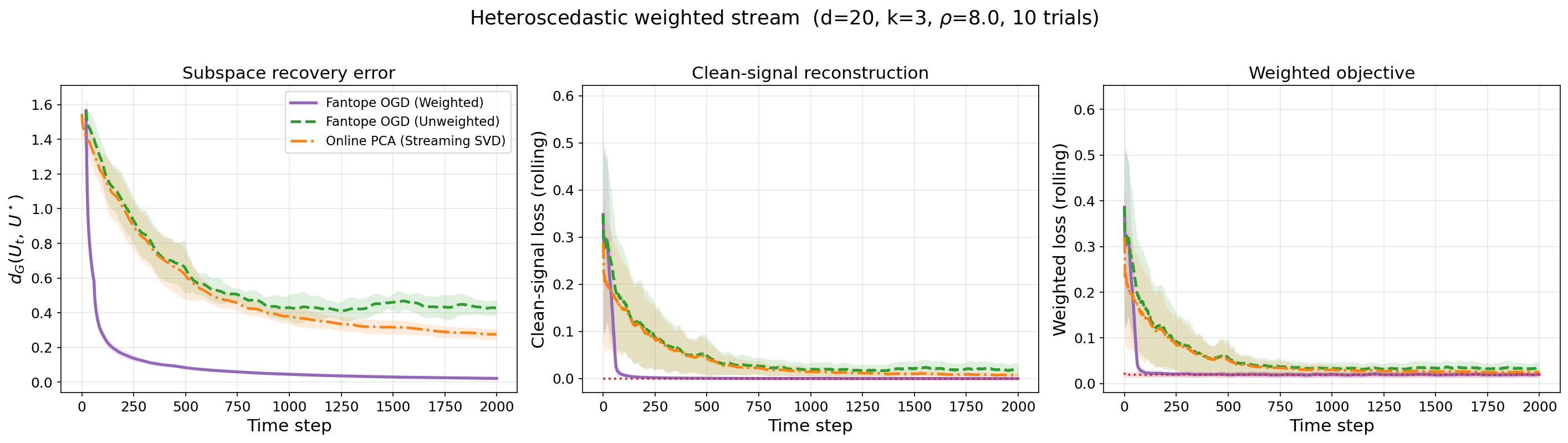}
    \caption{Heteroscedastic stream with $d=20$, $k=3$, and
    $\rho=8$, averaged over ten trials. The panels show chordal error
    relative to the generating subspace, rolling clean-signal
    reconstruction loss, and rolling weighted projection loss.
    Shaded regions show variation across trials.}
    \label{fig:weighted-hetero}
\end{figure}

\textbf{Heteroscedastic noise.}
The generating signal lies in the reliable coordinate block; another block
has larger noise. We assign inverse-variance weights, normalized so the
largest is one, and set $\rho=\sigma_{\mathrm{hi}}/\sigma_{\mathrm{lo}}$.
For the trajectory plot at $\rho=8$, $\sigma_{\mathrm{lo}}=0.10$. In
\Cref{fig:weighted-hetero}, the weighted Fantope variant has lower extracted
subspace error and clean-signal residual than the unweighted variant and
online PCA. The reported final subspace errors are approximately $0.02$,
$0.43$, and $0.28$, respectively, for this setup.

\begin{figure}[H]
    \centering
    \includegraphics[width=\textwidth]{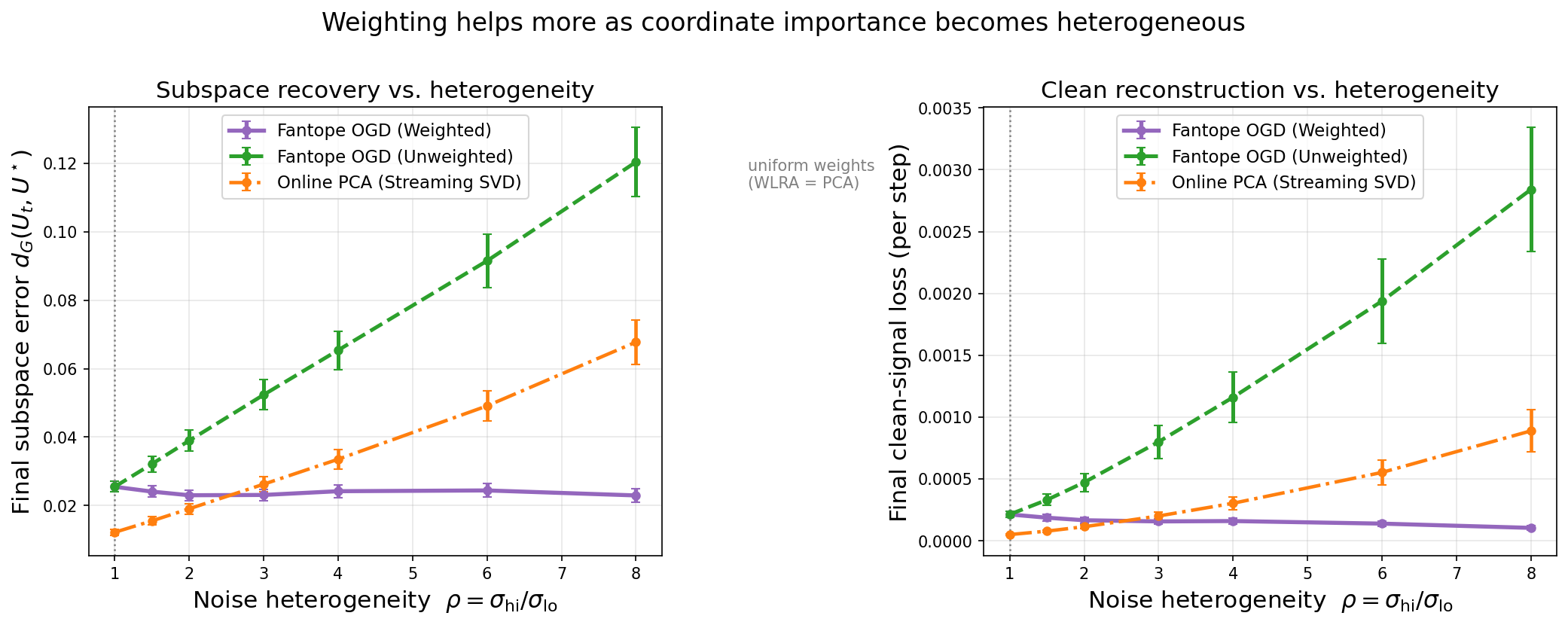}
    \caption{Final subspace error and clean-signal reconstruction loss
    as the noise standard-deviation ratio
    $\rho=\sigma_{\mathrm{hi}}/\sigma_{\mathrm{lo}}$ varies. Points show
    averages over ten independently generated streams, and error bars
    show variation across trials. At $\rho=1$, the coordinate weights
    are uniform.}
    \label{fig:weighted-sweep}
\end{figure}

We also vary $\rho$ using $\sigma_{\mathrm{lo}}=0.05$; hence this sweep does
not hold every noise setting from the trajectory experiment fixed. In
\Cref{fig:weighted-sweep}, the weighted and unweighted Fantope variants are
similar at $\rho=1$, whereas the weighted variant has a smaller extracted
subspace error as noise becomes more unequal in this tested range. Each
plotted value first averages the relevant diagnostic over the final $20\%$ of
one stream and then averages that value over ten streams.

\begin{figure}[H]
    \centering
    \includegraphics[width=\textwidth]{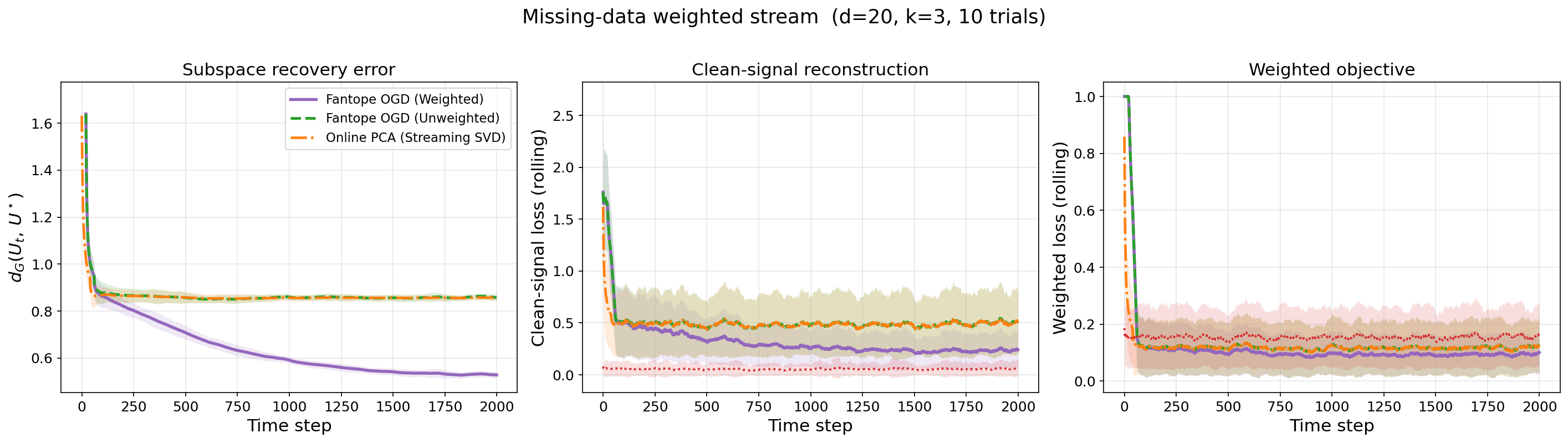}
    \caption{Missing-observation stream with $d=20$ and $k=3$,
    averaged over ten trials. Reliable and unreliable coordinates
    are observed with probabilities $0.95$ and $0.2$, respectively.
    The panels show chordal subspace error, rolling clean-signal
    reconstruction loss, and rolling weighted projection loss.}
    \label{fig:weighted-missing}
\end{figure}

\textbf{Missing observations.}
We observe reliable-block coordinates independently with probability $0.95$
and unreliable-block coordinates with probability $0.2$. Missing entries are
zero-filled and assigned diagonal weight $10^{-3}$, while observed entries
receive weight $1$. Weighted Fantope OGD uses these weights; the unweighted
variant and streaming SVD use the zero-filled observations without coordinate
weights. In \Cref{fig:weighted-missing}, the weighted variant has lower
extracted-basis chordal error and clean-signal residual in this setup. Its
reported final values are approximately $0.53$ and $0.23$, versus $0.86$ and
$0.50$ for online PCA.

These experiments illustrate the empirical role of the weighted Fantope
surrogate when coordinates have unequal reliability. Missing coordinates
receive weight $10^{-3}$, giving $\kappa=1000$ and satisfying strict
positivity. The limitations concern the extracted-subspace projection
metrics and anchor-prefix initialization; these diagnostics do not directly
establish standard WLRA regret or the fractional-surrogate guarantee.
\end{document}